\documentclass[11pt]{article}
\baselineskip
\pdfoutput=1
\usepackage{amssymb}
\usepackage{graphicx}
\usepackage{latexsym}
\usepackage[font=small,labelfont=bf]{caption}
\usepackage[left=1in, top=2.1cm]{geometry}
\usepackage{collref}
\usepackage{epsfig}
\usepackage{amsmath}
\usepackage{sidecap}
\usepackage{bbm}
\usepackage{color}
\usepackage{hyperref}
\usepackage{multirow}
\usepackage{epstopdf}
\usepackage{tabularx}
\usepackage{booktabs}
\usepackage{cite}
\DeclareMathOperator{\sign}{sign}
\DeclareMathOperator{\arccosh}{arcCosh}
\DeclareMathOperator{\Tr}{Tr}
\DeclareMathOperator{\tr}{Tr}
\DeclareMathOperator{\Real}{Re}

\DeclareMathOperator{\diag}{diag}

\newcommand{\N}{\mathbb{N}}

\newcommand{\C}{\mathbb{C}}
\newcommand{\SU}{\mathrm{SU}}
\newcommand{\SO}{\mathrm{SO}}
\newcommand{\Sp}{\mathrm{Sp}}
\newcommand{\U}{\mathrm{U}}

\newcommand{\ide}{\mathbbm{1}}
\newcommand{\Dc}{D_{\mbox{\tiny{\rm cont}}}}
\newcommand{\Dst}{D_{\mbox{\tiny{st}}}}

\newcommand{\transpose}[1]{{#1}^{{\tiny{T}}}}

\newcommand{\nconf}{n_{\mbox{\tiny{conf}}}}

\newcommand{\eq}{\begin{equation}}
\newcommand{\en}{\end{equation}}
\newcommand{\eqar}{\begin{eqnarray}}
\newcommand{\enar}{\end{eqnarray}}

\begin{document}

\begin{titlepage}
\renewcommand\thefootnote{\mbox{$\fnsymbol{footnote}$}}
\begin{center}
{\Large\bf Strong dynamics with matter in multiple representations: $\SU(4)$ gauge theory with fundamental and sextet fermions}
\end{center}
\vskip1.0cm
\centerline{Guido~Cossu$^{a,b,c}$, Luigi~Del~Debbio$^{a}$, Marco~Panero$^{d,e}$, and David~Preti$^{e}$}
\vskip1.0cm
\centerline{\sl $^a$Higgs Centre for Theoretical Physics, School of Physics \& Astronomy and 
University of Edinburgh}
\centerline{\sl EH9 3FD, Edinburgh, United Kingdom}
\vskip0.5cm
\centerline{\sl $^b$Alan Turing Institute}
\centerline{\sl NW1 2DB, London, United Kingdom}
\vskip0.5cm
\centerline{\sl $^c$Ascent Robotics}
\centerline{\sl 150-0012, Shibuja-ku, Tokyo, Japan}
\vskip0.5cm
\centerline{\sl $^d$Department of Physics, University of Turin and $^e$INFN, Turin}
\centerline{\sl Via Pietro Giuria 1, I-10125 Turin, Italy}
\vskip0.5cm
\begin{center}
{\sl  E-mail:} \hskip 1mm \\ \href{mailto:guido.cossu@ed.ac.uk}{{\tt guido.cossu@ed.ac.uk}},\\ \href{mailto:luigi.del.debbio@ed.ac.uk}{{\tt luigi.del.debbio@ed.ac.uk}},\\ \href{mailto:marco.panero@unito.it}{{\tt marco.panero@unito.it}}, \\ \href{mailto:david.preti@to.infn.it}{{\tt david.preti@to.infn.it}}
\end{center}
\vskip1.0cm
\begin{abstract}
\noindent We present a non-perturbative lattice study of $\SU(4)$ gauge theory with two flavors of fermions in the fundamental representation and two in the two-index antisymmetric representation: a theory closely related to a minimal partial-compositeness model for physics beyond the Standard Model, that was proposed by G.~Ferretti. We discuss the phase structure of the lattice theory and report results for various observables of interest, including the masses of states obtained from different combinations of valence fermions and the spectrum of the Dirac operator. Finally, we comment on the extension of this type of studies to other partial-compositeness models (including, in particular, one that was recently suggested by H.~Gertov \emph{et. al.}), which could admit lighter top-quark partners, highlighting some key features of our lattice simulation algorithm, that make it suitable for such generalizations.
\end{abstract}
\end{titlepage}

\section{Introduction}
\label{sec:int}

The experimental observation of a particle compatible with the Standard-Model Higgs boson at the Large Hadron Collider (LHC)~\cite{Aad:2012tfa, Chatrchyan:2012xdj} and the lack of evidence of any New Physics are putting very tight constraints on theories beyond the Standard Model. Nevertheless, for all its shortcomings, it remains very hard to imagine that the Standard Model be the correct description of Nature up to energies much higher than the TeV scale.

An unsatisfactory aspect of the Standard Model is the fact that, among its parameters, it features a large number of Yukawa couplings, which cannot be derived from first principles, and which give rise to broadly separated masses for the fermions. Also in the fermionic sector, it does not account for the experimental evidence of neutrino oscillations~\cite{Fukuda:1998mi, Ahmad:2002jz}, implying that these particles are not massless (although it can be easily extended to accommodate massive neutrinos just by adding a handful further parameters, at least if they are Dirac particles). Even more remarkably, the Standard Model fails spectacularly at predicting $95\%$ of the observed energy budget of the Universe~\cite{Ade:2015xua}, because it does not provide any explanation for Dark Matter or Dark Energy. Other unsatisfactory aspects of the Standard Model include the absence of unification of the gauge interactions, the ``strong-$\mathcal{CP}$ problem'' of quantum chromodynamics (QCD), and the fact that it does not include a proper quantum formulation of gravity. Finally, as is well known, one of the major theoretical puzzles in the Standard Model is the lightness of the Higgs boson: being the only fundamental scalar in the theory, its mass receives contributions (of opposite signs) from quantum fluctuations at all energies up to the Planck scale, but their sum turns out to be surprisingly (``unnaturally'') small in comparison to the latter scale; for a recent review, see ref.~\cite{Dawson:2018dcd}.

At least for the last of these issues, i.e. the ``naturalness problem'', supersymmetry provides a conceptually very elegant solution: the (nearly) perfect cancellation of the contributions to the Higgs boson mass from quantum fluctuations of different fields is a consequence of the (only softly broken) symmetry relating bosonic and fermionic species in the theory. From a formal point of view, it is also worth remarking that supersymmetry is the only type of symmetry combining spacetime and internal degrees of freedom in a non-trivial way~\cite{Wess:1974tw}, evading the Coleman-Mandula theorem~\cite{Coleman:1967ad}, and its experimental observation in elementary particle physics would be a major scientific discovery. In practice, however, its simplest realization in a framework compatible with the particle content of the Standard Model (the minimal supersymmetric Standard Model, MSSM), in which supersymmetry is necessarily broken, is far less aesthetically appealing: in particular, the MSSM has more than a hundred fundamental parameters, which, like their analogues in the non-supersymmetric Standard Model, cannot be derived from first principles. Despite the lack of predictive power due to this large number of free parameters, the MSSM (like most other New Physics models) generically predicts the existence of a host of new particles, including, in particular, four further Higgs particles, in addition to the Standard-Model one. All experimental searches in this direction so far, however, have come away empty-handed, indicating that supersymmetry, if exists, probably lies at an energy scale out of the reach of current accelerators.

Another popular theoretical framework that could explain the small mass of the Higgs boson is the one in which this particle is not considered as elementary, but rather as a composite state of some new, strongly coupled, elementary degrees of freedom, so that its lightness could be interpreted in terms of a Nambu-Goldstone mechanics---much like the pion, the lightest physical state in the QCD spectrum, is (nearly) massless because it can be interpreted as the Nambu-Goldstone boson associated with the breaking of chiral symmetry. This idea, dating back to more than thirty years ago~\cite{Kaplan:1983fs, Kaplan:1983sm}, has been studied in a large number of works~\cite{Hill:2002ap}: the simplest models realizing this scenario can be constrained by severe phenomenological tests~\cite{Peskin:1990zt} and have been falsified by now, but more refined implementations of this idea remain theoretically attractive and could still be viable candidates for New Physics beyond the Standard Model.

Partial-compositeness models, in which some additional fermionic fields from this new strongly coupled sector are linearly coupled to the top quark, are particularly appealing~\cite{Kaplan:1991dc, Contino:2010rs}. In this respect, a systematic, group-theoretical classification of the four-dimensional fermionic gauge theories providing an ultraviolet (UV) completion of composite-Higgs models was presented in ref.~\cite{Ferretti:2013kya}, imposing the requirements related to the existence of a custodial symmetry, and the presence of top-quark partners. The simplest UV-complete model of this type was then discussed in ref.~\cite{Ferretti:2014qta}: it is a theory based on local invariance under an $\SU(4)$ ``hypercolor'' group, featuring five flavors of massless Majorana fermions in the two-index antisymmetric representation, and three flavors of Dirac fermions in the fundamental representation of the gauge group. In the infrared limit, the formation of a condensate for the Majorana fermions in the two-index antisymmetric representation induces dynamical chiral-symmetry breaking according to the pattern $\SU(5) \to \SO(5)$, and a composite state, embodying the Standard-Model Higgs boson doublet, arises then from the $\SU(5)/\SO(5)$ coset~\cite{Dugan:1984hq}. The Dirac fermions in the fundamental representation bind with the Majorana fermions to form hypercolor-singlet states, that are interpreted as partners of the top quark, whereas the other massive Standard-Model fermions acquire their masses from quadratic coupling to the Higgs. This theory does not violate current experimental bounds e.g. on the decays of the $Z$ boson, and is a viable UV-complete model for New Physics.

Since the crucial phenomena of chiral-symmetry breaking and hypercolor confinement in the model proposed in ref.~\cite{Ferretti:2014qta} are intrinsically non-perturbative in nature, a theoretical study of this theory from first principles requires lattice calculations. For technical reasons (related to the computational cost of the fermionic-matter content of the theory), however, it is more convenient to study first a closely related theory, with two flavors of Dirac fermions in the two-index antisymmetric representation of $\SU(4)$, and two flavors of Dirac fermions in the fundamental representation of the gauge group. With such matter contents, the theory will undergo a different symmetry-breaking pattern (in particular, one which \emph{can not} accommodate a state with quantum numbers compatible with those of the Standard-Model Higgs boson); nevertheless, it remains an interesting theoretical laboratory, in which the main features of the actual model discussed in ref.~\cite{Ferretti:2014qta} can be studied, at least at a qualitative or semi-quantitative level.

With this motivation, in the present work we present a detailed numerical investigation of the $\SU(4)$ lattice gauge theory with two flavors of Dirac fermions in the two-index antisymmetric representation and two flavors of Dirac fermions in the fundamental representation of the gauge group, which recently has also been studied in a series of works~\cite{Ayyar:2017qdf, Ayyar:2018zuk, Ayyar:2018ppa, Ayyar:2019exp}. The structure of this article is the following: in section~\ref{sec:pheno}, we review the main features of the Ferretti model, in section \ref{sec:rmt}, we analyze in detail the symmetries of the Dirac operator (both in the continuum and in various lattice discretizations) in the two-index antisymmetric representation, and their implications for the spectrum supported by random matrix theory expectations. Next, in section~\ref{sec:alg} we discuss the features of a hybrid Monte~Carlo algorithm working with fermions in different representations, and in section~\ref{sec:meas} we present our results, both as algorithmic checks and as first exploratory steps into the theory under consideration. Section~\ref{sec:extension} deals with the generalization of this type of studies to non-minimal partial. The concluding section~\ref{sec:concl} presents a summary of this work, while the appendices~\ref{app:Notation}, \ref{app:commutator_proofs}, and~\ref{app:HMC_forces} respectively include our conventions for notations, the detailed proofs of some identities discussed in section~\ref{sec:rmt}. and technical details about our hybrid Monte~Carlo algorithm.
  
\section{Overview of the model}
\label{sec:pheno}
Let us briefly review the model described in ref.~\cite{Ferretti:2014qta} which we refer to as ``Ferretti model''. The UV completion is a gauge theory with $G_{\rm HC}=\SU(4)$ ``hypercolor'' gauge group, coupled to five Weyl fermions $\psi^I_{mn}$ in the two-index antisymmetric representation of the hypercolor group (i.e. the dimension $\bf{6}$ representation, that, in the following, we also call ``sextet'' representation: for a summary of group and group-representation properties, see, for instance, ref.~\cite[appendix]{Mykkanen:2012ri}) and three Dirac fermions written in terms of pair of Weyl fermions $\chi^a_m$,$\bar{\chi}^{a'}_m$ in the fundamental representation of the hypercolor group. 
Hence, in the field definition the indices $I,a,a'$ run over the flavor and read respectively $I=1,\dots,5$, whereas $a,a'=1,\dots,3$; on the other hand, $m,n = 1,\dots,4$ denote hypercolor indices. The global internal symmetry of the theory is 
\begin{gather}
    G_{\rm F}=\SU(5) \times \SU(3) \times \SU(3)' \times \U(1)_X \times \U(1)' \, . 
\end{gather}
The charges of the various fields are listed in table~\ref{TAB:SU4content}. 
\begin{table}[h!]
    \centering
    \begin{tabular}{c|c|c|c|c|c|c|}
       \multicolumn{1}{c}{} & \multicolumn{1}{c}{$G_{\rm HC}$}& \multicolumn{5}{c}{$G_{\rm F}$}\\
       \multicolumn{1}{c}{} & \multicolumn{1}{c}{$\overbrace{\phantom{aaaaa}}$}& \multicolumn{5}{c}{$\overbrace{\phantom{aaaaaaaaaaaaaaaaaaaaaaaaaaaaaaaaaa}}$}\\
        & $\SU(4)$ & $\SU(5) $& $\SU(3)$ & $\SU(3)'$ & $\U(1)_X$ & $\U(1)'$  \\
        \hline
      $\psi$ & $\mathbf{6}$ & $\mathbf{5}$ & $\mathbf{1}$ & $\mathbf{1}$ & $0$ & $-1$ \\
      $\chi $ &$\mathbf{4}$ & $\mathbf{1}$ & $\mathbf{3}$ & $\mathbf{1}$ & $-1/3$ & $5/3$\\
      $\tilde\chi $ & $\bar{\mathbf{4}}$ & $\mathbf{1}$ & $\mathbf{1}$ & $\bar{\mathbf{3}}$ & $1/3$ & $5/3$ \\
      \hline
    \end{tabular}
    \caption{\small $G_{\rm HC}$ is the hypercolor gauge group and $G_{\rm F}$ the global symmetry group before symmetry breaking.}\label{TAB:SU4content}
  \end{table}
  
The symmetry-breaking pattern of the model can be described as
\begin{gather}
    G_{\rm F}/H_{\rm F} = \left ( \frac{\SU(5)}{\SO(5)}\right) \times \left ( \frac{\SU(3)\times\SU(3)'}{\SU(3)_c} \right ) \times \left ( \frac{\U(1)\times \U(1)'}{\U(1)_X} \right )\, ,
\end{gather}
and is realized by the bilinear fermionic condensates $\langle \epsilon^{mnpq}\psi^I_{mn}\psi^J_{pq}\rangle \propto \delta^{IJ}$, and $\langle \bar{\chi}^{a'}_m \chi^a_m \rangle \propto \delta^{a' a}$. The symmetry-breaking pattern $G_{\rm F}/H_{\rm F}$ is compatible with a custodial symmetry, described by the $G_{\rm cus}$ group, such that $H_{\rm F} \supset G_{\rm cus} \supset G_{\rm SM}$, with $G_{\rm cus}=\SU(3)_c\times\SU(2)_L \times \SU(2)_R \times \U(1)_X$ and $G_{\rm SM} = \SU(3)_c \times \SU(2)_L \times \U(1)_Y$ is the Standard Model gauge group. More in detail, the electroweak gauge group $\SU(2)_L \times U(1)_Y$ is embedded in the unbroken $\SO(5)$, by considering the subgroup $\SO(4) \simeq \SU(2)_L\times \SU(2)_R$, identifying $\U(1)_R$ as the subgroup of $\SU(2)_R$ generated by the third generator $T^3_R$, and setting the hypercharge $Y=T^3_R + X$. The $14$ Nambu-Goldstone bosons in the $\SU(5)/\SO(5)$ coset can be classified according to their SM $\SU(2)_L \times U(1)_R$ charges as: 
\begin{gather}
{\bf 14} \to {\bf 1}_0 + {\bf 2}_{\pm 1/2} + {\bf 3}_0 + {\bf 3}_{\pm 1} = (\eta, H ,\phi_0,\phi_{\pm})\, ,
\end{gather}
where the field $H$ can be interpreted as the Higgs field. Indeed this field is a doublet under $\SU(2)_L$ and can therefore be written as a two-component complex field $H=(H_+,H_0)$. The spin-$1/2$ states appear as a triplet of the hypercolor theory, and are natural candidates to play the r\^ole of top-quark partners: in the effective field theory description of the low-energy dynamics, the latter are introduced as a Dirac fermion field $\Psi$ transforming in the $({\bf 5},{\bf 3})_{2/3}$ of $H_{\rm F}$. 
Such a field can be realized within the Standard Model, by decomposing the $({\bf 5},{\bf 3})_{2/3}$ multiplet as
\begin{gather}
    ({\bf 5},{\bf 3})_{2/3} \to ({\bf 3},{\bf 2})_{7/6} + ({\bf 3},{\bf 2})_{1/6} + ({\bf 3},{\bf 1})_{2/3} \, ,
\end{gather}
where the numbers on the right-hand side label the irreducible representations of $G_{\rm SM}$. The Nambu-Goldstone bosons can be combined into a $\Pi$ field 
\begin{gather}
\Pi = H + H^{\dag} + \phi_0 + \phi_+ + \phi^{\dag}_+ ,
\end{gather}
from which one can define
\begin{gather}
\label{Sigma_definition}
\Sigma = \exp \left ( \frac{i\Pi}{f} \right ),
\end{gather}
with $\Pi$ a real symmetric matrix. The matrix $\Sigma$ defined in eq.~(\ref{Sigma_definition}), however, transforms non-linearly under a transformation $g \in \SO(5)$, so it is convenient to consider the field $U=\Sigma\Sigma^T=\exp ( 2i \Pi / f)$, which transforms linearly: $U \to g U g^T$.

The couplings to vector bosons are obtained from the chiral Lagrangian
\begin{gather}
   \mathcal{L}  \supset \frac{f^2}{16}\tr \left ( (D_{\mu}U)^{\dag}D^{\mu}U \right ) 
\end{gather}
where the derivative is promoted to the covariant derivative, i.e.
\begin{gather}
D_{\mu} U = \partial_{\mu} U - igW^a_{\mu}[T^a_L,U] - ig'B_{\mu}[T^3_R,U]\, . 
\end{gather}

The mass term for fermions is 
\begin{gather}
\mathcal{L} \supset M\bar{\Psi}\Psi + \lambda_q f \bar{\hat{q}}_L \Sigma \Psi_R + \lambda_t f \bar{\hat{t}}_R\Sigma^* \Psi_L\, ,
\end{gather}
where $\hat{q}_L$ and $\hat{t}_R$ are the spurionic embedding of the SM quarks in the ${\bf 5}$ and $\bar{\bf 5}$ representations of $\SU(5)$, respectively.

An important feature of such a model is the vacuum misalignment, which is responsible for electro-weak symmetry breaking. In particular, the SM fermionic couplings are responsible for negative contributions to the Coleman-Weinberg potential, which are necessary to generate a non-vanishing vacuum expectation value for the $H_0$ component. Following ref.~\cite{Ferretti:2014qta}, we set $H_0 = h/\sqrt{2}$, while all other fields are set to zero. Then, the coupling of the $h$ field to the SM gauge bosons and fermions reads
\begin{align}
\tr \left [  U(h)W_{\mu}U(h)^{\dag}W_{\mu} \right ] & = \frac{1}{2} [1+\cos(2h/f)]W^c_{\mu}W^c_{\mu}\, , \\
\bar{\hat{q}}_L U(h) \hat{t}_R + \bar{\hat{t}}_R U(h)^* \hat{q}_L & = \frac{1}{\sqrt{2}}\sin(2h/f)(\bar{t}_Lt_R + \bar{t}_R t_L)\, .
\end{align}
The potential can thus be parametrized by the two low-energy constants $\alpha$ and $\beta$ as 
\begin{gather}
V(g) \propto \alpha \cos\left(\frac{2h}{f}\right) - \beta \sin^2\left(\frac{2h}{f}\right)
\end{gather}
and a suitable electro-weak-breaking minimum can be obtained at $\cos ( 2 \langle h \rangle /f ) = -\alpha/(2\beta)$ for $|\alpha/\beta| \lesssim 2$. These two constants can be computed as described in ref.~\cite{Golterman:2015zwa}. In particular, one has
\begin{gather}
2\beta = - y^2 C_{\rm top} \, ,
\end{gather}
which, in principle can be computed on the lattice, as well as all the other low-energy constants relevant for the infrared physics of the theory.

\section{Symmetries of the Dirac operator for fermions in the sextet representation}
\label{sec:rmt}

In order to construct the two-index antisymmetric representation for a generic $\SU(N)$ group, we introduce a set $\left\{ e^{(a,b)} \right\}$ of $N(N-1)/2$ real and antisymmetric matrices of size $N \times N$, which we label by strictly increasing pairs of indices $1 \le a < b \le N$. We sort the set of $(a,b)$ pairs starting from $a=1$ and $b=2$, and then increasing $b$ and letting $a$ run from $1$ to $b-1$, so that the sorted list of $(a,b)$ pairs reads $(1,2)$, $(1,3)$, $(2,3)$, $(1,4), (2,4)$, $(3,4)$, \dots , $(N-1,N)$. The elements of the $e^{(a,b)}$ matrices are defined by
\begin{equation}
\left( e^{(a,b)} \right)_{p q} = \frac{1}{\sqrt{2}} ( \delta_{p,a}\delta_{q,b} - \delta_{p,b}\delta_{q,a}).
\end{equation}
Then, given a generic element $u$ of the $\SU(N)$ group in the fundamental representation, the corresponding group element in the two-index antisymmetric representation is a complex-valued matrix of size $(N(N-1)/2) \times (N(N-1)/2)$, whose entries are defined as
\begin{equation}
\label{2as}
U_{(a,b) (c,d)} = \tr \left( e^{(a,b)\, {\tiny{T}}}\, u\, e^{(c,d)}\, \transpose{u} \right).
\end{equation}
It is then trivial to work out the explicit form of an arbitrary generator in the two-index antisymmetric representation, that we denote as $T^a_{\rm 2AS}$, for example, by defining an infinitesimal real parameter $\epsilon$, taking $u$ to be the group element infinitesimally close to the $N \times N$ identity matrix $u=\ide + i \epsilon t^a + O(\epsilon^2)$, and extracting the components of $T^a_{\rm 2AS}$ as the coefficients of the terms linear in $i \epsilon$ in the resulting expression for $U - \ide$ (where now $\ide$ denotes the $(N(N-1)/2) \times (N(N-1)/2)$ identity matrix).

For the purposes of this work, let us focus on the $\SU(4)$ group, whose generators in both the fundamental and in the two-index antisymmetric representation are reported in Appendix~\ref{app:Notation}. Consider now the totally antisymmetric four-index tensor $\epsilon_{abcd}$, with $\epsilon_{1\,2\,3\,4}=1$. Interpreting its indices pairwise, 
it can be used to construct a $6 \times 6$ matrix $W$, acting on the antisymmetric two-index representation of the $\SU(4)$ generators, whose rows (and columns) are labelled by the sorted $(a,b)$ (and $(c,d)$) pairs introduced above. The elements of $W$ are defined as
\begin{equation}
\label{W_entries}
W_{(a,b)\, (c,d)} = \epsilon_{abcd}.
\end{equation}
Remembering that, in our conventions, the indices from $1$ to $6$ of the antisymmetric two-index representation of 
$\SU(4)$ are associated with the sorted pairs $(1,2)$, $(1,3)$, $(2,3)$, $(1,4)$, $(2,4)$, $(3,4)$, in that order, $W$ takes the form  
\begin{equation}
\begin{small}
    \label{W}
W = \left(
\begin{array}{cccccc}
0 & 0 & 0 & 0 & 0 & 1 \\
0 & 0 & 0 & 0 & -1 & 0 \\
0 & 0 & 0 & 1 & 0 & 0 \\
0 & 0 & 1 & 0 & 0 & 0 \\
0 & -1 & 0 & 0 & 0 & 0 \\
1 & 0 & 0 & 0 & 0 & 0
\end{array}
\right).\end{small}
\end{equation}
Note that $W$ is real, symmetric, and unitary, hence it squares to the identity matrix. It is easy to check that all generators in the antisymmetric two-index representation of $\SU(4)$ satisfy
\begin{equation}
\label{W_and_T_a_star}
W^{-1} T^a_{\rm 2AS} W = -KT^a_{\rm 2AS},
\end{equation}
where $K$ denotes the complex-conjugation operator, defined by $K\alpha=\alpha^*$ for every $\alpha \in \C$.

Having set our notations for the generators of the $\SU(4)$ algebra in their antisymmetric two-index representation and the $\gamma$ matrices (for their explicit forms, see Appendix~\ref{app:Notation}), let us now introduce the Euclidean Dirac operator for a fermionic Dirac field of (real) bare mass $m$, transforming under the antisymmetric two-index color representation in a theory with $\SU(4)$ gauge symmetry.
In the continuum, the Euclidean Dirac operator reads:
\begin{equation}
\label{continuum_Dirac_operator}
\Dc = \gamma_\mu D_\mu + m = \gamma_\mu \left( \partial_\mu +ig A_\mu^a T^a_{\rm 2AS} \right) + m.
\end{equation}
Note that the kinetic ($\gamma_\mu D_\mu$) part of $\Dc$ is anti-Hermitian, whereas the mass term $m$ is Hermitian, so that, in general, $\Dc$ is neither Hermitian, nor anti-Hermitian. However, the anti-commutation relations $\left\{ \gamma_5, \gamma_\mu \right\}=0$ imply that the $\gamma_5 \Dc$ operator is Hermitian:
\begin{equation}
\label{continuum_gamma5_D_Hermiticity}
\left( \gamma_5 \Dc \right)^\dagger = \Dc^\dagger \gamma_5^\dagger = \left( -\gamma_\mu D_\mu +m \right)\gamma_5 = \gamma_5 \left( \gamma_\mu D_\mu +m \right) = \gamma_5 \Dc.
\end{equation}
Let us introduce the notion of ``anti-unitary operator'': given a complex Hilbert space $\mathcal{H}$ with inner product $\langle \dots , \dots \rangle$, an invertible mapping
\begin{equation}
\mathcal{U} : \mathcal{H} \to \mathcal{H}, \qquad \phi \to \mathcal{U}(\phi)
\end{equation}
(where $\phi$ denotes an arbitrary element of $\mathcal{H}$) is said to be ``anti-unitary'' if it is antilinear
\begin{equation}
\label{antilinearity}
\mathcal{U}\left(\alpha \phi + \beta \rho\right) = \alpha^\star \mathcal{U}(\phi) +\beta^\star \mathcal{U}(\rho)
\end{equation}
and satisfies
\begin{equation}
\label{antiunitarity}
\langle \mathcal{U}(\phi), \mathcal{U}(\rho) \rangle = \langle \phi, \rho \rangle^\star\end{equation}
for every $\phi$ and $\rho$ in $\mathcal{H}$ and for every $a$ and $b$ in $\C$. It is possible to prove that, given a unitary operator $\mathcal{V}$, the $\mathcal{V}K$ operator is anti-unitary, and that, conversely, every anti-unitary operator $\mathcal{U}$ can be written as
\begin{equation}
    \label{op_antiunit}
\mathcal{U} = \mathcal{V} K,
\end{equation}
where $\mathcal{V}$ is a unitary operator.

Let us introduce the charge conjugation $\mathcal{C}$ and define the operator $A$ as
\begin{equation}
\label{A_definition}
A = W \mathcal{C} \gamma_5 K.
\end{equation}
The combination $W \mathcal{C} \gamma_5$ appearing in eq.~(\ref{A_definition}) is a unitary operator, so it follows from eq.~(\ref{op_antiunit}) that $A$ is anti-unitary. Moreover, it is trivial to show that $A$ squares to minus the identity, because
\begin{equation}
\label{A_square}
A^2 = W \mathcal{C} \gamma_5 K W \mathcal{C} \gamma_5 K = W \mathcal{C} \gamma_5 W^\star \mathcal{C}^\star \gamma_5^\star = W^2 \mathcal{C}^2 \gamma_5^2 = - \ide,
\end{equation}
having used the facts that $W$ (acting only on the color indices) commutes with $\mathcal{C}$ and $\gamma_5$ (which act only on the spinor indices), that $W$, $\mathcal{C}$ and $\gamma_5$ are real, that $\mathcal{C}$ commutes with $\gamma_5$, and that $W$, $\gamma_5$ and $K$ square to the identity, whereas $\mathcal{C}$ squares to minus the identity.

From the aforementioned properties of $W$, $\mathcal{C}$, $\gamma_5$, and $A$ it also follows that 
\begin{eqnarray}
    [A, \gamma_5 \Dc] &=& 0.
\end{eqnarray}
A detailed proof of the above relation is provided in Appendix~\ref{app:commutator_proofs}.

Now, let us introduce the Dirac operator for the lattice discretization of the theory with fermions in the antisymmetric two-index representation, on a hypercubic spacetime lattice of spacing $a$. Its matrix elements in the Wilson formulation\footnote{The following arguments continue to hold also in the presence of improvement terms with the same symmetries.} are of the form
\begin{equation}
    \label{WilsonD}
    (D)_{x,y} =    \frac{1}{a} \left\{ \delta_{x,y} - \kappa \sum_{\mu=1}^4 \left[ (\ide - \gamma_\mu) U_\mu(x) \delta_{x+a\hat{\mu},y} +  (\ide + \gamma_\mu) U_\mu^{\dagger}(y) \delta_{x-a\hat{\mu},y}\right] \right\}  
\end{equation}
Thus, one also has:
\begin{equation}
\label{gamma5_WilsonD_elements}
(\gamma_5 D)_{x,y} = \frac{1}{a} \left\{ \gamma_5 \delta_{x,y} - \kappa \sum_{\mu=1}^4 \left[ (\gamma_5 - \gamma_5 \gamma_\mu) U_\mu(x) \delta_{x+a\hat{\mu},y} +  (\gamma_5 + \gamma_5 \gamma_\mu) U_\mu^\dagger(y) \delta_{x-a\hat{\mu},y}\right] \right\}.
\end{equation}
Defining the four, unitary, ``positive-shift'' operators $P_\mu$, that act trivially on all internal degrees of freedom and have real matrix elements between sites $x$ and $y$ given by
\begin{equation}
(P_\mu)_{x,y}=\delta_{x+a\hat{\mu},y},
\end{equation}
(while their inverses have elements $(P_\mu)^{-1}_{x,y}=\delta_{x-a\hat{\mu},y}$), and the local 
``positively-oriented, parallel-transporter'' operators $U_\mu$ (having matrix elements $U_\mu(x) \delta_{x,y}$ 
between sites $x$ and $y$), the Wilson Dirac operator can be written as
\begin{equation}
D = \frac{1}{a} \left\{ \ide - \kappa \sum_{\mu=1}^4 \left[ ( \ide - \gamma_\mu ) (U_\mu P_\mu) +  ( \ide + \gamma_\mu) (U_\mu P_\mu)^\dagger \right] \right\}.
\end{equation}
We now prove that the $\gamma_5 D$ lattice operator commutes with $A$, exactly as its continuum counterpart $\gamma_5 \Dc$ does. In order to prove this statement, we first study the transformation properties of the $U_\mu(x)$ link variables under complex conjugation. When $D$ is the Wilson Dirac operator for fermions in the antisymmetric two-index representation, a generic link variable $U_\mu(x)$ can be written as the exponential of $i$ times a linear combination with real coefficients (that is convenient to write as $agA_\mu^a(x)$) of the $T^a_{\rm 2AS}$ generators defined by eq.~(\ref{2as}) and explicitly reported in Appendix~\ref{app:Notation}:
\begin{equation}
U_\mu(x) = \exp \left( i a g A_\mu^a(x) T^a_{\rm 2AS} \right) = \sum_{n=0}^\infty \frac{(iag)^n}{n!} \left( A_\mu^a(x) T^a_{\rm 2AS} \right)^n.
\end{equation}
As a consequence:
\begin{equation}
K U_\mu(x) = U_\mu^\star(x) = \exp \left( -i a g A_\mu^a(x) T^a_{\rm 2AS} \right) = \sum_{n=0}^\infty \frac{(-iag)^n}{n!} \left( A_\mu^a(x) T^{a \star}_{\rm 2AS} \right)^n.
\end{equation}
Using eq.~(\ref{W_and_T_a_star}), the latter equation can be rewritten as
\begin{eqnarray}
\label{KUmu}
K U_\mu(x) &=& \sum_{n=0}^\infty \frac{(iag)^n}{n!} \left( - A_\mu^a(x) W^{-1} T^a_{\rm 2AS} W \right)^n = W^{-1} \left\{ \sum_{n=0}^\infty \frac{(iag)^n}{n!} \left( A_\mu^a(x) T^a_{\rm 2AS} \right)^n \right\} W \nonumber \\
&=& W^{-1} U_\mu(x) W.
\end{eqnarray}
From the transpose of the latter identity, using the fact that $W$ is symmetric and equal to its inverse, 
it follows that $U_\mu^\dagger(x)=W^{-1} \left( U_\mu^\dagger(x) \right)^\star W$, so 
that $K U_\mu^\dagger(x) = W^{-1} U_\mu^\dagger(x) W$. This is actually a trivial implication of eq.~(\ref{KUmu}), 
since $U_\mu(x)$ is an element of a unitary group, so $U_\mu^\dagger(x)=U_\mu^{-1}(x)$ is still a group element, in the same representation. 
As a consequence, the Wilson Dirac operator $D$ is such that 
\begin{eqnarray}
    \label{comm_Wilson}
    [A, \gamma_5 D] &=& 0.
\end{eqnarray} 
with $A^2=-\ide$: this is a property that the Wilson Dirac operator shares with the continuum Dirac operator. A detailed proof of eq.~(\ref{comm_Wilson}) is provided in Appendix~\ref{app:commutator_proofs}.

Eq.~(\ref{comm_Wilson}) implies that $\gamma_5 D$ can always be rewritten as a matrix whose elements are real quaternions of the form
\begin{equation}
q_0 + i \vec{\sigma} \cdot \vec{q},
\end{equation}
where $q_0$ and the components of $\vec{q}$ are real. As a consequence, the eigenvalues of $\gamma_5 D$ are pairwise-degenerate.

A second, more interesting, consequence is that certain universal features of the spectrum of 
eigenvalues of $\gamma_5 D$ can be described by the chiral Gau{\ss}ian symplectic ensemble (chSE) in random matrix theory---see ref.~\cite{Verbaarschot:2000dy} for an excellent review. In particular, the unfolded density of spacings $s$ between subsequent eigenvalues of $\gamma_5 D$ is expected to follow the Wigner surmise
\begin{equation}
\label{Wigner_surmise}
P(s) = N_\beta s^\beta \exp (- c_\beta s^2), \qquad \mbox{with}~N_\beta = 2 \frac{\Gamma^{\beta+1}\left(\frac{\beta}{2}+1\right)}{\Gamma^{\beta+2}\left(\frac{\beta+1}{2}\right)}, \quad c_\beta = \frac{\Gamma^{2}\left(\frac{\beta}{2}+1\right)}{\Gamma^{2}\left(\frac{\beta+1}{2}\right)},
\end{equation}
for the Dyson index corresponding to the symplectic ensemble, $\beta=4$. This is expected to hold for the unfolded density of spacings, in which the spacing between subsequent eigenvalues of $\gamma_5 D$ in one gauge-field configuration is rescaled by the local spectral density (obtained from an average over all configurations).

Note that, for the continuum and Wilson Dirac operators for fundamental $\SU(4)$ fermions, no global anti-unitary symmetry like the one encoded in eq.~(\ref{comm_Wilson}) exists. As a consequence, the unfolded density of spacings between eigenvalues of the Wilson Dirac operator for fermions in the fundamental representation of the $\SU(4)$ gauge group is expected to be described by the Wigner surmise for the chUE, i.e. by eq.~(\ref{Wigner_surmise}) with Dyson index $\beta=2$.

In passing, we also note that in the staggered formulation of the lattice Dirac operator $\Dst$ defined as
\begin{equation}
    \Dst = m \ide + \frac{1}{2a} \sum_{\mu=1}^4 \eta_\mu \left[ (U_\mu P_\mu) - (U_\mu P_\mu)^\dagger \right],
\end{equation}
where $\eta_\mu$ has elements between sites $x$ and $y$ defined as
\begin{equation}
    (\eta_\mu)_{x,y} = \delta_{x,y} (-1)^{\sum_{\nu<\mu} x_{\nu}},
\end{equation}
and where $\gamma_5$ is replaced by $\epsilon$, having elements
\begin{equation}
    \epsilon_{x,y} = \delta_{x,y} (-1)^{\sum_{\mu=1}^4 x_{\mu}},
\end{equation}
the analogue of $\gamma_5 \Dc$ is
\begin{equation}
    \epsilon \Dst = m \epsilon + \frac{1}{2a} \sum_{\mu=1}^4 \epsilon \eta_\mu \left[ (U_\mu P_\mu) - (U_\mu P_\mu)^\dagger \right].
\end{equation}
Now, consider the antiunitary operator
\begin{equation}
B = WK,
\end{equation}
which squares to the identity:
\begin{equation}
B^2 = (WK)^2 = W^2 = \ide.
\end{equation}
Analogously to the continuum and Wilson formulation, also in this case it is possible to show that 
\begin{equation}
\label{B_epsilonDst_commutator_vanishes}
[B, \epsilon \Dst] =0.
\end{equation}
As a consequence of the above relation (whose demonstration is provided in Appendix~\ref{app:commutator_proofs}), the staggered Dirac operator $\Dst$ is such that $\epsilon \Dst$ commutes with the antiunitary operator $B$, which squares to $\ide$. This property implies that $\epsilon \Dst$ can always be rewritten as a matrix whose elements are real, and that its universal spectral properties are described in terms of the chiral Gau{\ss}ian orthogonal ensemble (chOE) of random matrix theory. In particular, the unfolded eigenvalue spacing distribution is expected to be approximated by the Wigner surmise defined in eq.~(\ref{Wigner_surmise}), but with $\beta=1$, instead of $4$ (as for the continuum and Wilson Dirac operators). This difference between the anti-unitary symmetries of the staggered and the continuum Dirac operators is, in fact, unsurprising, given that a similar situation also occurs for the $\SU(2)$ gauge group~\cite{Verbaarschot:1994qf, Bruckmann:2008xr}, and the convergence of the staggered-spectrum results to the correct continuum limit occurs in a subtle way~\cite{Follana:2006zz}. The investigation of the restoration of the continuum symmetry in the staggered discretization of fermions in the sextet representation of the $\SU(4)$ group for $a \to 0$, however, would require a dedicated investigation and lies clearly beyond the scope of the present study.

\section{Lattice-calculation setup}
\label{sec:alg}

The simulations for this project were performed using a hybrid Monte~Carlo (HMC) algorithm implemented with the GRID lattice QCD library~\cite{Boyle:2016lbp}. As discussed above, given the exploratory nature of this work, we considered an approximation of the Ferretti model, reducing its matter content down to two fundamental and two sextet fermions. This prescription greatly simplifies the computational cost of the theory allowing to use a two-flavor pseudofermion action in the two representations. While this matter content does not yield the same symmetry breaking pattern as in the original model, this theory still represents an interesting theoretical framework with rich non-perturbative dynamics, analogous to the one proposed in ref.~\cite{Ferretti:2014qta}. Moreover, the simulation code we developed admits a rational hybrid Monte~Carlo implementation that allows to simulate any number of dynamical flavors in a generic representation.

As in a standard HMC algorithm, the main steps are the following:
\begin{enumerate}
    \item generation of pseudofermion fields from a heat-bath distribution;
    \item dynamical evolution of the gauge field configuration according to a fictitious Hamiltonian with randomly chosen initial momenta for each link;
    \item ``accept-reject'' step, to correct for possible errors in the integration of the equation of motion of the previous step.
\end{enumerate}
While several sophisticated techniques can considerably improve the algorithmic performance (in particular for the inversion of the Dirac operator), for the purposes of this work we limited ourself to a conjugate gradient solver, without preconditioning. Simulations of the theory on a much larger scale would, of course, require a careful optimization of the setup, which is not discussed in this work.

\subsection{HMC with fermions in multiple representations}
Earlier works addressing the simulation of gauge theories with dynamical fermions in a generic representation include ref.~\cite{DelDebbio:2008zf} and subsequent publications by the same authors. The exploration of models with fermions in multiple representations, however, is still ongoing, mainly due to the variety of different models and to the computational effort their study requires. Among recent works, we would like to mention ref.~\cite{Ayyar:2017qdf}, which presents a substantial set of numerical results (including continuum and chiral extrapolations for various physical quantities), which are obtained from an algorithm similar to the one presented in detail in this section.

Let us write the gauge link variable defined in a generic representation $R$ as 
\begin{gather}
    U_{\mu}(x)= \exp \left \{ i \alpha^a(x)_{\mu} T_R^a \right \}.  
\end{gather}
In order to define the molecular-dynamics (MD) force for both gauge and fermions, let us define the variation of the link variable as 
\begin{gather}
\delta(U_{\mu}(x)) = \delta(\alpha_{\mu}(x))U_{\mu}(x) \quad \quad \text{with} \quad \quad \delta(\alpha_{\mu}(x))=i\delta(\alpha_{\mu}^a(x))T_R^a
\end{gather}
and the conjugate momentum associated with each fundamental link    
\begin{gather}
\pi(x,\mu)=i\pi^a(x,\mu)T_F^a .
\end{gather}
Note that the full dependence on the representation is encoded into the generators $T_R$, meaning that the algebra weights (i.e. the gauge field components) are the same in any representation of the gauge group. Generalizing the same idea as in ref.~\cite{DelDebbio:2008zf}, we consider the following Hamiltonian:   
\begin{gather}
\label{eq:hamiltonian_tot}
    H = H_{\pi} + H_g + \sum_R^{N_{\rm rep}} H^R_{f},
\end{gather}
where 
\begin{itemize}
    \item $H_{\pi}$ is the kinetic contribution from the conjugate momenta associated with links in the fundamental representation,
    \item $H_g$ is the pure gauge contribution, also based on gauge fields in the fundamental representation, while
    \item $H^R_{f}$ is the fermionic contribution, which can be in an arbitrary representation. 
\end{itemize} 
In the present case, the latter is chosen to be $H^F_f + H^{\rm 2AS}_f$. These terms are formally defined in the same way, except that in $H^F_f$ the links and the pseudofermion fields are in the fundamental representation, while in $H^{\rm 2AS}_f$ the same links are ``promoted'' to the two-index antisymmetric representation by eq.~(\ref{2as}), and 
the pseudofermions are generated by a different heat-bath distribution.\footnote{Restricting to the $\SU(4)$ gauge group, for the fundamental representation links appear as $4\times4$ matrices and pseudofermions as $4$-component vectors, while for the two-index antisymmetric representation links are $6 \times 6$ matrices and pseudofermions are $6$-component vectors.}

\begin{figure}[t!] 
    \centering
    \includegraphics[width=0.45\textwidth]{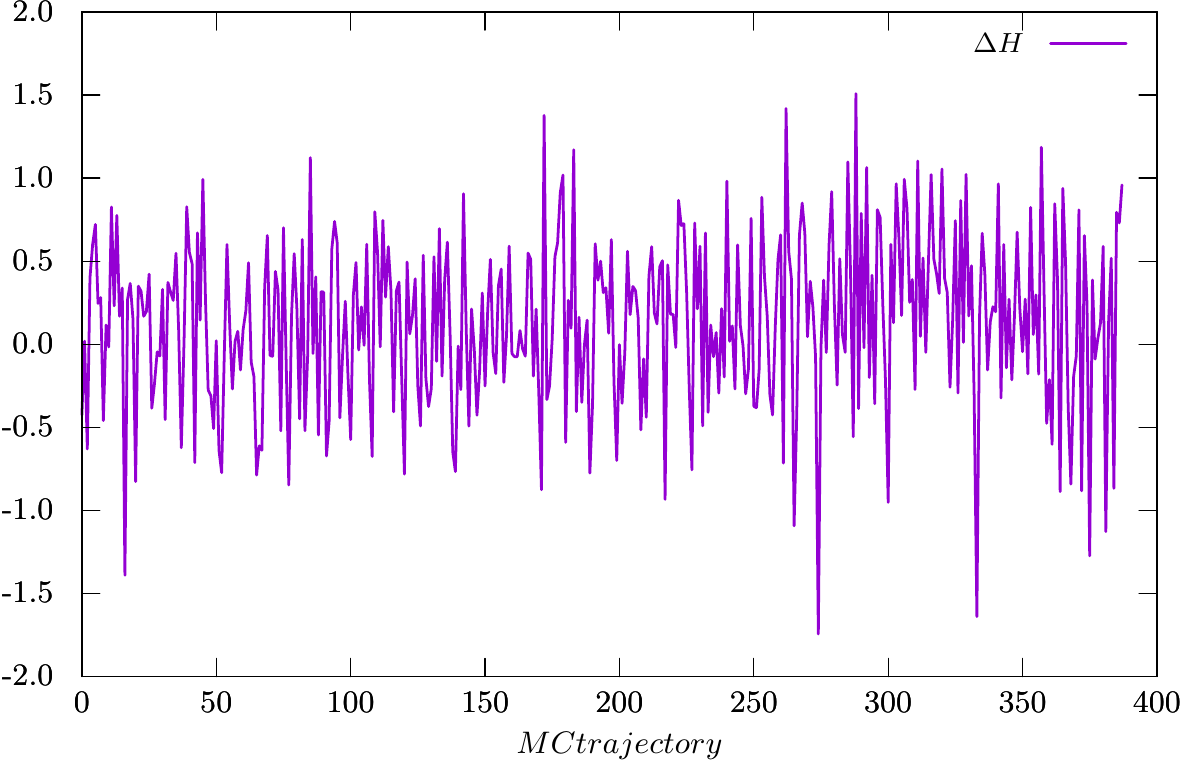}
    \includegraphics[width=0.45\textwidth]{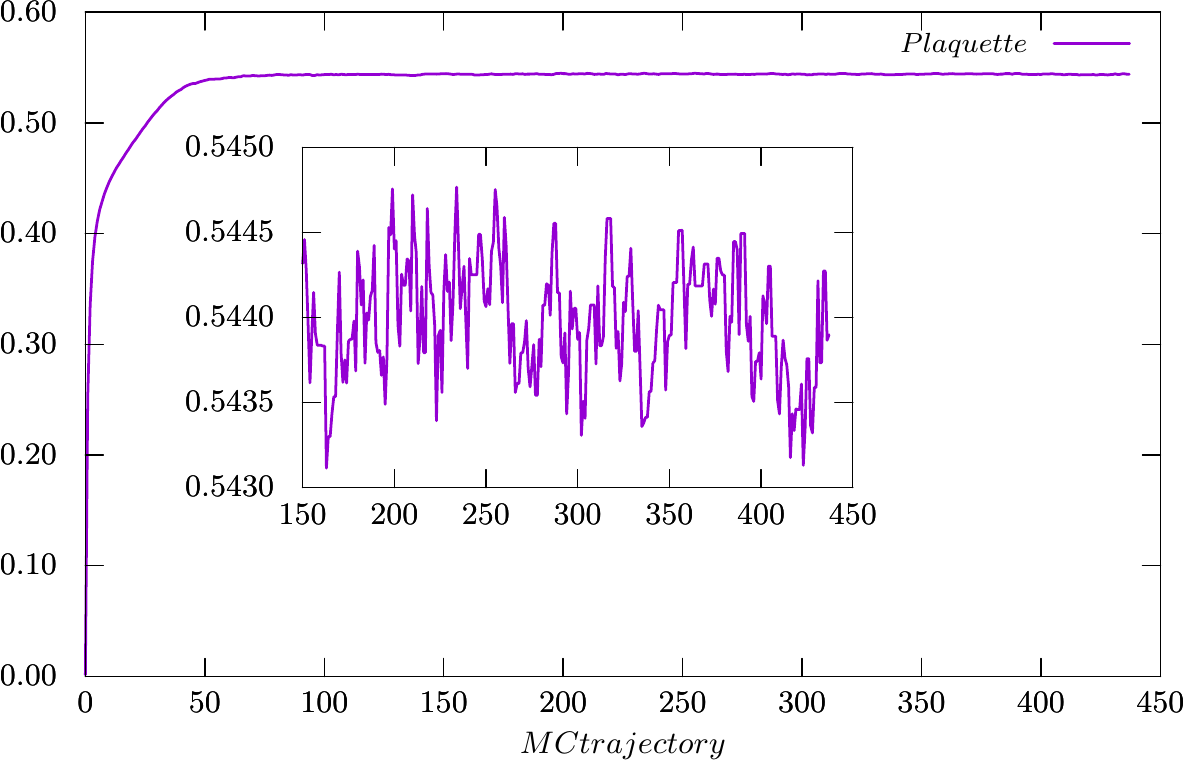}
    \caption{Plaquette (right) and $\Delta H$ (left) as a function of the MD trajectories for $\beta=10.0$ and $am_4=am_6=-0.55$.}
    \label{fig:dH} 
\end{figure}

More in detail, the terms appearing in eq.(\ref{eq:hamiltonian_tot}) are:
\begin{align}
    H_{\pi} & = \frac{1}{2} \sum_{x,\mu} \left ( \pi(x,\mu)),\pi(x,\mu) \right ) = \frac{1}{2}T_F \sum_{x,a,\mu} \pi^a(x,\mu)^2 \, ,\\ 
    H_g & = S_g = \frac{\beta}{N_c} \sum_x \sum_{\mu < \nu} \Real \tr \left (1 - \mathcal{P}_{\mu\nu}(x) \right ) \, , \\
    H^R_f & = S^R_f = \sum_x \phi^{\dag;R}(x)[D^{\dag;R}\, D^R]^{-1}\phi^R(x) \,  . \label{eq:Hrf}
\end{align}
We emphasize that the superscript $R$ means that eq.~(\ref{eq:Hrf}) holds for an arbitrary representation $R$ (Sanity checks of the alforithm are showed in Fig.~\ref{fig:dH}, while in Table~\ref{tab:quenched} the plaquette observable is showed to approach correctly the quenched results from Ref.~\cite{Bali:2013kia} in the infinite mass limit).\\
For this project we consider the discretized Dirac operator $D$ (dropping the superscript $R$) as the Wilson operator with the $\mathcal{O}(a)$ clover improvement 
and bare fermion mass $m$ (in unit of lattice spacing):
\begin{gather}
    D = D_{\rm Wilson} + D_{\rm clover},
\end{gather}
where the matrix element of the Wilson operator has been already introduced in eq.~(\ref{WilsonD}) and the improvement term reads
\begin{align}
    \label{WilsonD_elements}
[D_{\rm clover}^R]_{xy} &= \frac{ia}{2}c_{\rm sw}(g_0^2)\kappa^R \sum_{\mu,\nu} \tilde{F}^R_{\mu \nu}(x)\sigma_{\mu \nu}\delta_{xy} \, .
\end{align}
Let us express the fermion masses in terms of the hopping parameter  
\begin{align}
 \kappa=\frac{1}{2(am_0 + 4)} \quad \quad \text{and} \quad \quad am^R_q=am_0-am^R_{\rm crit}=\frac{1}{2}\left(\frac{1}{\kappa} - \frac{1}{\kappa_{\rm crit}^R} \right )
\end{align}
and $\sigma_{\mu \nu}=(i/2)[\gamma_{\mu},\gamma_{\nu}]$. We stress that the critical value of the bare mass (or, equivalently, of the hopping parameter) which corresponds to a vanishing renormalized mass depends on the representation.

The gauge part entering the fermionic $\mathcal{O}(a)$ improvement is given by
\begin{align}
\hat{F}_{\mu \nu}(x)= \frac{1}{8}\left [ \mathcal{Q}_{\mu \nu}(x) - \mathcal{Q}_{\nu \mu}(x) \right ] \quad \quad \text{with} \quad \mathcal{Q}_{\mu \nu}(x)=\mathcal{Q}^{\dag}_{\nu \mu}(x) 
\end{align}
where $\mathcal{Q}_{\mu \nu}$ is the clover combination of plaquettes around the point $x$, while the improvement coefficient $c_{\rm sw}$ can be expanded perturbatively as
\begin{gather}
c_{\rm sw}(g_0^2) = 1 + c^{(1);R}_{\rm sw}g_0^2 + \mathcal{O}(g_0^4) 
\end{gather}
In this work, $c_{\rm sw}(g_0^2)$ is fixed to its tree-level value.\footnote{For a calculation of the $\mathcal{O}(g_0^2)$ improvement terms, see ref.~\cite{Musberg:2013foa}.}
Denoting the molecular-dynamics integration time by $\tau$, the equations of motion can be written as 
\begin{align}
\frac{d}{d\tau}U_{\mu}(x) &= \pi(x,\mu)U_\mu(x)\, \\
\frac{d}{d\tau}\pi(x,\mu) &= F(x,\mu)\, ,
\end{align}
where the dynamics of the gauge link is governed by the force $F(x,\mu)$ which reads 
\begin{gather}
    F(x,\mu) = F_g(x,\mu) + \sum_R^{N_{\rm rep}}F^R_f(x,\mu).
\end{gather}
The force terms entering the HMC Hamilton equations are implicitly defined through  
\begin{align}
\delta S_g &= -(\delta \alpha,F_g)\\
\delta S_f &= -(\delta \alpha,F_f).
\end{align}

\begin{figure}[t!] 
    \centering
    \includegraphics[width=0.45\textwidth]{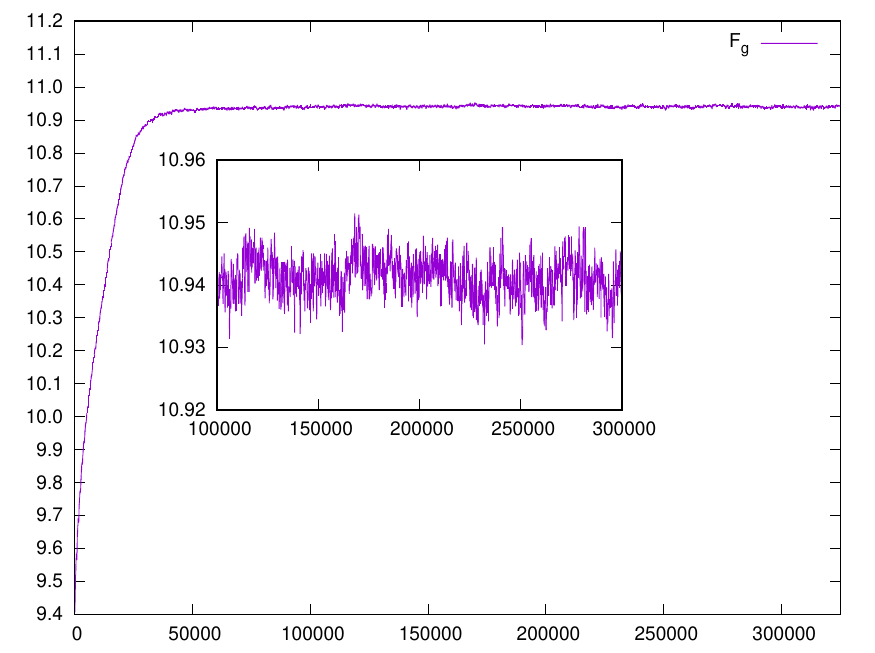}
    \includegraphics[width=0.45\textwidth]{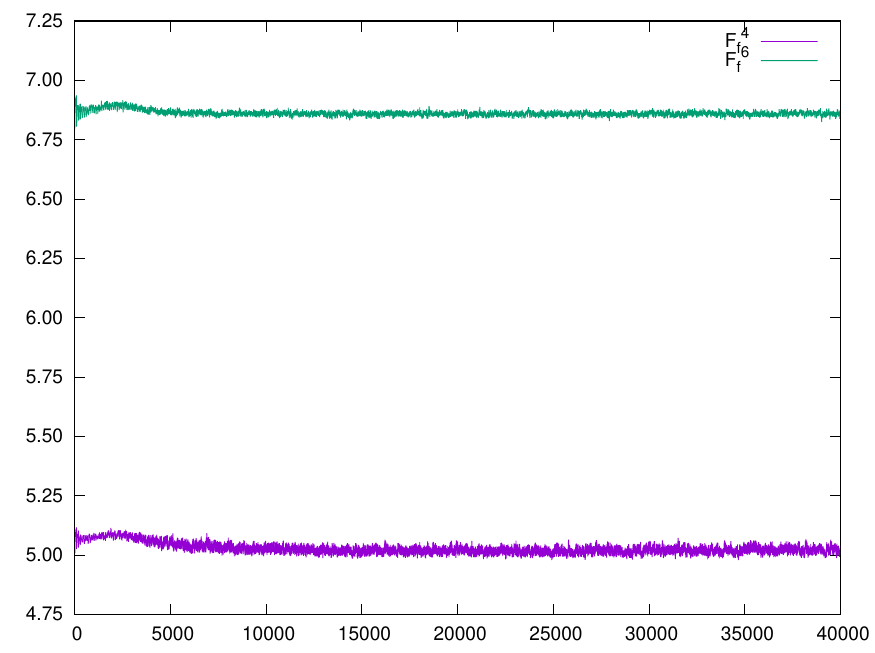}
     \caption{Left panel: Example of Monte~Carlo history of the gauge force. Right panel: Example of Monte~Carlo history of fermionic forces in the fundamental and two-index antisymmetric representations are displayed on the right. 
     Note that we are employing a multi-level integration scheme with a relative factor among fermionic and gauge forces of $8$.}
    \label{fig:MDForces} 
\end{figure}

\begin{SCfigure}
        \centering
        \includegraphics[width=0.45\textwidth]{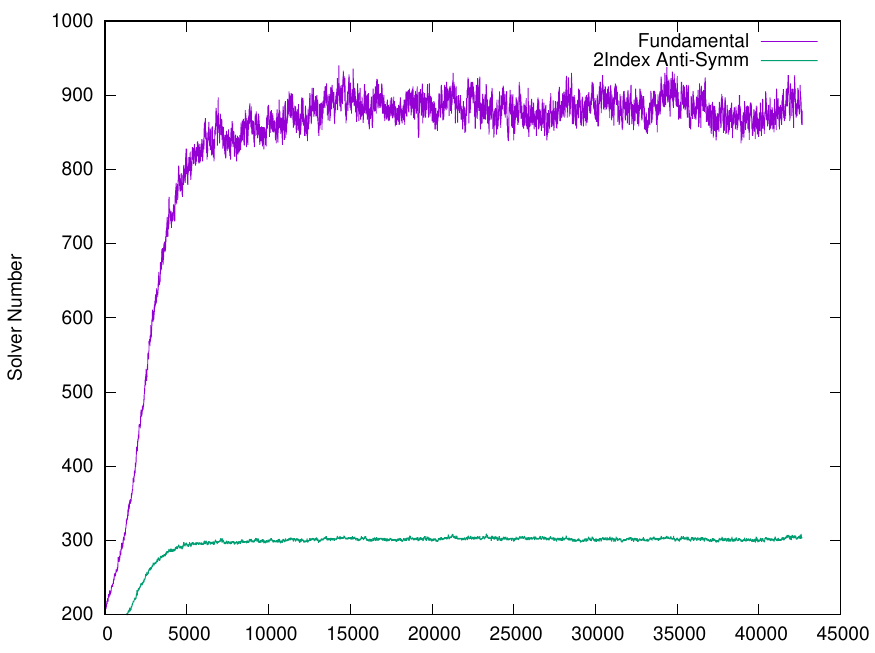}
     \caption{Comparison between conjugate-gradient-solver iterations for the fundamental and the two-index antisymmetric representation at degenerate bare fermion masses $am_4=am_6=-0.55$. As expected, the Dirac operator in the fundamental representation at a fixed value of the bare mass has smaller eigenvalues than the one in the sextet representation. The Dirac operator for the fundamental representation is then more ill-conditioned than its sextet counterpart and requires more solver iteration to reach the same residual.}
        \label{fig:MDSolvers} 
    \end{SCfigure}

The variation of the gauge action (which is defined in terms of fundamental link variables) reads 
\begin{gather}
    \delta S_g = \frac{\beta}{N} \sum_x \sum_{\mu,\nu} \delta \alpha_{\mu}^a(x) \Real \tr \left[ iT_F^a U_{\mu}(x)V^{\dagger}_{\mu}(x) \right]
\end{gather}    
with $V_{\mu}(x)$ the sum of the forward and backward staples around the link $U_{\mu}(x)$. 
The fermionic force is more intricate to derive. Dropping the $R$ superscript and the site index to avoid cumbersome notation, the fermionic action variation is
\begin{gather}
\delta S_f = \kappa \sum_x \phi^{\dag} (D^{\dag} D)^{-1} \delta (D^{\dag} D) (D^{\dag} D) \phi ;
\label{fermionic_variation}
\end{gather}
defining the modified pseudofermion fields 
\begin{align}
    X &=(D^{\dag}D)^{-1} \phi \\ 
    Y &=DX \, ,
\end{align}
eq.~(\ref{fermionic_variation}) simplifies to 
\begin{gather}
    \delta S_f = \kappa \sum_x \left( X^{\dag}(\delta D)Y + Y^{\dag}(\delta D)X \right) \, .
\label{fermionic_variation2}
\end{gather}
In the case of the Wilson action (i.e. $D=D_{\rm Wilson}$), from eq.~(\ref{fermionic_variation2}) we have
\begin{align}
\, \, \delta S_f^{\rm Wilson} = i \kappa \sum_x \tr \sum_{\mu}  \delta \alpha^a_{\mu}(x)T_R^a&  \left [  U_{\mu}(x)Y(x+\hat{\mu})X^{\dag}(x)(\ide + \gamma_{\mu})-Y(x)X^{\dag}(x+\hat{\mu})U^{\dag}_{\mu}(x)(\ide - \gamma_{\mu}) \right . \nonumber \\
                        &  \left . - X(x)Y^{\dag}(x+\hat{\mu})U^{\dag}_{\mu}(x)(\ide + \gamma_{\mu})+U_{\mu}(x)X(x+\hat{\mu})Y^{\dag}(x)(\ide - \gamma_{\mu}) \right ] \nonumber \\
                        = i \kappa \sum_x \sum_{\mu} \tr_{\rm color} & \left \{ \delta  \alpha^a_{\mu}(x)T_R^a \left [ U_{\mu}(x)W_{\mu}(x) - h.c. \right ] \right \}
\end{align}
with 
\begin{gather}
W_{\mu}(x) = \tr_{\rm spin} \left [ Y(x+\hat{\mu})X^{\dag}(x) (\ide + \gamma_{\mu}) + X(x+\hat{\mu})Y^{\dag}(x)(\ide - \gamma_{\mu}) \right ]\, .
\end{gather}
On the other hand, the variation of the clover term defined in eq.~(\ref{WilsonD_elements}) reads  
\begin{gather}
    \delta S_f^{\rm clov} = - \frac{i}{16} c_{\rm sw}(g_0^2)\kappa \sum_x \sum_{\mu,\nu}\tr_{\rm color} \left [ i \delta \alpha^a_{\mu}(x) T_R^a U_{\mu}(x)C_{\mu}(x) + i C^{\dag}_{\mu}(x)U^{\dag}_{\mu}(x)\delta \alpha^a_{\mu}(x)T_R^a \right].
\end{gather}
The derivation of the above formula is reported in appendix~\ref{app:HMC_forces}. All equations above hold for a generic representation $R$; the dependence on the representation only enters $\delta \alpha_{\mu}(x)$ (the forces for a specific set of bare parameters are displayed in Fig.~\ref{fig:MDForces} and the CG solver interations in Fig.~\ref{fig:MDSolvers}). In this way the MD equations can be easily generalized to arbitrary matter content, including for fields in multiple representations. 

	\begin{table}
    \centering
    \begin{tabular}{cllll}
     \toprule
     \rm{Ensemble} & $\beta$ & $am_{4}$ & $am_{6}$ & $\langle P \rangle$ \\
       \midrule
  $A27$               & $11.028$     & $-0.10$               & $-0.10$               & $0.5989(1)$    \\ 
  $A28$               & $11.028$     & $1.00$                & $1.00$                & $0.5867(1)$    \\ 
  $A29$               & $11.028$     & $100.00$              & $100.00$              & $0.5789(2)$    \\ 
  \midrule
  \rm{ref.~\cite{Bali:2013kia}} & $11.028$     & $-$               & $-$             & $0.578794(2)$    \\ 
  \bottomrule
  \end{tabular}
  \caption{Benchmark comparison of the value of the average plaquette in the infinitely-heavy-fermion limit to the quenched results for $\SU(4)$ from ref.~\cite{Bali:2013kia}}
  \label{tab:quenched}
\end{table}

\section{Observables}
\label{sec:meas}

Having discussed our results for elementary algorithmic quantities that can be monitored in the lattice simulation (such as plaquette expectation values, Monte~Carlo histories of forces involved in the HMC algorithm, etc.), in this section we present our results from the computation of Dirac spectra, as discussed in section~\ref{sec:rmt}, and of basic phenomenological observables which can be extracted from two-point correlation functions of ``meson-like'' and ``baryon-like'' states. With this terminology inspired by hadron physics, we respectively indicate hypercolor-singlet states built from a fermion and an anti-fermion, and from fermions only. We discuss in detail how these correlation functions can be built on the lattice and we provide numerical results for the ``meson-like'' states, while we restrict ourselves to a theoretical treatment of the ``baryon-like'' correlators, whose study is postponed to future works.\\
In particular, we focus on quantities providing information on the critical line of the theory. For this purpose, the best-suited quantities are the fermion masses defined in terms of the partially conserved axial current (PCAC), the masses of the ``pion-like'' states, that are interpreted as the pseudo-Nambu-Goldstone bosons associated with the breaking of chiral symmetry, and the distribution of the smallest eigenvalue of the Dirac operator, which is expected to get smaller when one approaches the critical line. Monitoring these quantities allows one to map out the phase structure of this lattice theory with clover-Wilson fermions in different representations, which is a necessary step before embarking in exhaustive investigation of its phenomenology---a task that we leave for future work.\\
Detailed results of the present study are shown in the figures and in the tables included here.\\

\subsection{Unfolded distributions of Dirac-spectrum spacings}
The analytical motivation for the study of unfolded distributions of the spacings between subsequent eigenvalues of the Dirac operator is discussed in detail in section~\ref{sec:rmt}.

In our computation we define the unfolded density of eigenvalue spacings as follows. First, we compute the spectrum of $\gamma_5 D$ on a set of $\nconf$ configurations, then we sort all non-degenerate eigenvalues in increasing order, labeling each of them by a positive integer that represents the eigenvalue position in the list. Then, the spacing $s$ between subsequent eigenvalues in each configuration $c$ is defined to be proportional to the difference of their positions in the list:
\begin{equation}
s = \frac{n_{i+1}^{(c)}-n_{i}^{(c)}}{\mathcal{N}},
\end{equation}
where the normalization factor $1/\mathcal{N}$ is fixed by requiring the average value of $s$ to be equal to one, and the unfolded density of spacings, also normalized to one, is obtained by dividing the real non-negative half-axis into intervals of width $\delta s$, and counting how many values of $s$ are found in a generic interval $[k\delta s, (k+1)\delta s]$, with $k \in \N$.

\begin{figure}[t!] 
    \centering
    \includegraphics[width=0.49\linewidth]{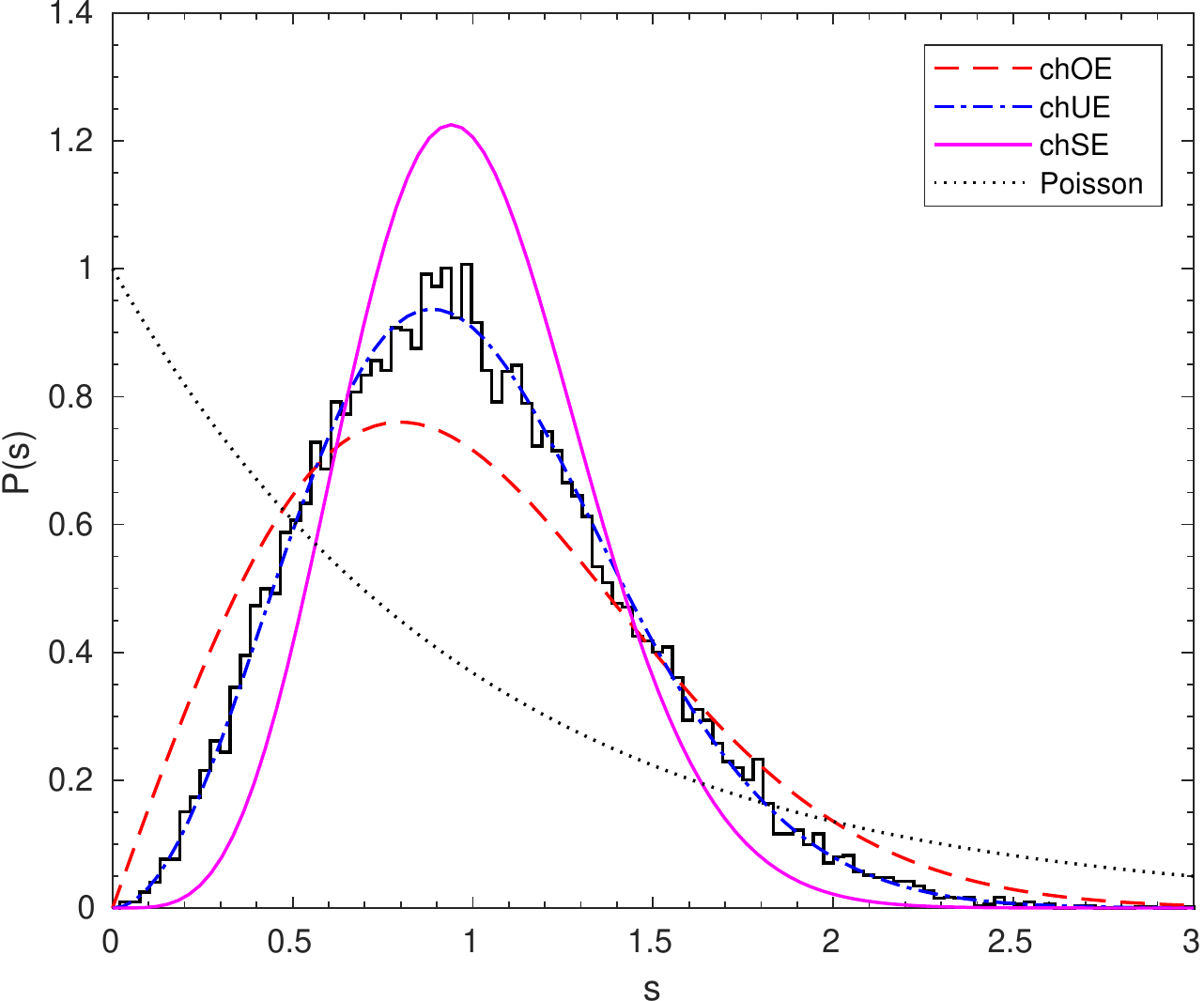}
    \includegraphics[width=0.49\linewidth]{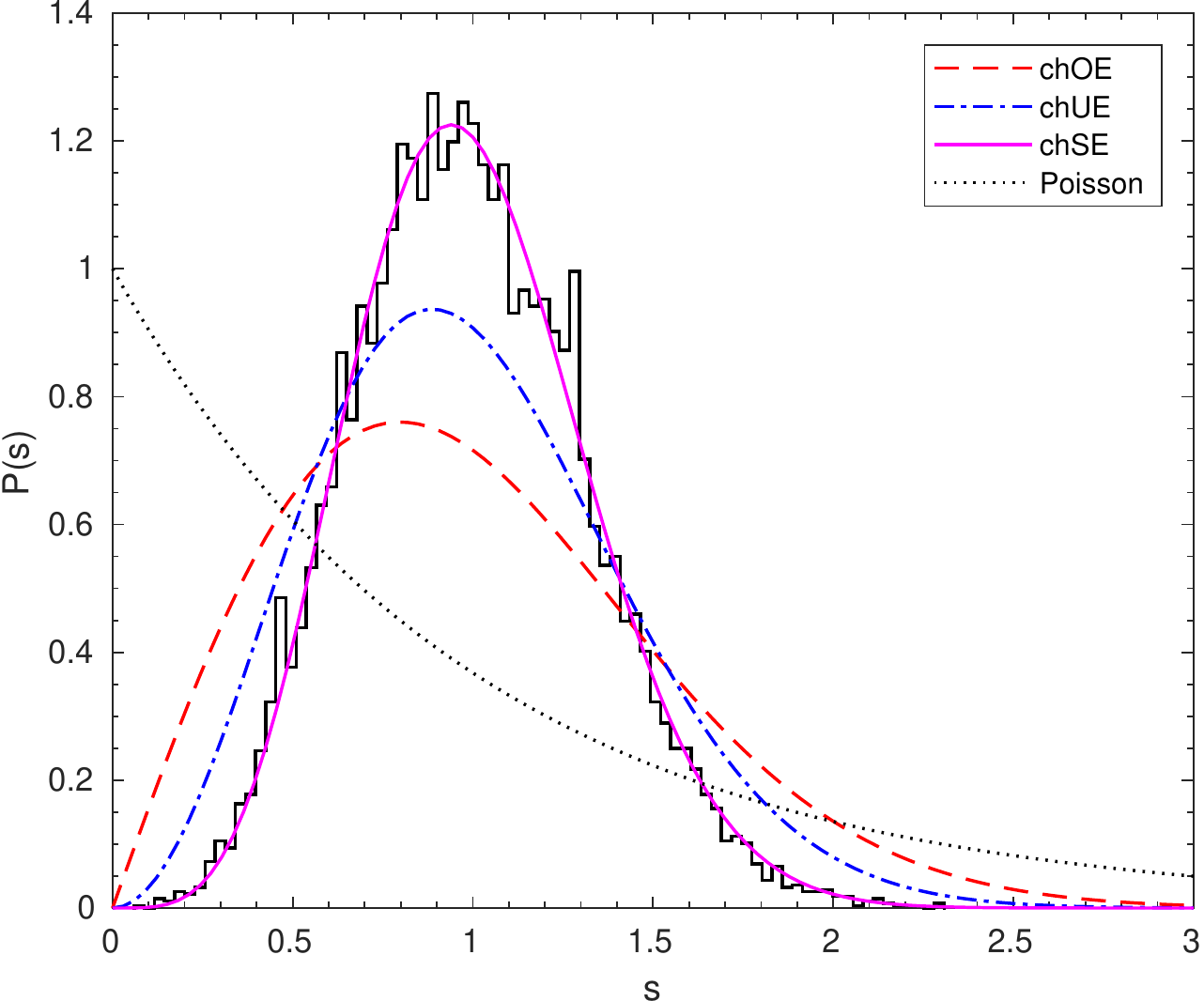}
    \caption{Unfolded density of eigenvalue spacings obtained from spectra of the Wilson Dirac operator with a clover improvement term, for the fundamental (left-hand-side panel) and sextet (right-hand-side panel) representations of the $\SU(4)$ gauge group. The results were obtained from quenched simulations on a lattice of size $(L/a)^4=4$ for $\beta=10.0$ and $am_4=am_6=-0.2$. The numerical results are consistent with the Wigner surmise according to the symmetries of the operator, i.e. the chUE curve for the fundamental representation, and the chSE for the two-index antisymmetric representation. For completeness, the plots also show the curves corresponding to the chOE, and the Poissonian distribution that would be expected, if, instead of the eigenvalues of an operator, one were considering the spacings between uniformly distributed random real numbers.}
\label{fig:chRMT_4x4x4x4} 
\end{figure}

Fig.~\ref{fig:chRMT_4x4x4x4} shows our results for the unfolded density of eigenvalue spacings that we extracted from an ensemble of spectra of the Wilson Dirac operator with clover improvement term, that we use in this work, which shares the same global anti-unitary symmetries as the continuum Dirac operator. The results were obtained from quenched simulations on a lattice of hypervolume $L^4=(4a)^4$ at $\beta=10.0$, with the choice $am_4=am_6=-0.2$.

The results for the fundamental representation (in the left panel of the figure) and for the two-index antisymmetric representation (right panel) are in complete agreement with the predictions from the Wigner surmise, eq.~(\ref{Wigner_surmise}), for the expected Dyson indices ($\beta=2$ for the fundamental representation and $\beta=4$ for the sextet representation). For completeness, we also show the analytical predictions for the chOE, as well as the exponential distribution that would correspond to the unfolded spacing obtained from uniformly distributed random real numbers.

\begin{figure}[t!] 
    \centering
    \includegraphics[width=0.47\linewidth]{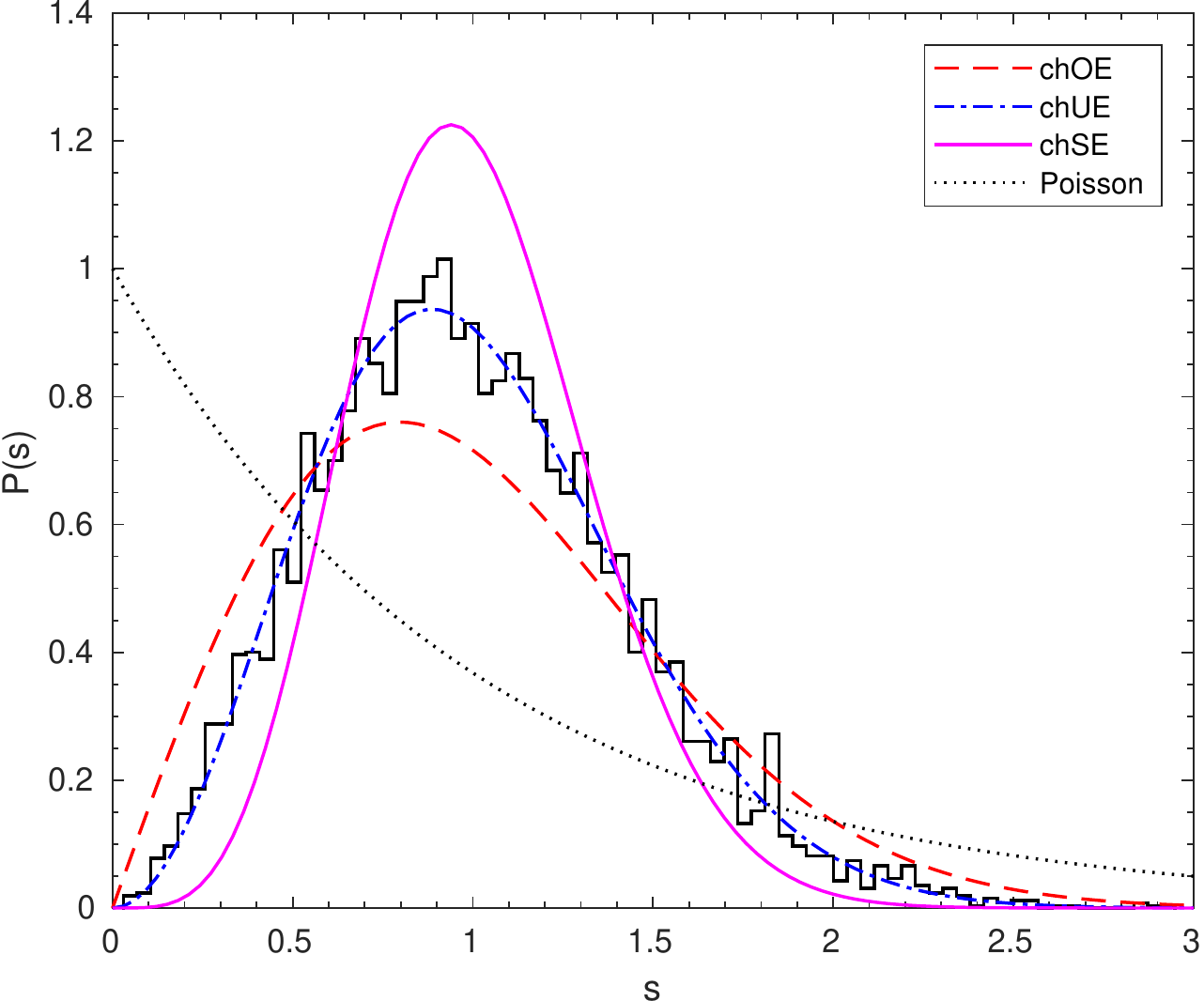}
    \includegraphics[width=0.49\linewidth]{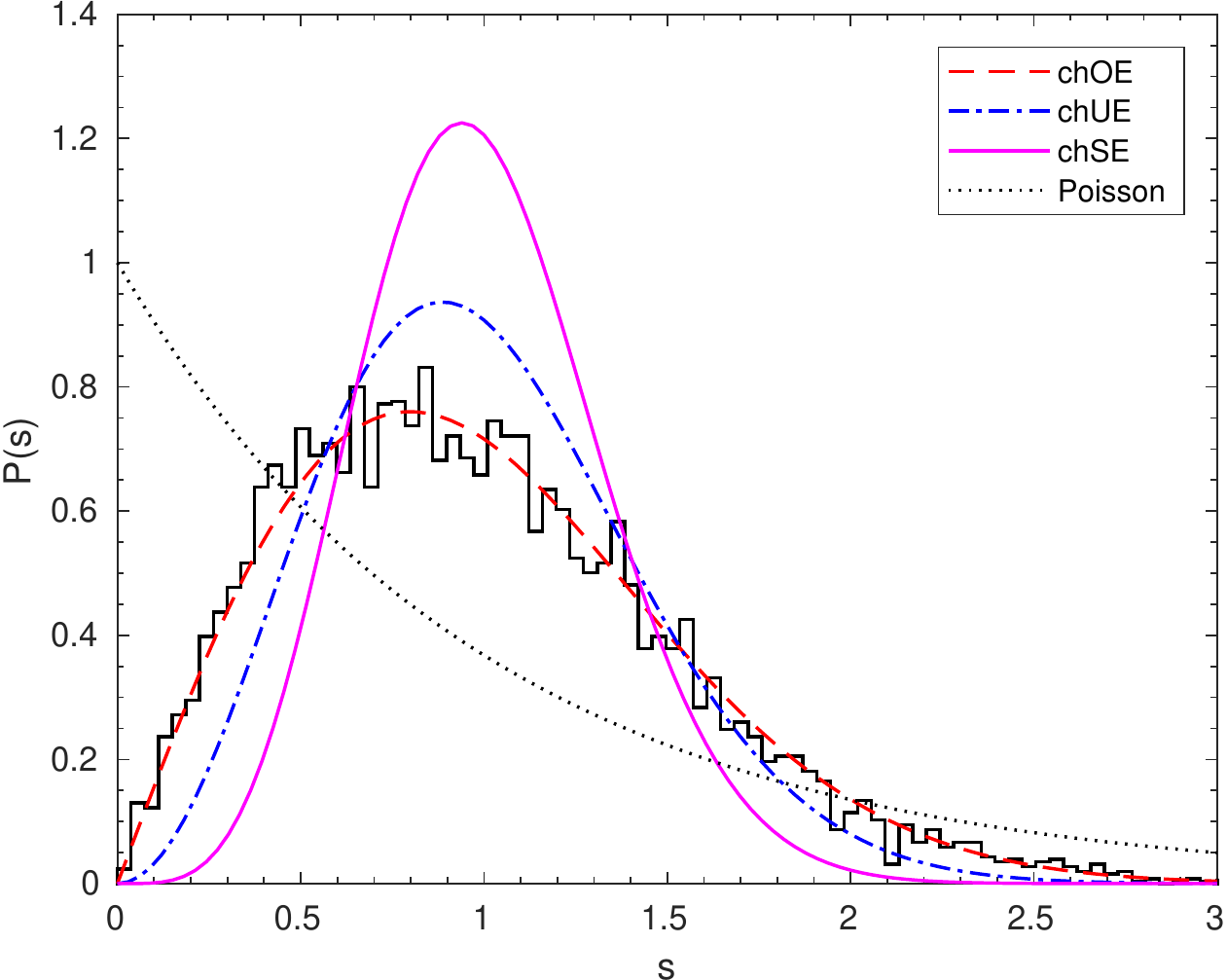}
    \caption{Same as in fig.~\ref{fig:chRMT_4x4x4x4}, but for the unfolded density of eigenvalue spacings obtained from spectra of the staggered Dirac operator in the fundamental (left-hand-side panel) and sextet (right-hand-side panel) representations of the $\SU(4)$ gauge group. The results were obtained from quenched simulations on a lattice of hypervolume $(L/a)^4=4$. Also in this case, the numerical data follow the analytical curves predicted by the Wigner surmise, according to the anti-unitary symmetries of the operator: the results for the fundamental representation are in very good agreement with the prediction for the chUE, and those for the sextet representation match the prediction for the chOE.}
\label{fig:chRMT_4x4x4x4_STAGGERED3} 
\end{figure}

Similarly, fig.~\ref{fig:chRMT_4x4x4x4_STAGGERED3} shows the results that we obtained from the same type of analysis, but using the staggered Dirac operator. As discussed in section~\ref{sec:rmt}, the global anti-unitary symmetries of this operator for fermions in the two-index antisymmetric representation are \emph{different} from those of the continuum Dirac operator, and this is confirmed by our numerical results shown in the right-hand side of this plot, which follow the chOE.

\subsection{Meson-like observables}
For a generic $\SU(N)$ gauge group and Lorentz structure $\Gamma_A$, the fermion bilinear with flavor indices $f_1$,$f_2$ in has the form 
\begin{gather}
O_A(x)=\overline{\psi}_{f_1}(x)\Gamma_A \psi_{f_2}(x)\, \quad \text{with} \quad \Gamma_A \in \left\{\ide,\gamma_5,\gamma_{\mu},\gamma_{\mu}\gamma_5,\sigma_{\mu\nu}\right\}, \quad \text{and} \quad f_1\neq f_2.
\end{gather}
where the fermion field $\psi$ can be in any representation of the gauge group. The two-point function can be written as
\begin{gather}
\langle O_A(x)\overline{O}_B(y)\rangle=\langle \overline{\psi}_{f_2}(x)\Gamma_A\psi_{f_1}(x)\overline{\psi}_{f_1}(y)\Gamma_B\psi_{f_2}(y)\rangle.
\end{gather}
Using Wick's contractions, the above equation can be rewritten as 
\begin{align}
\langle O_A(x)\overline{O}_B(y)\rangle & =  -\Gamma_A^{ij}\Gamma_B^{kl} \langle\psi_{f_1}^l(y)\overline{\psi}_{f_1}^i(x) \rangle \langle \psi_{f_2}^j(x)\bar{\psi}_{f_2}^k(y)\rangle \nonumber \\ 
& = - \tr [\Gamma_A S(x,y)_{f_2} \Gamma_B S(y,x)_{f_1}],
\end{align}
where $S$ denotes the fermion propagator in coordinate space. Its $\gamma_5$-Hermiticity $S^{\dagger}(y,x)=\gamma_5 S(x,y) \gamma_5$ implies 
\begin{gather}
\langle O_A(x)\overline{O}_B(y)\rangle = - \tr [\gamma_5\Gamma_A S(x,y)_{f_2} \Gamma_B \gamma_5 S^{\dagger}(x,y)_{f_1}].
\end{gather}
This structure holds for fermions in any representation. In fact for a generic representation $R$ we have $R\otimes\overline{R}=\ide \oplus \dots $, i.e. it is always possible to identify a hypercolor-singlet made of a fermion-antifermion pair.

\subsection{Baryon-like observables}
Let us refer to fermionic fields in the fundamental representation as $q^a_i(x)$, where $a=1,\dots, N$ is a hypercolor index while $i$ is a Dirac index, and to fields in the two-index antisymmetric representation as $\mathcal{Q}^{a b}_j(x)$ with spin $j$ and $a,b=1,\dots,N$. In order to avoid cumbersome notation we map the two-index into a single one $(a,b) \to \alpha=1,\dots,N(N-1)/2$ as discussed in section~\ref{sec:rmt}, i.e. by sorting the two-index pairs as $(1,2)$, $(1,3)$, $(2,3)$, $(1,4), (2,4)$, $(3,4)$, \dots , $(N-1,N)$.\\
It is a trivial consequence of group-representation theory that the minimum number of fermions in the fundamental representation of the $\SU(N)$ gauge group to construct a hypercolor-singlet state is $N$. In the current context, this corresponds to ``baryon-like'' states formed by four (fundamental) fermions, with a $qqqq$ structure.\footnote{Note that the analogy with the baryons of quantum chromodynamics is not complete: most notably, these $qqqq$ are bosons, rather than fermions.} Similarly, hypercolor-singlet states can also be built from three fermions in the two-index antisymmetric representation fermions $\mathcal{QQQ}$. A further, ``hybrid'' type of color-singlet states can be built by combining fermions in both representations, as in $qq\mathcal{Q}$. In the present work we restrict ourselves to the study of this three-fermion baryon, which, playing the r\^ole of the top-quark partner in the model under investigation, is particularly interesting. Such a state is often referred to as a ``chimera baryon''. The simplest interpolating operator for this state\footnote{Here we only consider a nucleon-like state, but the same analysis can be extended to states with different quantum numbers.} can be written as
\begin{gather}
O_{N_{\pm}}(x) = \epsilon_{abcd} P_{\pm} \Gamma_A q_a(x)( q_b^T(x) \Gamma_B \mathcal{Q}_{c d}(x)) \\
\overline{O}_{N_{\pm}}(x) = \epsilon_{abcd} (\overline{q}_a(x) \Gamma_B  \overline{\mathcal{Q}}_{bc}^T(x)) \overline{q}_d(x)  \Gamma_A P_{\pm}  
\end{gather} 
where $P_{\pm}=(\ide\pm\gamma_0)/2$ projects onto the desired isospin channel, and $(\Gamma_A,\Gamma_B)$ define the spin content of the baryon. For the channel with angular momentum and parity quantum numbers $J^P=1/2^+$, common choices are $(\Gamma_A,\Gamma_B) \in \left\{ (\ide,\mathcal{C}\gamma_5),(\gamma_5,\mathcal{C}),(\ide,i\gamma_0\mathcal{C}\gamma_5) \right\}$, where $\mathcal{C}=\gamma_0\gamma_2$ denotes the charge-conjugation matrix. The two-point contraction for these three-fermion objects can be written as
\begin{eqnarray}
\langle O_{N_{\pm}}(x) \overline{O}_{N_{\pm}}(y)\rangle &=& -\epsilon_{abcd} \, \epsilon_{a'b'c'd'} \,  [\overline{q}_a'^i(y) \Gamma_B^{ij}\overline{\mathcal{Q}}_{b'c'}^j(y)]\overline{q}_d'^k(y)\Gamma_A^{kl} \, P_{\pm}^{lm}\Gamma_A^{mn}q_{a}^n(x)(q_{b}^o(y)\Gamma_B^{op}\mathcal{Q}_{cd}^p(x)) \nonumber \\
&=& \epsilon_{abcd} \, \epsilon_{a'b'c'd'} \, \Gamma_B^{ij}(\Gamma_A P_{\pm}\Gamma_A)^{kn}\Gamma_B^{op} \mathcal{K}_{bcb'c'}^{pj}[S_{ad'}^{nk}S_{ba'}^{oi} - S_{aa'}^{ni}S_{bd'}^{ok}] \nonumber \\
&=& \epsilon_{abcd} \, \epsilon_{a'b'c'd'} \, \{ \tr[(\Gamma_A P_{\pm} \Gamma_A) S_{ad'}] \, \tr[ S_{ba'} (\Gamma_B \mathcal{K}_{bcb'c'} \Gamma_B^T)^T] + \nonumber \\
&& \qquad \qquad \qquad \! -\tr[(\Gamma_AP_{\pm}\Gamma_A)S_{aa'}(\Gamma_B\mathcal{K}_{bcb'c'}\Gamma_B^T)^TS_{bd'}] \},  
\label{eq:baryon_tr}
\end{eqnarray}
where $S^{ab}_{ij}$ is the fermionic propagator in the fundamental representation and $\mathcal{K}^{abcd}_{ij}$ is the one in the two-index antisymmetric representation, for the hypercolor indices $(a,b)$ and $(c,d)$.\\
By exchanging color indices, eq.~(\ref{eq:baryon_tr}) can be recast into the form
\begin{eqnarray}
    \langle O_{N_{\pm}}(x) \overline{O}_{N_{\pm}}(y)\rangle&=&\epsilon_{abcd} \, \epsilon_{a'b'c'd'} \, \{ \tr[(\Gamma_A P_{\pm} \Gamma_A) S_{aa'}] \, \tr[S_{bb'}(\Gamma_B \mathcal{K}_{cdc'd'} \Gamma_B^T)^T ] + \\ \nonumber
&& \qquad \qquad \qquad + \tr[(\Gamma_AP_{\pm}\Gamma_A)S_{aa'}(\Gamma_B\mathcal{K}_{bcb'c'}\Gamma_B^T)^TS_{dd'}] \}.
\label{eq:baryon_final}
\end{eqnarray}
Eq.~(\ref{eq:baryon_final}) is formally identical to the one relevant for the nucleon in quantum chromodynamics, where all quark fields are in the fundamental representation of the $\SU(3)$ gauge group. It is well known that two-point functions interpolating baryonic states are typically very noisy, compared to the ones for mesons: this is mostly due to the presence of an additional propagator with respect to the mesonic case. To extract a clear signal from these correlation functions, several techniques have been developed (see ref.~\cite{Leinweber:2004it} and references therein). In the theory investigated in this work, the problem is expected to be even more severe, due to the presence of the propagators in the sextet representation, hence we postpone a systematic study of baryon spectroscopy to a future publication.

\begin{figure}[t!] 
    \centering
	\includegraphics[width=0.50\linewidth]{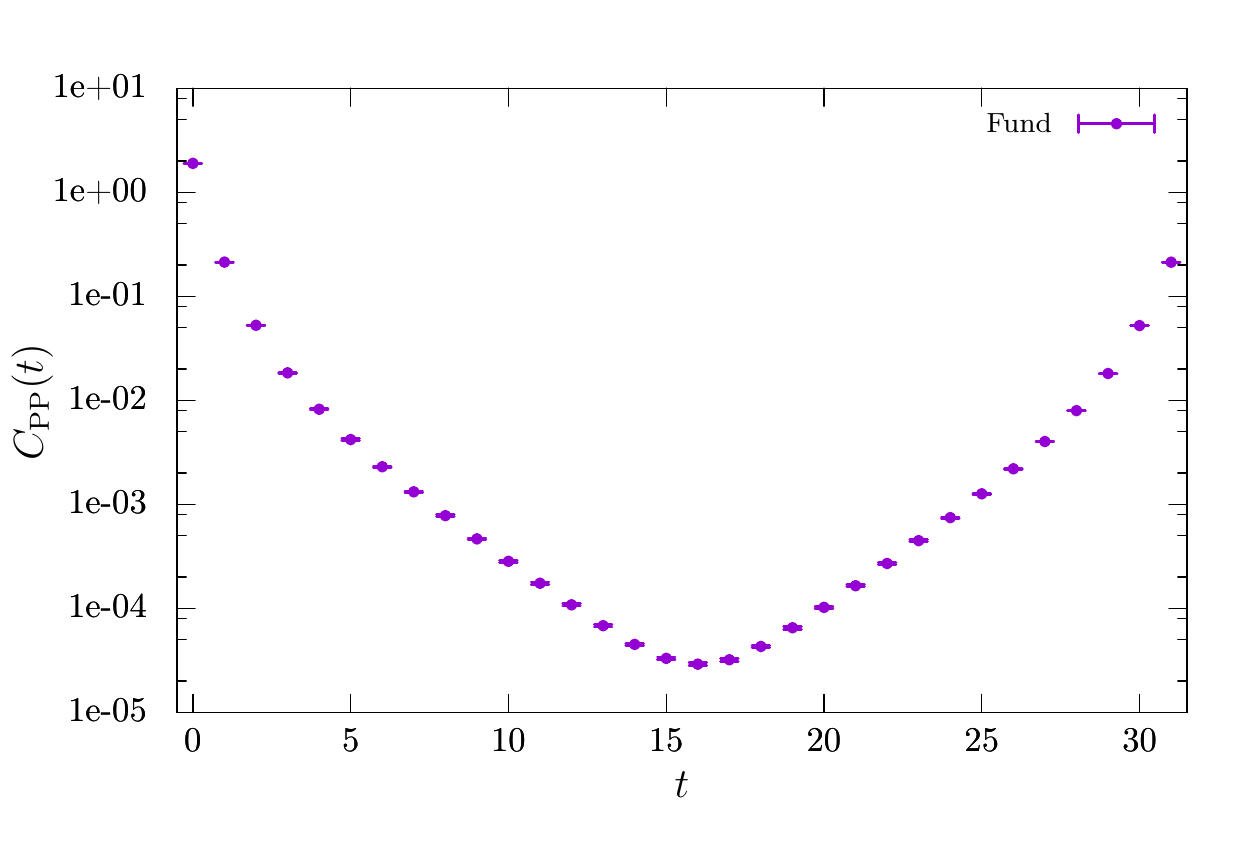}
	\hspace{-0.5cm}
	\includegraphics[width=0.50\linewidth]{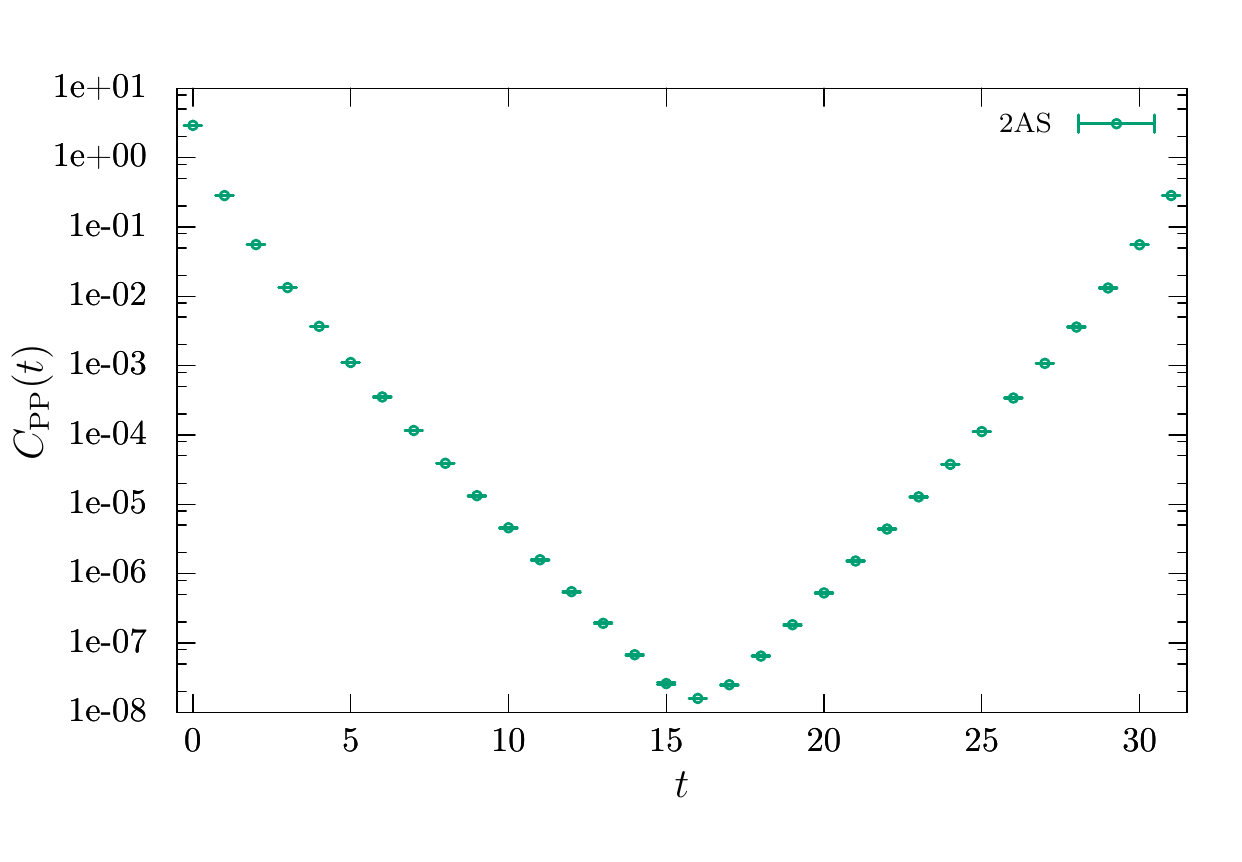} 
	\includegraphics[width=0.48\linewidth]{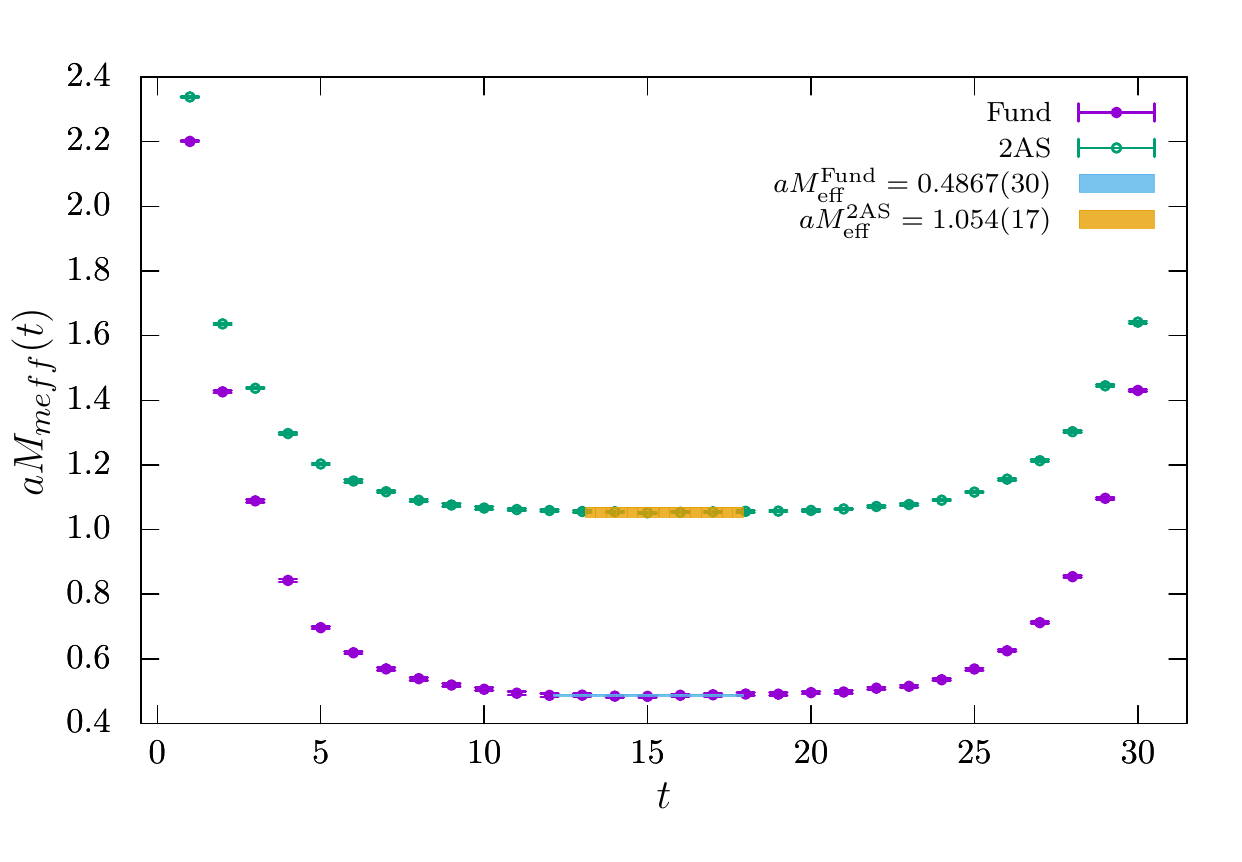}
	\hspace{-0.3cm}     
	\includegraphics[width=0.48\linewidth]{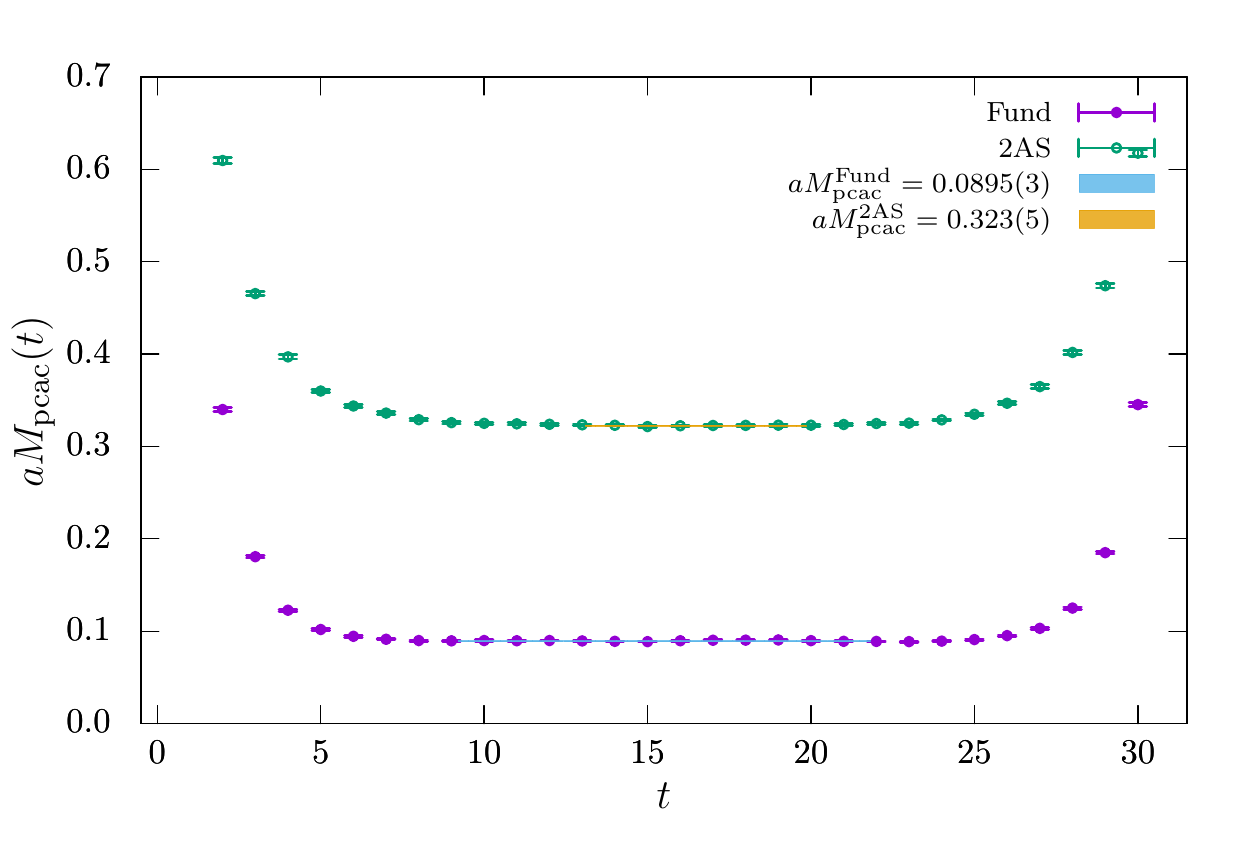}
	\vspace{-0.4cm}
    \caption{On the top panels we show an example of the pseudoscalar correlators obtained with degenerate bare fermion masses $am_4=am_6=-0.55$, on the bottom left panels the effective masses of ``pion-like'' states, for both representations 
    for the pseudoscalar correlator in unit of lattice spacing while on the bottom right we display the PCAC fermion mass. }
    \label{fig:PP_example} 
\end{figure}

\subsection{Extraction of effective masses}
Once the correlators are computed we project to zero-momentum by summing on the space directions $\mathbf{x}$ as 
\begin{gather}
C(t)=\sum_{\mathbf{x}} \langle \overline{O}(\mathbf{x},t)O(\mathbf{x},0)\rangle .
\end{gather}
The masses of pseudoscalar (``pion-like'') and vector (``$\rho$-like'') states are respectively extracted from the asymptotic behavior of the $C_{PP}(t)$ and $C_{V_{i}V_{i}}(t)$ correlators. For large Euclidean-time separation, the former behave like
\begin{gather}
\label{CPP_correlator}
    C_{PP}(t) \propto \exp\left \{ -M_{PP}t \right\} + \text{contribution from excited states}.
\end{gather}
In addition, in a system of finite Euclidean-time extent $L_t$, where fermionic fields obey anti-periodic boundary conditions in the Euclidean-time direction, the correlator above also receives contributions from the periodic copies of the operators, resulting in additional terms like $\exp\left \{ -M_{PP} (L_t -t) \right\}$, etc. on the right-hand side of eq.~(\ref{CPP_correlator}).

The mass of the ``meson-like'' states is thus obtained by fitting the decay of the correlators at sufficiently large $t$, including the effect of the first periodic copy of the operators. That is, we define
\begin{gather}
aM_{\rm eff}= \arccosh \left [\frac{C_{PP}(t+a) + C_{PP}(t-a)}{2C_{PP}(t)} \right]. 
\end{gather}
The same analysis is applied to the correlator involving the $i$-th component of the vector current $C_{V_iV_i}(t)$. In order to study the distance from the critical line of the theory we also consider the PCAC fermion mass defined through the non-anomalous axial Ward identity 
\begin{figure}[t!] 
    \centering
    \includegraphics[width=0.6\textwidth]{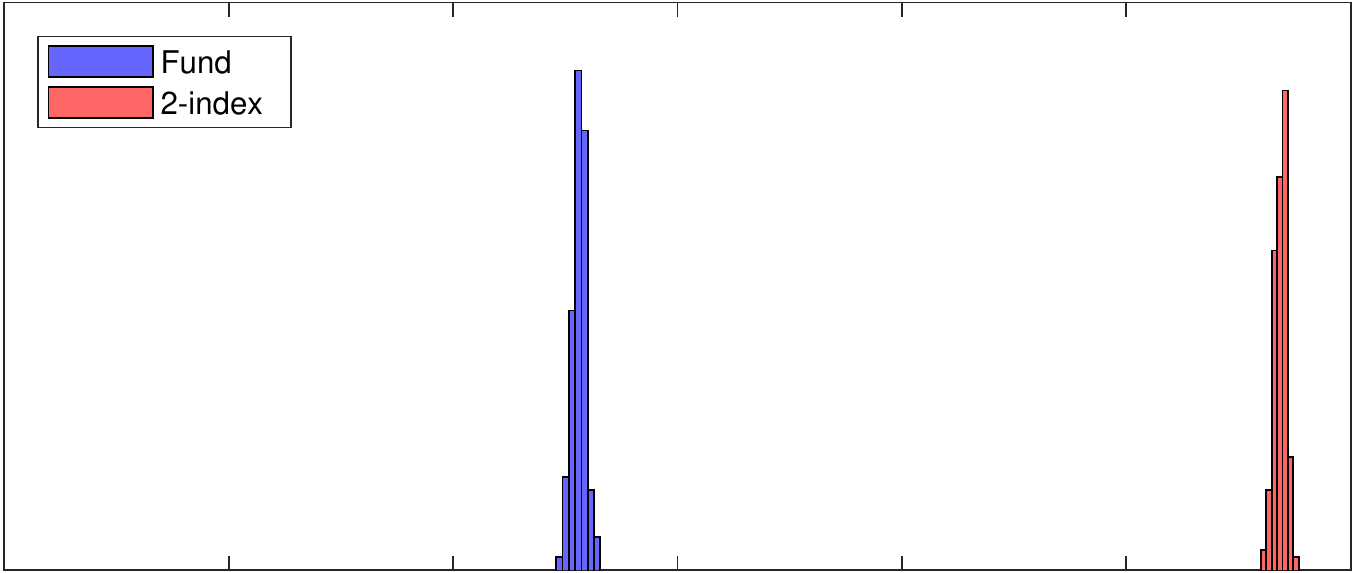}
   \, \includegraphics[width=0.6\textwidth]{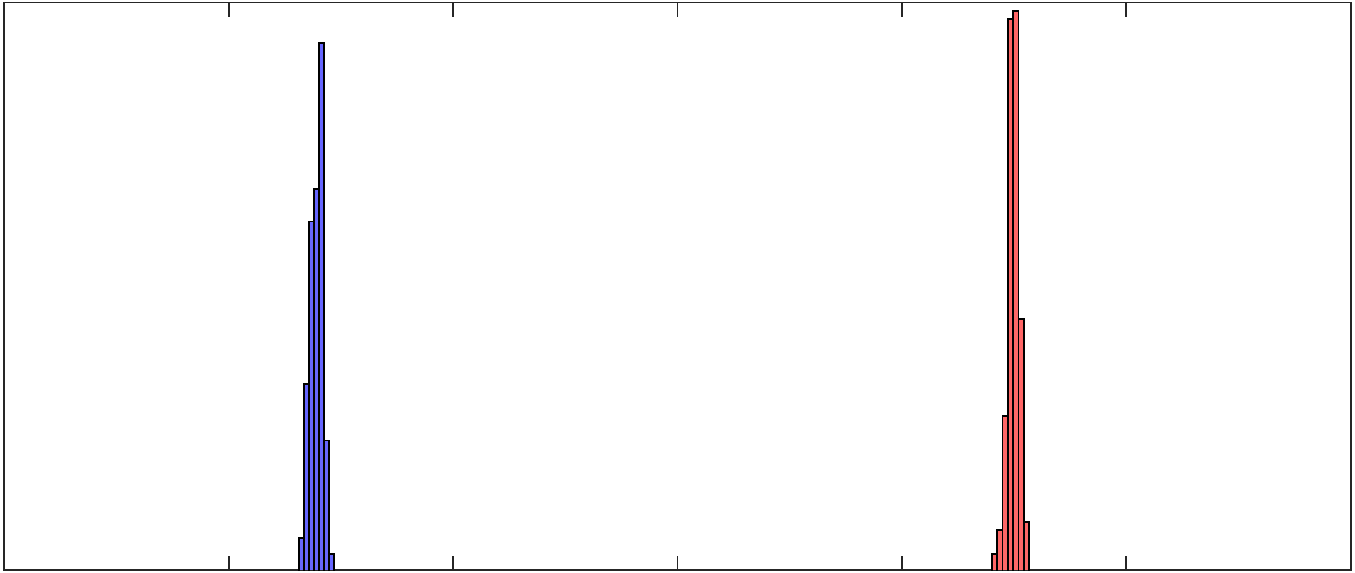}
   \, \includegraphics[width=0.615\textwidth]{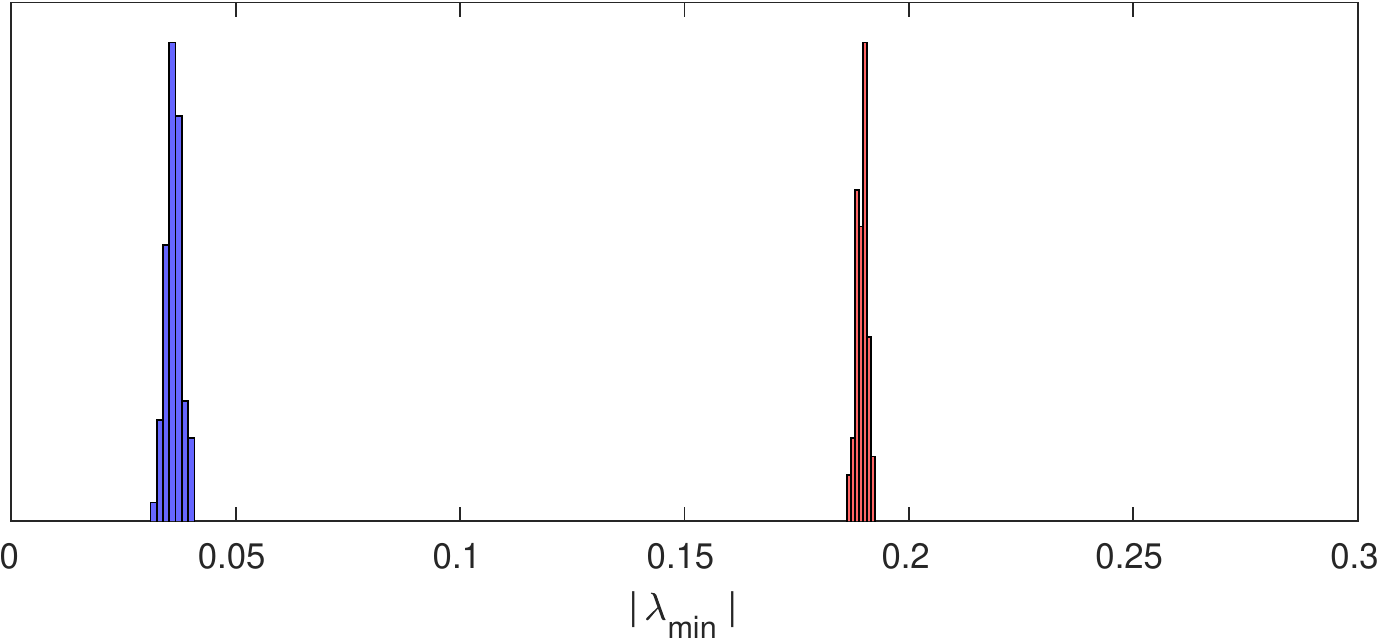}
    \caption{Distribution of the minimum eigenvalue of $\gamma_5 D$ in the fundamental (blue) and two-index antisymmetric(red). The critical line is approached from the top to the bottom corresponding to bare fermion masses $am_4=am_6=[-0.50,-0.55,-0.58]$. 
	We observe that for degenerate bare fermion masses the fundamental representation is lighter than the two-index antisymmetric one, i.e. $\langle|\lambda_{\rm min}^{\rm Fund}|\rangle < \langle|\lambda_{\rm min}^{\rm 2AS}|\rangle$. 
	This result is consistent with the hierarchies found in PCAC quark masses and in ``pion-like'' effective masses displayed in fig.~\ref{fig:PP_example}. This observation is also compatible with the results reported in ref.~\cite{Ayyar:2017qdf}.}
    \label{fig:lambdamin_example} 
\end{figure}
\begin{gather}
am_{PCAC}= \frac{\tilde{\partial}_0 C_{AP(t)}}{2C_{PP}(t)}
\end{gather}
with $\tilde{\partial}_0=(\partial_0 + \partial^*_0)/2$ the symmetric derivative in the time-direction. Note that the PCAC fermion mass approaches to the continuum limit linearly in the lattice spacing.
$\mathcal{O}(a)$ effect would be removed by considering the improved axial correlator $C^{\rm I}_{AP}(t) = C_{AP}(t) + c_A(g_0) \, \tilde{\partial}_0 C_{PP}(t)$, with the (currently) unknown coefficient $c_A(g_0)$ which 
depend on both number of colors and the representation of fermions. 
The top panels of fig.~\ref{fig:PP_example} illustrates the typical hyperbolic-cosine shape of the pseudoscalar correlator in both representations, while bottom panels in fig.~\ref{fig:PP_example} show fits to plateau region for the extraction of the two correspondent effective masses. Similar plot are provided for the PCAC fermion mass on the bottom right of fig.~\ref{fig:PP_example}. 

\subsection{Spectral observables and scale setting}
As discussed in sec.~\ref{sec:rmt}, a very interesting observable to probe the chiral regime of the theory is provided 
by the study of the low lying spectra of the Dirac operator in both representations under investigation. In this section rather than the Dirac operator itself, we prefer to consider the hermitian operator $\gamma_5 D$, since the latter is Hermitian and hence has a real spectrum. On finite lattice the smallest eigenvalues of the Dirac operator 
defines a spectral gap 
\begin{gather}
|\lambda_{\text{min}}| = \text{min}\{ |\lambda| : \lambda \text{ is an eigenvalue of } \gamma_5D \}\, .
\end{gather}
As a further control on the critical line of the theory we observe the scaling of $|\lambda_{\text{min}}|$ with the bare mass.  An example of showing the drift of the smallest eigenvalues is showed in fig.~\ref{fig:lambdamin_example}. 
We note here that at degenerate bare fermion masses the spectral gap is much larger in the two-index representation respect to the fundamental one. This picture is consistent with both the PCAC fermion masses and pion masses.
	\begin{figure}[t!] 
		\centering
		\includegraphics[width=0.85\linewidth]{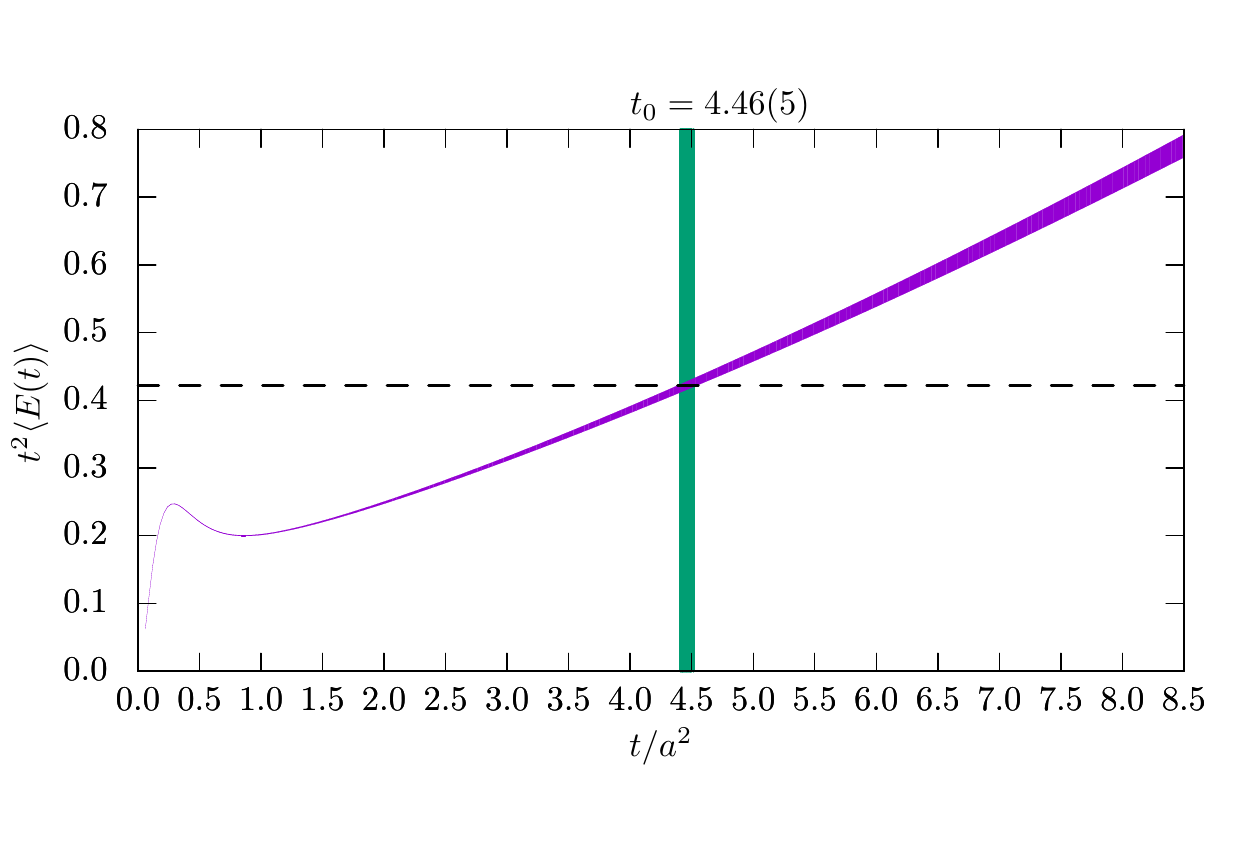}
		\vspace{-1.0cm}
		\caption{The purple band shows $t^2\langle E(t)\rangle$ with its uncertainty, while the vertical green 
		band is the value of $t_0$ implicitly defined by requiring $t_0^2\langle E(t_0)\rangle=0.421875$. 
		As expected after a first transient induced by the plaquette discretization
		of the energy, the plotted quantity enters a polynomial regime.}
		\label{fig:t0_example} 
	\end{figure}
The scale is set using the Wilson flow introduced in ref.~\cite{Luscher:2010iy}. 
The reference scale $t_0$ is implicitly defined via the relation (generalized to $\SU(N)$ as in ref.~\cite{Ce:2016awn})
\begin{gather}
\left . t^2 \langle E(t) \rangle \right |_{t_0} = 0.1125\frac{(N^2-1)}{N} \, , 
\label{eq:floweq}
\end{gather}
where the action density $E(t)=\frac{1}{4}G^a_{\mu \nu}(t)G^a_{\mu \nu}(t)$ is constructed from the plaquette,
formed by gauge links at flow time $t$. The r.h.s. of eq.~\ref{eq:floweq} is chosen to be a dimensionless number according to perturbative expansion at small $t$, reducing to $0.3$ for $N=3$.\footnote{Note that for $N=4$ the r.h.s. reads $0.421875$, which differs from the one employed in ref.~\cite{DeGrand:2015lna}.} Note that the (Gau{\ss}ian) smearing radius of the Wilson flow scales with the flow time as $\sqrt{8t}$. 
Hence, in order to avoid over-smearing we imposed $t\leq L^2/32$, with $L$ as the shortest direction in our lattice. 
An example of fit used to extract the value of $t_0/a^2$ is displayed in 
fig.~\ref{fig:t0_example}. We observe that for values of $\beta<10.0$, where we expect a bulk phase transition the scale 
cannot be set since the reference scale is reached too fast and within the initial transient regime. 
This is a further confirmation pointing to an unphysical phase fully dominated by cutoff effects. 
However assessing the nature of such a phase would requires further investigations on larger volumes and more values 
of the bare gauge coupling.

\subsection{Discussion}
\label{subsec:discussion}

The results presented here deserve some comments.

First of all, our data confirm that the simulation code that we used, featuring a Wilson Dirac operator with a clover improvement term, is a robust tool to explore the phase structure of this theory. Beside reproducing well-known results in the quenched limit, it also passes all other required algorithmic and physics consistency checks, and turns out to be efficient and easy to generalize to arbitrary matter content.

Our investigation of the spectrum of the Dirac operator in the $\SU(4)$ theory with matter in  $2+2$ different representations confirms the non-trivial implications of the global anti-unitary symmetries of sextet fermions, and proves that the spectral properties of the continuum operator are correctly reproduced in our lattice simulation. Moreover, the distribution of the \emph{lowest} eigenvalue of the Hermitian $\gamma_5 D$ operator, which is a useful probe to study the chiral limit of the theory, follows what is expected from general arguments (e.g. the absolute value of the lowest eigenvalue of $\gamma_5 D$ for fundamental fermions is always smaller than for sextet fermions, etc.).

Similarly, the investigation of ``meson-like'' hypercolor-singlet states that is summarized in tables~\ref{tab:runtable_meas1},~\ref{tab:runtable_meas1b}, and~\ref{tab:runtable_meas2} provides useful information about the non-perturbative dynamics of this theory, and, again, confirms that states built from fermions in the two-index antisymmetric representation are generally heavier than those from fundamental valence fermions. Also, the mass hierarchies between pseudoscalar and vector states follow a pattern similar to the one familiar from quantum chromodynamics, and are consistent with how our results for PCAC masses for the fermions scale.

The plaquette expectation values reported in tables~\ref{tab:runtable1} and~\ref{tab:runtable2} appear to reveal the presence of a rather large strong-coupling phase, likely dominated by quite severe, unphysical discretization effects: an important piece of information for future studies of this model with this lattice discretization. We also note a significant shift of the lines (or ``surfaces'') of constant physics with respect to the analysis reported in ref.~\cite{DeGrand:2015lna} and in subsequent works by that group; however, it should be emphasized that any possible discrepancy between the parameters in our work and theirs does not imply that these studies are inconsistent with each other, simply because they are based on different lattice discretizations, and, by virtue of universality, only continuum-extrapolated physical results should agree. For our scale setting in terms of the $t_0$ parameter, see also table~\ref{tab:runtable_meas2}.

\vspace{-0.2cm}
\section{Generalization to other partial-compositeness models}
\label{sec:extension}

While the numerical study reported in this work is restricted to (a slightly simplified version of) the theory proposed in ref.~\cite{Ferretti:2014qta}, it should be remarked that this is only one in a broad class of partial-compositeness models potentially relevant to describe the electroweak-symmetry breaking mechanism and physics at the TeV scale. Hence, it would be interesting to study also other strong-dynamics models, with low-energy symmetries compatible with those of the Standard Model, but based on other gauge groups and/or with a different matter content.

In fact, the simulation code that we used in this work is very versatile and the exploration of the phase structure and physical observables that was carried out here could be easily repeated for other models.

As we mentioned, the model originally proposed in ref.~\cite{Ferretti:2014qta} features five Weyl fermions in the sextet representation, but in the present study we considered a closely related theory, which instead has \emph{two Dirac} fermions in the sextet representation. Beside being simpler to simulate, the motivation underlying this choice is that the model with two sextet Dirac fermions (and two fundamental ones), which is an excellent proxy for the original model, has also been studied in other recent works~\cite{Ayyar:2017qdf, Ayyar:2018zuk, Ayyar:2018ppa, Ayyar:2019exp}, and, as usual, testing the universality of physical results obtained with a different lattice regularization is an important requirement of a lattice calculation. However, as our code includes numerical rational hybrid Monte~Carlo routines, it can be used to repeat the calculation for any number of fermion flavors, in arbitrary combinations of multiple representations. The generalization to larger values of the number of hypercolor charges, too, is already implemented in our code, and the computational-cost scaling with this parameter does not involve particular subtleties (see, e.g., ref.~\cite[section~3]{Lucini:2012gg}).

Furthermore, our code can be readily adapted to different types of gauge groups. In this respect, a novel and interesting proposal for a different strongly coupled New Physics model has been recently put forward in ref.~\cite{Gertov:2019yqo}. Like in the model that we considered here~\cite{Ferretti:2014qta}, the idea underlying the construction of this model is that the contributions to the Higgs boson mass from its Yukawa coupling to the top quark can be partially compensated for by the presence of sufficiently light top partners. However, in contrast to the proposal of ref.~\cite{Ferretti:2014qta}, the model discussed in ref.~\cite{Gertov:2019yqo} is characterized by local invariance under a symplectic, rather than a special unitary, group.

More specifically, the model described in ref.~\cite{Gertov:2019yqo} is based on the $\SU(4)/\Sp(4)$ symmetry-breaking scheme~\cite{Kaplan:1983sm, Katz:2005au} and its ultraviolet completion is a vector gauge theory with local internal invariance under the $\Sp(6)$ group. In addition to the gauge bosons, the field content of the theory includes ten fermions in the fundamental representation, and one in the adjoint representation of the gauge group. The choice of this internal symmetry and matter fields comes from the requirements of a global symmetry sufficiently large to include the gauge group of the Standard Model, the existence of a non-linearly realized symmetry that could protect the mass of the Higgs boson from arbitrarily large quantum corrections, and the presence of massless fermions compatible with the 't~Hooft anomaly-matching conditions. As discussed in ref.~\cite{Gertov:2019yqo}, this model is expected to present a rich low-energy phenomenology, which could include top-quark partners, scalar particles, and color-charged fermions. These features make it an interesting target for non-perturbative lattice calculations---a research program that could be a natural generalization of the present work.

It is worth remarking that the lattice investigation of $\Sp(2N)$ gauge theories with dynamical fermions has already begun~\cite{Bennett:2017kga}, and extending this type of calculations to the model described in ref.~\cite{Gertov:2019yqo} should be feasible with a minor effort with the technology already developed for the current project.

\vspace{-0.2cm}
\section{Concluding remarks and future perspectives}
\label{sec:concl}

In the present article, we reported our results of a non-perturbative lattice investigation of a non-Abelian $\SU(4)$ gauge theory with two dynamical flavors of fundamental Dirac fermions, and two dynamical flavors of Dirac fermions in the two-index antisymmetric representation. As discussed in the introduction, the main motivation to study this model is its close proximity to the simplest UV-complete partial-compositeness model, that was introduced in ref.~\cite{Ferretti:2014qta}, and that may provide a solution to some of the tantalizing conundrums of the present state of affairs in theoretical elementary particle physics: in particular, it features a composite Higgs boson and a partially composite top quark. While the model studied in the present work has slightly different matter content with respect to the one advocated in ref.~\cite{Ferretti:2014qta}, it is expected to capture its main features at least semi-quantitatively, and to provide useful guidance for future studies.

We carried out our Monte~Carlo calculations by adapting existing code to a setup with fermionic matter in multiple, arbitrary representations; moreover, this code already supports rational hybrid Monte~Carlo routines, so that an extension to an arbitrary number of fermion flavors would be straightforward. At the technical level, our lattice discretization of the continuum theory is based on a Wilson Dirac operator with clover improvement term. Our setup is, thus, slightly different\footnote{In our setup we do not use any smeared action, while in ref.~\cite{DeGrand:2015lna} and subsequent works nHYP smearing for the fermion actions and an NDS gauge action are employed.} with respect to the one used in ref.~\cite{DeGrand:2015lna} and in later works by that group~\cite{Ayyar:2017qdf, Ayyar:2018zuk, Ayyar:2018ppa, Ayyar:2018glg}.

As discussed in detail in subsection~\ref{subsec:discussion}, the results that we presented here provide a clear picture of the phase structure of this lattice theory, and confirm important properties related to its global symmetries, as well as its non-perturbative dynamics. While this could already provide a useful roadmap for further lattice investigation of this model, it should be pointed out that the results of the very recent paper~\cite{Ayyar:2018glg} appear to rule out the viability of this model for a partial-compositeness scenario: they indicate that the renormalized overlap factors relevant for the mixing of ``chimera'' states with the top quark are too small, and disfavor its phenomenological relevance for New Physics. The possibility that this problem could be evaded through a four-fermion coupling enhanced at low energies by a large, negative anomalous dimension was also ruled out, in particular in view of the QCD-like, rather than conformal, behavior of the spectroscopy of this theory, which our present results also confirm.

As we pointed out in section~\ref{sec:extension}, however, an interesting alternative partial-compositeness model has been recently proposed in ref.~\cite{Gertov:2019yqo}, and the simulation algorithm that we used in the present study is sufficiently versatile to use it for the study of this model, too. The lattice investigation of strongly coupled models for New Physics, (see ref.~\cite{Witzel:2019jbe} for a very recent review) remains an active research field.

\noindent{\bf Acknowledgments}\\
The simulations were run on the supercomputers of the Consorzio Interuniversitario per il Calcolo Automatico dell'Italia Nord Orientale (CINECA). We thank Peter~Boyle for support at the early stages of this project and Stefano~Piemonte for discussions. LDD is supported by an STFC Consolidated Grant, ST/P0000630/1, and a Royal Society Wolfson Research Merit Award, WM140078. GC acknowledges funding by Intel and an STFC IAA award, and the support by STFC, grants ST/L000458/1 and ST/P002447/1.  

\vspace{-0.2cm}
\begin{appendix}
\section{Group-theory conventions}
\label{app:Notation}
\renewcommand{\theequation}{A.\arabic{equation}}
\setcounter{equation}{0}

\vspace{-0.2cm}
\subsection{$\SU(N)$ group}
In our conventions, the generators of the algebra of the generic $\SU(N)$ group, in a generic irreducible representation labeled by $R$, are represented as Hermitian and traceless matrices $T^a_R$, with $1 \le a \le N^2-1$. They satisfy the relations
\begin{equation}
\label{Lie_algebra}
[T^a_R,T^b_R] = i f^{abc} T^c_R,
\end{equation}
where the $f^{abc}$ are the structure constants that define the Lie algebra; the relations in eq.~(\ref{Lie_algebra}) are satisfied by the generators in every representation. The $f^{abc}$ structure constants are totally antisymmetric under the exchange of any pair of indices, so that
\begin{equation}
\label{fabc_antisymmetry}
f^{a^\prime b^\prime c^\prime} = \sign (P^{a^\prime b^\prime c^\prime}_{abc}) f^{abc} ,
\end{equation}
where $P^{a^\prime b^\prime c^\prime}_{abc}$ denotes the permutation mapping the ordered set of indices $abc$ to $a^\prime b^\prime c^\prime$. For the algebra of generators of the $\SU(4)$ group, the non-vanishing structure constants are
\begin{eqnarray}
\label{su4_structure_constants}
f^{ 1\, 2 \, 3 } \!\!\!\!&=&\!\!\!\! 1, \nonumber \\
f^{ 1\, 4 \, 7 } \!\!\!\!&=&\!\!\!\! f^{ 1\, 6 \, 5 } = f^{ 1\, 9 \, 12} = f^{ 1\, 11\, 10} = f^{ 2\, 4 \, 6 } = f^{ 2\, 5 \, 7 } = f^{ 2\, 9 \, 11} = f^{ 2\, 10\, 12} = f^{ 3\, 4 \, 5 } = f^{ 3\, 7 \, 6 } = f^{ 3\, 9 \, 10} \nonumber \\
                 \!\!\!\!&=&\!\!\!\! f^{ 3\, 12\, 11} = f^{ 4\, 9 \, 14} = f^{ 4\, 13\, 10} = f^{ 5\, 9 \, 13} = f^{ 5\, 10\, 14} = f^{ 6\, 11\, 14} = f^{ 6\, 13\, 12} = f^{ 7\, 11\, 13} \nonumber \\
                 \!\!\!\!&=&\!\!\!\! f^{ 7\, 12\, 14} = 1/2, \nonumber \\
f^{ 4\, 5 \, 8 } \!\!\!\!&=&\!\!\!\! f^{ 6\, 7 \, 8 } = \sqrt{3}/2, \nonumber \\
f^{ 8\,  9\, 10} \!\!\!\!&=&\!\!\!\! f^{ 8\, 11\, 12} = 1/\sqrt{12}, \nonumber \\
f^{ 8\, 14\, 13} \!\!\!\!&=&\!\!\!\! 1/\sqrt{3}, \nonumber \\
f^{ 9\, 10\, 15} \!\!\!\!&=&\!\!\!\! f^{11\, 12\, 15} = f^{13\, 14\, 15} = \sqrt{2/3},
\end{eqnarray}
and those related to them by permutations of the indices. For the algebra of generators of the $\SU(3)$ group, the non-vanishing $f^{abc}$ components are given by the subset of eqs.~(\ref{su4_structure_constants}), in which $a$, $b$ and $c$ are less than or equal to $8$ (and permutations thereof). In turn, for the generators of the $\SU(2)$ group, the non-vanishing structure constants are those with indices not larger than $3$, i.e. $f^{123}=1$, and their permutations (that is, $f^{abc}=\epsilon^{abc}$, the totally antisymmetric Levi-Civita symbol of three elements).

In the following, we will focus on the two irreducible representations considered in this work, namely the fundamental and the antisymmetric two-index representation.

We denote the $\SU(N)$ generators in the fundamental representation as $T^a_F$: they are $N \times N$ complex matrices normalized according to
\begin{equation}
\Tr \left( T^a_F T^b_F \right) = \frac{1}{2} \delta_{a,b}.
\end{equation}
In this representation, the $(N-1)$ Cartan generators are chosen to be the matrices $T_F^{k(k+2)}$, with $1 \le k < N$, and their explicit form is
\begin{equation}
\label{Cartan_generators}
T^{k(k+2)}_F=\frac{1}{\sqrt{2k(k+1)}} \diag ( \underbrace{1, 1, \dots , 1}_{k~\mbox{\small{terms}}}, -k, \underbrace{0, \dots, 0}_{N-k-1~\mbox{\small{terms}}} ).
\end{equation}
The non-diagonal generators are defined as follows: for every value of $1 \le k < N$, every natural number $n$ such that $k^2-1 < n < k(k+2)$ can always be written either as $n=k^2-2+2j$ or as $n=k^2-1+2j$, where the integer $j$ satisfies $1 \le j \le k$. Then:
\begin{equation}
\label{non-diagonal_generators}
(T^{k^2-2+2j}_F)_{p q}= \frac{1}{2} ( \delta_{p, j} \delta_{q, k+1} + \delta_{p, k+1} \delta_{q, j}), \qquad (T^{k^2-1+2j}_F)_{p q}=-\frac{i}{2} ( \delta_{p, j} \delta_{q, k+1} - \delta_{p, k+1} \delta_{q, j}).
\end{equation}
Note that, with these conventions, the generators of the $\SU(2)$ group in the fundamental representation are 
proportional to the Pauli matrices, $T^a=\sigma^a/2$:
\begin{equation}
\label{SU2_fund}
T^1_F = \frac{1}{2} \left(
\begin{array}{cc}
0 & 1 \\
1 & 0
\end{array}
\right), \quad 
T^2_F = \frac{1}{2} \left(
\begin{array}{cc}
0 & -i \\
i & 0
\end{array}
\right), \quad
T^3_F = \frac{1}{2} \left(
\begin{array}{cc}
1 & 0\\
0 & -1
\end{array}
\right), 
\end{equation}
while those for the $\SU(3)$ group are proportional to the Gell-Mann matrices, $t^a=\lambda^a/2$:
\begin{eqnarray}
\label{SU3_fund}
&& T^1_F = \frac{1}{2}\left(
\begin{array}{ccc}
0 & 1 & 0 \\
1 & 0 & 0 \\
0 & 0 & 0 
\end{array}
\right), \quad
T^2_F = \frac{1}{2}\left(
\begin{array}{ccc}
0 & -i & 0 \\
i & 0 & 0 \\
0 & 0 & 0 
\end{array}
\right), \quad
T^3_F = \frac{1}{2} \left(
\begin{array}{ccc}
1 & 0 & 0 \\
0 & -1 & 0 \\
0 & 0 & 0 
\end{array}
\right), \nonumber \\
&& T^4_F = \frac{1}{2} \left(
\begin{array}{ccc}
0 & 0 & 1 \\
0 & 0 & 0 \\
1 & 0 & 0 
\end{array}
\right), \quad
T^5_F = \frac{1}{2} \left(
\begin{array}{ccc}
0 & 0 & -i \\
0 & 0 & 0 \\
i & 0 & 0 
\end{array}
\right), \quad
T^6_F = \frac{1}{2}\left(
\begin{array}{ccc}
0 & 0 & 0 \\
0 & 0 & 1 \\
0 & 1 & 0 
\end{array}
\right), \nonumber \\
&& T^7_F = \frac{1}{2}\left(
\begin{array}{ccc}
0 & 0 & 0 \\
0 & 0 & -i \\
0 & i & 0 
\end{array}
\right), \quad
T^8_F = \frac{1}{2\sqrt{3}} \left(
\begin{array}{ccc}
1 & 0 & 0 \\
0 & 1 & 0 \\
0 & 0 & -2 
\end{array}
\right).
\end{eqnarray}
Finally, for the $\SU(4)$ generators in the fundamental representation, we have:
\begin{small}
\begin{eqnarray}
\label{SU4_fund}
&& T^1_F = \frac{1}{2}\left(
\begin{array}{cccc}
0 & 1 & 0 & 0 \\
1 & 0 & 0 & 0 \\
0 & 0 & 0 & 0 \\
0 & 0 & 0 & 0
\end{array}
\right), \,\,
T^2_F = \frac{1}{2}\left(
\begin{array}{cccc}
0 & -i & 0 & 0 \\
i & 0 & 0 & 0 \\
0 & 0 & 0 & 0 \\
0 & 0 & 0 & 0
\end{array}
\right), \,\,
T^3_F = \frac{1}{2} \left(
\begin{array}{ccccc}
1 & 0 & 0 & 0 \\
0 & -1 & 0 & 0 \\
0 & 0 & 0 & 0 \\
0 & 0 & 0 & 0
\end{array}
\right), \nonumber \\
&& T^4_F = \frac{1}{2} \left(
\begin{array}{cccc}
0 & 0 & 1 & 0 \\
0 & 0 & 0 & 0 \\
1 & 0 & 0 & 0 \\
0 & 0 & 0 & 0
\end{array}
\right), \,\,
T^5_F = \frac{1}{2} \left(
\begin{array}{cccc}
0 & 0 & -i & 0 \\
0 & 0 & 0 & 0 \\
i & 0 & 0 & 0 \\
0 & 0 & 0 & 0
\end{array}
\right), \,\,
T^6_F = \frac{1}{2}\left(
\begin{array}{cccc}
0 & 0 & 0 & 0 \\
0 & 0 & 1 & 0 \\
0 & 1 & 0 & 0 \\
0 & 0 & 0 & 0
\end{array}
\right), \nonumber \\
&& T^7_F = \frac{1}{2}\left(
\begin{array}{cccc}
0 & 0 & 0 & 0 \\
0 & 0 & -i & 0 \\
0 & i & 0 & 0 \\
0 & 0 & 0 & 0
\end{array}
\right), \,\,
T^8_F = \frac{1}{2\sqrt{3}} \left(
\begin{array}{cccc}
1 & 0 & 0 & 0 \\
0 & 1 & 0 & 0 \\
0 & 0 & -2 & 0 \\
0 & 0 & 0 & 0
\end{array}
\right), \,\,
T^9_F = \frac{1}{2} \left(
\begin{array}{cccc}
0 & 0 & 0 & 1 \\
0 & 0 & 0 & 0 \\
0 & 0 & 0 & 0 \\
1 & 0 & 0 & 0
\end{array}
\right), \nonumber \\
&& T^{10}_F = \frac{1}{2} \left(
\begin{array}{cccc}
0 & 0 & 0 & -i \\
0 & 0 & 0 & 0 \\
0 & 0 & 0 & 0 \\
i & 0 & 0 & 0
\end{array}
\right), \,\,
T^{11}_F = \frac{1}{2} \left(
\begin{array}{cccc}
0 & 0 & 0 & 0 \\
0 & 0 & 0 & 1 \\
0 & 0 & 0 & 0 \\
0 & 1 & 0 & 0
\end{array}
\right), \,\,
T^{12}_F = \frac{1}{2} \left(
\begin{array}{cccc}
0 & 0 & 0 & 0 \\
0 & 0 & 0 & -i \\
0 & 0 & 0 & 0 \\
0 & i & 0 & 0
\end{array}
\right), \nonumber \\
&& T^{13}_F = \frac{1}{2} \left(
\begin{array}{cccc}
0 & 0 & 0 & 0 \\
0 & 0 & 0 & 0 \\
0 & 0 & 0 & 1 \\
0 & 0 & 1 & 0
\end{array}
\right), \,\,
T^{14}_F = \frac{1}{2} \left(
\begin{array}{cccc}
0 & 0 & 0 & 0 \\
0 & 0 & 0 & 0 \\
0 & 0 & 0 & -i \\
0 & 0 & i & 0
\end{array}
\right), \,\,
T^{15}_F = \frac{1}{2\sqrt{6}} \left(
\begin{array}{cccc}
1 & 0 & 0 & 0 \\
0 & 1 & 0 & 0 \\
0 & 0 & 1 & 0 \\
0 & 0 & 0 & -3
\end{array}
\right). \qquad
\end{eqnarray}
\end{small}
Note that, with the conventions defined in eqs.~(\ref{Cartan_generators}) and (\ref{non-diagonal_generators}), the generators of $\SU(1 < Q \le N)$ are given by the sub-matrices obtained by taking the elements in the first $Q$ rows and in the first $Q$ columns of the first $Q^2-1$ generators of $\SU(N)$.

Explicitly, for the generators of the $\SU(4)$ group in the two-index antisymmetric representation, one obtains:
\begin{footnotesize}
\begin{eqnarray}
\label{SU4_2a}
&& \hspace{-20mm} T^1_{\rm 2AS} = \frac{1}{2}\left(
\begin{array}{cccccc}
0 & 0 & 0 & 0 & 0 & 0 \\
0 & 0 & 1 & 0 & 0 & 0 \\
0 & 1 & 0 & 0 & 0 & 0 \\
0 & 0 & 0 & 0 & 1 & 0 \\
0 & 0 & 0 & 1 & 0 & 0 \\
0 & 0 & 0 & 0 & 0 & 0
\end{array}
\right), \,\,\,\,\,\,\,\,\,\,\,
T^2_{\rm 2AS} = \frac{1}{2}\left(
\begin{array}{cccccc}
0 & 0 & 0 & 0 & 0 & 0 \\
0 & 0 & -i & 0 & 0 & 0 \\
0 & i & 0 & 0 & 0 & 0 \\
0 & 0 & 0 & 0 & -i & 0 \\
0 & 0 & 0 & i & 0 & 0 \\
0 & 0 & 0 & 0 & 0 & 0
\end{array}
\right), \,\,\,\,\,\,\,\,\,\,\,\,\,\,
T^3_{\rm 2AS} = \frac{1}{2}\left(
\begin{array}{cccccc}
0 & 0 & 0 & 0 & 0 & 0 \\
0 & 1 & 0 & 0 & 0 & 0 \\
0 & 0 & -1 & 0 & 0 & 0 \\
0 & 0 & 0 & 1 & 0 & 0 \\
0 & 0 & 0 & 0 & -1 & 0 \\
0 & 0 & 0 & 0 & 0 & 0
\end{array}
\right), \nonumber \\
&& \hspace{-20mm} T^4_{\rm 2AS} = \frac{1}{2}\left(
\begin{array}{cccccc}
0 & 0 & -1 & 0 & 0 & 0 \\
0 & 0 & 0 & 0 & 0 & 0 \\
-1 & 0 & 0 & 0 & 0 & 0 \\
0 & 0 & 0 & 0 & 0 & 1 \\
0 & 0 & 0 & 0 & 0 & 0 \\
0 & 0 & 0 & 1 & 0 & 0
\end{array}
\right), \,\,
T^5_{\rm 2AS} = \frac{1}{2}\left(
\begin{array}{cccccc}
0 & 0 & i & 0 & 0 & 0 \\
0 & 0 & 0 & 0 & 0 & 0 \\
-i & 0 & 0 & 0 & 0 & 0 \\
0 & 0 & 0 & 0 & 0 & -i \\
0 & 0 & 0 & 0 & 0 & 0 \\
0 & 0 & 0 & i & 0 & 0
\end{array}
\right), \,\,\,\,\,\,\,\,\,\,\,\,\,\,
T^6_{\rm 2AS} = \frac{1}{2}\left(
\begin{array}{cccccc}
0 & 1 & 0 & 0 & 0 & 0 \\
1 & 0 & 0 & 0 & 0 & 0 \\
0 & 0 & 0 & 0 & 0 & 0 \\
0 & 0 & 0 & 0 & 0 & 0 \\
0 & 0 & 0 & 0 & 0 & 1 \\
0 & 0 & 0 & 0 & 1 & 0
\end{array}
\right), \nonumber \\
&& \hspace{-20mm} T^7_{\rm 2AS} = \frac{1}{2}\left(
\begin{array}{cccccc}
0 & -i & 0 & 0 & 0 & 0 \\
i & 0 & 0 & 0 & 0 & 0 \\
0 & 0 & 0 & 0 & 0 & 0 \\
0 & 0 & 0 & 0 & 0 & 0 \\
0 & 0 & 0 & 0 & 0 & -i \\
0 & 0 & 0 & 0 & i & 0
\end{array}
\right), \,\,
T^8_{\rm 2AS} = \frac{1}{\sqrt{12}}\left(
\begin{array}{cccccc}
2 & 0 & 0 & 0 & 0 & 0 \\
0 & -1 & 0 & 0 & 0 & 0 \\
0 & 0 & -1 & 0 & 0 & 0 \\
0 & 0 & 0 & 1 & 0 & 0 \\
0 & 0 & 0 & 0 & 1 & 0 \\
0 & 0 & 0 & 0 & 0 & -2
\end{array}
\right), \,\,
T^9_{\rm 2AS} = \frac{1}{2}\left(
\begin{array}{cccccc}
0 & 0 & 0 & 0 & -1 & 0 \\
0 & 0 & 0 & 0 & 0 & -1 \\
0 & 0 & 0 & 0 & 0 & 0 \\
0 & 0 & 0 & 0 & 0 & 0 \\
-1 & 0 & 0 & 0 & 0 & 0 \\
0 & -1 & 0 & 0 & 0 & 0
\end{array}
\right), \nonumber \\
&& \hspace{-20mm} T^{10}_{\rm 2AS} = \frac{1}{2}\left(
\begin{array}{cccccc}
0 & 0 & 0 & 0 & i & 0 \\
0 & 0 & 0 & 0 & 0 & i \\
0 & 0 & 0 & 0 & 0 & 0 \\
0 & 0 & 0 & 0 & 0 & 0 \\
-i & 0 & 0 & 0 & 0 & 0 \\
0 & -i & 0 & 0 & 0 & 0
\end{array}
\right), \,\,
T^{11}_{\rm 2AS} = \frac{1}{2}\left(
\begin{array}{cccccc}
0 & 0 & 0 & 1 & 0 & 0 \\
0 & 0 & 0 & 0 & 0 & 0 \\
0 & 0 & 0 & 0 & 0 & -1 \\
1 & 0 & 0 & 0 & 0 & 0 \\
0 & 0 & 0 & 0 & 0 & 0 \\
0 & 0 & -1 & 0 & 0 & 0
\end{array}
\right), \,\,\,\,\,\,\,\,
T^{12}_{\rm 2AS} = \frac{1}{2}\left(
\begin{array}{cccccc}
0 & 0 & 0 & -i & 0 & 0 \\
0 & 0 & 0 & 0 & 0 & 0 \\
0 & 0 & 0 & 0 & 0 & i \\
i & 0 & 0 & 0 & 0 & 0 \\
0 & 0 & 0 & 0 & 0 & 0 \\
0 & 0 & -i & 0 & 0 & 0
\end{array}
\right), \nonumber \\
&& \hspace{-20mm} T^{13}_{\rm 2AS} = \frac{1}{2}\left(
\begin{array}{cccccc}
0 & 0 & 0 & 0 & 0 & 0 \\
0 & 0 & 0 & 1 & 0 & 0 \\
0 & 0 & 0 & 0 & 1 & 0 \\
0 & 1 & 0 & 0 & 0 & 0 \\
0 & 0 & 1 & 0 & 0 & 0 \\
0 & 0 & 0 & 0 & 0 & 0
\end{array}
\right), \,\,\,\,\,\,\,\,\,
T^{14}_{\rm 2AS} = \frac{1}{2}\left(
\begin{array}{cccccc}
0 & 0 & 0 & 0 & 0 & 0 \\
0 & 0 & 0 & -i & 0 & 0 \\
0 & 0 & 0 & 0 & -i & 0 \\
0 & i & 0 & 0 & 0 & 0 \\
0 & 0 & i & 0 & 0 & 0 \\
0 & 0 & 0 & 0 & 0 & 0
\end{array}
\right), \,\,\,\,\,\,\,
T^{15}_{\rm 2AS} = \frac{1}{\sqrt{6}}\left(
\begin{array}{cccccc}
1 & 0 & 0 & 0 & 0 & 0 \\
0 & 1 & 0 & 0 & 0 & 0 \\
0 & 0 & 1 & 0 & 0 & 0 \\
0 & 0 & 0 & -1 & 0 & 0 \\
0 & 0 & 0 & 0 & -1 & 0 \\
0 & 0 & 0 & 0 & 0 & -1
\end{array}
\right). \nonumber \\
\end{eqnarray}
\end{footnotesize}

\subsection{Clifford-algebra matrices}

Let us now introduce the Euclidean gamma matrices $\gamma_1$, $\gamma_2$, $\gamma_3$, and $\gamma_4$, that generate a matrix representation of the Clifford algebra
\begin{equation}
\left\{ \gamma_\mu, \gamma_\nu \right\} = 2 \delta_{\mu,\nu} \ide .
\end{equation}
In our conventions, the Euclidean gamma matrices are Hermitian, traceless, and are defined as $\gamma_k = \sigma_2 \otimes \sigma_k$, for $k=1$, $2$ and $3$, where $\otimes$ denotes the tensor product, and $\sigma_k$ is a Pauli matrix, while $\gamma_4 = \sigma_1 \otimes \ide$:
\begin{eqnarray}
\label{Euclidean_gamma_matrices}
&& \gamma_1 = \left(
\begin{array}{cccc}
0 & 0 & 0 & -i \\
0 & 0 & -i & 0 \\
0 & i & 0 & 0 \\
i & 0 & 0 & 0 
\end{array}
\right), \qquad
\gamma_2 = \left(
\begin{array}{cccc}
0 & 0 & 0 & -1 \\
0 & 0 & 1 & 0 \\
0 & 1 & 0 & 0 \\
-1 & 0 & 0 & 0 
\end{array}
\right), \nonumber \\
&& \gamma_3 = \left(
\begin{array}{cccc}
0 & 0 & -i & 0 \\
0 & 0 & 0 & i \\
i & 0 & 0 & 0 \\
0 & -i & 0 & 0 
\end{array}
\right), \qquad
\gamma_4 = \left(
\begin{array}{cccc}
0 & 0 & 1 & 0 \\
0 & 0 & 0 & 1 \\
1 & 0 & 0 & 0 \\
0 & 1 & 0 & 0 
\end{array}
\right).
\end{eqnarray}
Note that $\gamma_1$ and $\gamma_3$ have purely imaginary entries, whereas $\gamma_2$ and $\gamma_4$ are real.

In addition to the four $\gamma_\mu$ matrices, we also introduce the $\gamma_5$ matrix, defined as $\gamma_5 = \gamma_1 \gamma_2 \gamma_3 \gamma_4$, which is Hermitian, traceless, squares to the identity, and anti-commutes with the $\gamma_\mu$ matrices: $\left\{ \gamma_5, \gamma_\mu \right\}=0$. In our conventions, it is real and diagonal, and its explicit form is $\gamma_5=\sigma_3 \otimes \ide$, namely:
\begin{equation}
\gamma_5 = \left(
\begin{array}{cccc}
1 & 0 & 0 & 0 \\
0 & 1 & 0 & 0 \\
0 & 0 & -1 & 0 \\
0 & 0 & 0 & -1 
\end{array}
\right).
\end{equation}
Moreover, we also introduce the $\mathcal{C}$ matrix (related to charge conjugation), defined as $\mathcal{C}=\gamma_2 \gamma_4$. As both $\gamma_2$ and $\gamma_4$ are Hermitian, square to the identity, and anti-commute with each other, $\mathcal{C}$ is anti-Hermitian, and $\mathcal{C}^2=-\ide$, so that $\mathcal{C}^{-1}=\mathcal{C}$. Moreover, $\mathcal{C}$ commutes with $\gamma_5$. In our conventions, $\mathcal{C}$ takes the form $\mathcal{C}=i \sigma_1 \otimes \sigma_2$:
\begin{equation}
\label{C_matrix}
\mathcal{C} = \left(
\begin{array}{cccc}
0 & 0 & 0 & 1 \\
0 & 0 & -1 & 0 \\
0 & 1 & 0 & 0 \\
-1 & 0 & 0 & 0
\end{array}
\right).
\end{equation}
$\mathcal{C}$ relates each of the four $\gamma_\mu$ matrices to its complex conjugate via
\begin{equation}
\label{C_and_gamma_mu_star}
\mathcal{C}^{-1} \gamma_\mu \mathcal{C} = - \gamma_\mu^\star.
\end{equation}

\section{Proof of commutation relations}
\label{app:commutator_proofs}
\renewcommand{\theequation}{B.\arabic{equation}}
\setcounter{equation}{0}

In this section, we present the proofs of some commutation relations introduced in section~\ref{sec:rmt}.

\subsection{Proof of the commutation relation $[A,\gamma_5 D_c]=0$}
We show here that $A$ commutes with $\gamma_5 \Dc$:
\begin{eqnarray}
[A, \gamma_5 \Dc] &=& W \mathcal{C} \gamma_5 K \gamma_5 \Dc - \gamma_5 \Dc W \mathcal{C} \gamma_5 K \nonumber \\
&=& W \mathcal{C} \gamma_5 \gamma_5 \Dc^\star K - \gamma_5 \Dc W \mathcal{C} \gamma_5 K \nonumber \\
&=& W \mathcal{C} \Dc^\star K - \gamma_5 \Dc W \mathcal{C} \gamma_5 K \nonumber \\
&=& W \mathcal{C} \left(\gamma_\mu^\star \partial_\mu -ig \gamma_\mu^\star A_\mu^a T^{a \star}_{\rm 2AS} +m \right) K - \gamma_5 \Dc W \mathcal{C} \gamma_5 K \nonumber \\
&=& W \mathcal{C} \left(-\mathcal{C}^{-1}\gamma_\mu \mathcal{C} \partial_\mu -ig \mathcal{C}^{-1}\gamma_\mu \mathcal{C} A_\mu^a W^{-1} T^a_{\rm 2AS} W +m \right) K + \nonumber \\
&& - \gamma_5 \gamma_\mu \partial_\mu W \mathcal{C} \gamma_5 K -ig \gamma_5 \gamma_\mu A_\mu^a T^a_{\rm 2AS} W \mathcal{C} \gamma_5 K -m \gamma_5 W \mathcal{C} \gamma_5 K \nonumber \\
&=& - W \gamma_\mu \mathcal{C} \partial_\mu K -ig W \gamma_\mu \mathcal{C} A_\mu^a W^{-1} T^a_{\rm 2AS} W K +m W \mathcal{C} K + \nonumber \\
&& + W \gamma_\mu \partial_\mu \mathcal{C} K +ig \gamma_\mu A_\mu^a T^a_{\rm 2AS} W \mathcal{C} K -m \gamma_5 W \mathcal{C} \gamma_5 K \nonumber \\
&=& -ig W \gamma_\mu \mathcal{C} A_\mu^a W^{-1} T^a_{\rm 2AS} W K +ig \gamma_\mu A_\mu^a T^a_{\rm 2AS} W \mathcal{C} K \nonumber \\
&=& -ig \gamma_\mu \mathcal{C} A_\mu^a T^a_{\rm 2AS} W K +ig \gamma_\mu A_\mu^a T^a_{\rm 2AS} W \mathcal{C} K \nonumber \\
&=& -ig \gamma_\mu \mathcal{C} A_\mu^a T^a_{\rm 2AS} W K +ig \gamma_\mu \mathcal{C} A_\mu^a T^a_{\rm 2AS} W K \nonumber \\
&=& 0.
\end{eqnarray}

\subsection{Proof of the commutation relation $[A,\gamma_5 D]=0$}
We show here that as in the continuum case, $A$ commutes with hermitian Wilson Dirac operator $\gamma_5 D$:
The $[A,\gamma_5 D]$ commutator can be written as
\begin{eqnarray}
\hspace{-10mm} [A, \gamma_5 D] &=& W \mathcal{C} \gamma_5 K \gamma_5 D - \gamma_5 D W \mathcal{C} \gamma_5 K \nonumber \\
&=& \left( W \mathcal{C} \gamma_5^2 D^\star - \gamma_5 D W \mathcal{C} \gamma_5 \right) K\nonumber \\
&=& \left( W \mathcal{C} D^\star - \gamma_5 D W \mathcal{C} \gamma_5 \right) K \nonumber \\
&=& \frac{1}{a} \left\{ W \mathcal{C}  - \kappa \sum_{\mu=1}^4 \left[ (W \mathcal{C} - W \mathcal{C} \gamma_\mu^\star) (U_\mu^\star P_\mu) + (W \mathcal{C} + W \mathcal{C} \gamma_\mu^\star) (U_\mu^\star P_\mu)^\dagger \right] \right. \nonumber \\
&& \left. - \gamma_5 W \mathcal{C} \gamma_5 + \kappa \sum_{\mu=1}^4 \left[ (\gamma_5 - \gamma_5 \gamma_\mu) (U_\mu P_\mu)W \mathcal{C} \gamma_5 + (\gamma_5 + \gamma_5 \gamma_\mu) (U_\mu P_\mu)^\dagger W \mathcal{C} \gamma_5 \right]\right\} K. \nonumber \\
\end{eqnarray}
Using the fact that $\mathcal{C}$ commutes with $\gamma_5$, and that both of them (which act on spinor indices only) commmute with $W$ (which acts on color indices only), the latter expression reduces to
\begin{eqnarray}
\label{A_gamma5WilsonD_commutator}
\hspace{-10mm} [A, \gamma_5 D] &=& \frac{\kappa}{a} \sum_{\mu=1}^4 \left\{ \left[ -W \mathcal{C} U_\mu^\star P_\mu + \gamma_5 U_\mu P_\mu W \mathcal{C} \gamma_5 \right] + \left[ W \mathcal{C} \gamma_\mu^\star U_\mu^\star P_\mu - \gamma_5 \gamma_\mu U_\mu P_\mu W \mathcal{C} \gamma_5 \right] \phantom{(U_\mu^\dagger)^\star} \right. \nonumber \\
&& \left. + \left[ -W \mathcal{C} P_\mu^\dagger (U_\mu^\dagger)^\star + \gamma_5 P_\mu^\dagger U_\mu^\dagger W \mathcal{C} \gamma_5 \right] + \left[ -W \mathcal{C} \gamma_\mu^\star P_\mu^\dagger (U_\mu^\dagger)^\star + \gamma_5 \gamma_\mu P_\mu^\dagger U_\mu^\dagger W \mathcal{C} \gamma_5 \right] \right\} K. \nonumber \\
\end{eqnarray}
The pairs of terms in each square bracket sum up to zero: the second term in the first square bracket can be rewritten as
\begin{eqnarray}
\label{first-bracket_term}
\gamma_5 U_\mu P_\mu W \mathcal{C} \gamma_5 &=& U_\mu P_\mu W \gamma_5 \mathcal{C} \gamma_5 \nonumber \\
&=& W W^{-1} U_\mu P_\mu W \mathcal{C} \gamma_5^2 \nonumber \\
&=& W W^{-1} U_\mu P_\mu W \mathcal{C} \nonumber \\
&=& W U_\mu^\star P_\mu \mathcal{C} \nonumber \\
&=& W \mathcal{C} U_\mu^\star P_\mu,
\end{eqnarray}
while the second term in the second square bracket is
\begin{eqnarray}
\label{second-bracket_term}
- \gamma_5 \gamma_\mu U_\mu P_\mu W \mathcal{C} \gamma_5 &=& - U_\mu P_\mu W \gamma_5 \gamma_\mu \mathcal{C} \gamma_5 \nonumber \\
&=& U_\mu P_\mu W \gamma_\mu \gamma_5 \mathcal{C} \gamma_5 \nonumber \\
&=& U_\mu P_\mu W \gamma_\mu \mathcal{C} \gamma_5^2 \nonumber \\
&=& U_\mu P_\mu W \gamma_\mu \mathcal{C} \nonumber \\
&=& U_\mu P_\mu W \mathcal{C} \mathcal{C}^{-1} \gamma_\mu \mathcal{C} \nonumber \\
&=& - U_\mu P_\mu W \mathcal{C} \gamma_\mu^\star \nonumber \\
&=& - W W^{-1} U_\mu W P_\mu \mathcal{C} \gamma_\mu^\star \nonumber \\
&=& - W U_\mu^\star P_\mu \mathcal{C} \gamma_\mu^\star \nonumber \\
&=& - W \mathcal{C} \gamma_\mu^\star U_\mu^\star P_\mu.
\end{eqnarray}
In turn, the second term in the third bracket can be recast in the form
\begin{eqnarray}
\label{third-bracket_term}
\gamma_5 P_\mu^\dagger U_\mu^\dagger W \mathcal{C} \gamma_5 &=& P_\mu^\dagger U_\mu^\dagger W \gamma_5 \mathcal{C} \gamma_5 \nonumber \\
&=& P_\mu^\dagger W W^{-1} U_\mu^\dagger W \mathcal{C} \gamma_5^2 \nonumber \\
&=& P_\mu^\dagger W (U_\mu^\dagger)^\star \mathcal{C} \nonumber \\
&=& W \mathcal{C} P_\mu^\dagger (U_\mu^\dagger)^\star,
\end{eqnarray}
and the second term in the fourth bracket is equal to
\begin{eqnarray}
\label{fourth-bracket_term}
\gamma_5 \gamma_\mu P_\mu^\dagger U_\mu^\dagger W \mathcal{C} \gamma_5 &=& \gamma_5 \gamma_\mu P_\mu^\dagger W W^{-1} U_\mu^\dagger W \mathcal{C} \gamma_5 \nonumber \\
&=& \gamma_5 \gamma_\mu P_\mu^\dagger W (U_\mu^\dagger)^\star \mathcal{C} \gamma_5 \nonumber \\
&=& \gamma_5 \gamma_\mu \mathcal{C} \gamma_5 P_\mu^\dagger W (U_\mu^\dagger)^\star\nonumber \\
&=& \gamma_5 \mathcal{C}\mathcal{C}^{-1} \gamma_\mu \mathcal{C} \gamma_5 P_\mu^\dagger W (U_\mu^\dagger)^\star\nonumber \\
&=& -\gamma_5 \mathcal{C} \gamma_\mu^\star \gamma_5 P_\mu^\dagger W (U_\mu^\dagger)^\star\nonumber \\
&=& \gamma_5 \mathcal{C} \gamma_5 \gamma_\mu^\star P_\mu^\dagger W (U_\mu^\dagger)^\star\nonumber \\
&=& \gamma_5^2 \mathcal{C} \gamma_\mu^\star P_\mu^\dagger W (U_\mu^\dagger)^\star\nonumber \\
&=& \mathcal{C} \gamma_\mu^\star W P_\mu^\dagger (U_\mu^\dagger)^\star \nonumber \\
&=& W \mathcal{C} \gamma_\mu^\star P_\mu^\dagger (U_\mu^\dagger)^\star.
\end{eqnarray}
Using eqs.~(\ref{first-bracket_term}), (\ref{second-bracket_term}), (\ref{third-bracket_term}), and (\ref{fourth-bracket_term}) in eq.~(\ref{A_gamma5WilsonD_commutator}), one finds that
\begin{equation}
\label{A_gamma5WilsonD_commutator_vanishes}
[A, \gamma_5 D] =0.
\end{equation}

\subsection{Proof of the commutation relation $[B, \epsilon \Dst]=0$}
We show here that the operator $B$ defined in the main text commutes with $\epsilon \Dst$:
\begin{eqnarray}
\label{B_epsilonDst_commutator}
\hspace{-10mm} [B, \epsilon \Dst] &=& W K \epsilon \Dst - \epsilon \Dst W K \nonumber \\
&=& (W \epsilon \Dst^\star - \epsilon \Dst W) K \nonumber \\
&=& \left\{ W m \epsilon + \frac{1}{2a} \sum_{\mu=1}^4 \epsilon \eta_\mu W \left[ (U_\mu P_\mu)^\star - (U_\mu P_\mu)^{\dagger \star} \right] \right. \nonumber \\
&& \left. -m \epsilon W - \frac{1}{2a} \sum_{\mu=1}^4 \epsilon \eta_\mu \left[ (U_\mu P_\mu) - (U_\mu P_\mu)^{\dagger} \right] W \right\} K \nonumber \\
&=& \frac{1}{2a} \sum_{\mu=1}^4 \epsilon \eta_\mu \left( W U_\mu^\star P_\mu - W P^\dagger U_\mu^{\dagger \star} - U_\mu P_\mu W + P_\mu^\dagger U_\mu^\dagger W \right) K.
\end{eqnarray}
At this point, note that
\begin{eqnarray}
\label{WUmustarPmu}
W U_\mu^\star P_\mu &=& W U_\mu^\star W^{-1} W P_\mu \nonumber \\
&=& U_\mu W P_\mu \nonumber \\
&=& U_\mu P_\mu W
\end{eqnarray}
and that
\begin{eqnarray}
\label{WPmudaggerUmudaggerstar}
W P_\mu^\dagger U_\mu^{\dagger \star} &=& W P_\mu^\dagger W^{-1} U_\mu^{\dagger \star} W \nonumber \\
&=& WW^{-1} P_\mu^\dagger U_\mu^\dagger W \nonumber \\
&=& P_\mu^\dagger U_\mu^\dagger W.
\end{eqnarray}
Plugging equations~(\ref{WUmustarPmu}) and (\ref{WPmudaggerUmudaggerstar}) into eq.~(\ref{B_epsilonDst_commutator}) one finds:
\begin{eqnarray}
 [B, \epsilon \Dst] &=& 0.
\end{eqnarray}

\section{Derivative of the clover term}
\label{app:HMC_forces}
\renewcommand{\theequation}{C.\arabic{equation}}
\setcounter{equation}{0}

Inserting the explicit form of the clover term into eq.~(\ref{fermionic_variation2}) we have 
\begin{align}
    \delta S_f^{\rm clover} & = - \sum_x \frac{i}{2}c_{\rm sw}(g_0^2) \kappa \sum_{\mu,\nu} \left \{  X^{\dag}(x) (\delta\tilde{F}_{\mu \nu}(x))\sigma_{\mu \nu}Y(x) + Y^{\dag}(x)(\delta \tilde{F}_{\mu \nu}(x))\sigma_{\mu \nu}X(x)\right \} \nonumber \\ 
                            & = -\frac{i}{16}c_{\rm sw}(g_0^2)\kappa \sum_x \sum_{\mu,\nu}\tr \left [ (\delta \mathcal{Q}_{\mu \nu}(x) - \delta \mathcal{Q}^{\dag}_{\mu \nu}) \sigma_{\mu \nu}Y(x)X^{\dag}(x) + (\delta \mathcal{Q}_{\mu \nu}(x) - \delta \mathcal{Q}^{\dag}_{\mu \nu}) \sigma_{\mu \nu}X(x)Y^{\dag}(x) \right ] \nonumber \\
                            & = - \frac{i}{16} c_{\rm sw}(g_0^2)\kappa \sum_x \sum_{\mu,\nu}\tr \left [ \delta Q_{\mu \nu}\sigma_{\mu \nu}\Lambda(x) - h.c. \right ] \, ,
\end{align}
where 
\begin{gather}
    \Lambda(x) = Y(x)X^{\dag}(x) - X(x)Y^{\dag}(x) \, . 
\end{gather}
In order to write explicitly the variation of the clover plaquette, let us define the following upper ($C^{+}$) and lower ($C^-$) ``staple insertions'' as
\begin{align}
C^{1;+}_{\mu}(x) &= \sum_{\nu} \tr_{\rm spin}[\sigma_{\mu \nu} \Lambda(x+\hat{\mu})]U_{\nu}(x+\hat{\mu})U^{\dag}_{\nu}(x+\hat{\nu})U^{\dag}_{\nu}(x) \\
C^{2;+}_{\mu}(x) &= \sum_{\nu} U_{\nu}(x+\hat{\mu}) \tr_{\rm spin}[\sigma_{\mu \nu} \Lambda(x+\hat{\mu}+\hat{\nu})] U^{\dag}_{\nu}(x+\hat{\nu})U^{\dag}_{\nu}(x) \\
C^{3;+}_{\mu}(x) &= \sum_{\nu} U_{\nu}(x+\hat{\mu})U^{\dag}_{\nu}(x+\hat{\nu})\tr_{\rm spin}[\sigma_{\mu \nu} \Lambda(x+\hat{\nu})] U^{\dag}_{\nu}(x) \\
C^{4;+}_{\mu}(x) &= \sum_{\nu} U_{\nu}(x+\hat{\mu})U^{\dag}_{\nu}(x+\hat{\nu})U^{\dag}_{\nu}(x)\tr_{\rm spin}[\sigma_{\mu \nu} \Lambda(x)] \\
C^{1;-}_{\mu}(x) &= \sum_{\nu} \tr_{\rm spin}[\sigma_{\mu \nu}\Lambda(x+\hat{\mu})]U^{\dag}_{\nu}(x+\hat{\mu}-\hat{\nu})U^{\dag}_{\mu}(x-\hat{\nu})U_{\nu}(x-\hat{\nu}) \\
C^{2;-}_{\mu}(x) &= \sum_{\nu} U^{\dag}_{\nu}(x+\hat{\mu}-\hat{\nu})\tr_{\rm spin}[\sigma_{\mu \nu}\Lambda(x+\hat{\mu}-\hat{\nu})]U^{\dag}_{\mu}(x-\hat{\nu})U_{\nu}(x-\hat{\nu}) \\
C^{3;-}_{\mu}(x) &= \sum_{\nu} U^{\dag}_{\nu}(x+\hat{\mu}-\hat{\nu})U^{\dag}_{\mu}(x-\hat{\nu})\tr_{\rm spin}[\sigma_{\mu \nu}\Lambda(x-\hat{\nu})]U_{\nu}(x-\hat{\nu}) \\
C^{4;-}_{\mu}(x) &= \sum_{\nu} U^{\dag}_{\nu}(x+\hat{\mu}-\hat{\nu})U^{\dag}_{\mu}(x-\hat{\nu})U_{\nu}(x-\hat{\nu})\tr_{\rm spin}[\sigma_{\mu \nu}\Lambda(x)]
\end{align}
and then 
\begin{gather}
    C_{\mu}(x) = \sum_{s=1}^{4} \left [ C^{s;+}_{\mu}(x) - C^{s;-}_{\mu}(x) \right ]\, .
\end{gather}
Finally we have 
\begin{gather}
    \delta S_f^{\rm clov} = - \frac{i}{16} c_{\rm sw}(g_0^2)\kappa \sum_x \sum_{\mu,\nu}\tr_{\rm color} \left [ i \delta \alpha^a_{\mu}(x) T_R^a U_{\mu}(x)C_{\mu}(x) + i C^{\dag}_{\mu}(x)U^{\dag}_{\mu}(x)\delta \alpha^a_{\mu}(x)T_R^a \right].
\end{gather}
Note that the above equation holds for a generic representation $R$.

\end{appendix}

\begin{table}[htbp]
	\begin{scriptsize}
		\centering
		\begin{center}
			\begin{tabular}{cllllllcc}
				\toprule
				\rm{Ensemble} & $\beta$ & $am_{4}$  & $am_{6}$  & $\langle P \rangle$ & $|\lambda^{(4)}_{\rm min}|$ & $|\lambda^{(6)}_{\rm min}|$ & \rm{cnfg} & $\Delta\rm{MD}$ \\
				\midrule
				$A1$          & $10.00$ & $-0.20$   & $-0.20$   & $0.4597(8) $        & $0.6081(11)$                & $0.7561(11)$                & $415$     & $30$            \\
				$A2$          & $10.00$ & $-0.40$   & $-0.40$   & $0.5312(10)$        & $0.2633(12)$                & $0.4167(9)$                 & $320$     & $30$            \\
				$A3$          & $10.00$ & $-0.48$   & $-0.48$   & $0.5422(2)$         & $0.1601(16)$                & $0.3140(10)$                & $316$     & $50$            \\
				$A4$          & $10.00$ & $-0.45$   & $-0.48$   & $0.5403(2)$         & $0.1916(7)$                 & $0.3210(5)$                 & $331$     & $50$            \\
				$A5$          & $10.00$ & $-0.48$   & $-0.50$   & $0.5437(2)$         & $-$                         & $-$                         & $306$     & $50$            \\
				$A6$          & $10.00$ & $-0.50$   & $-0.50$   & $0.5446(2)$         & $0.1457(45)$                & $0.2899(8)$                 & $293$     & $50$            \\
				$A7$          & $10.00$ & $-0.50$   & $-0.52$   & $0.5453(4)$         & $-$                         & $-$                         & $90$      & $50$            \\  
				$A8$          & $10.00$ & $-0.53$   & $-0.53$   & $0.5482(2)$         & $0.1316(68)$                & $0.2576(38)$                & $274$     & $50$            \\
				$A9$          & $10.00$ & $-0.55$   & $-0.55$   & $0.5503(2)$         & $0.1278(62)$                & $0.2364(15)$                & $244$     & $50$            \\
				$A10$         & $10.00$ & $-0.55$   & $-0.58$   & $0.5521(2)$         & $-$                         & $-$                         & $228$     & $50$            \\
				$A12$         & $9.00$  & $-0.40$   & $-0.40$   & $0.35084(8)$        & $0.6260(4)$                 & $0.7680(3)$                 & $637$     & $50$            \\
				$A13$         & $9.00$  & $-0.45$   & $-0.45$   & $0.35182(8)$        & $0.5815(3)$                 & $0.7230(3)$                 & $632$     & $50$            \\
				$A14$         & $9.00$  & $-0.48$   & $-0.48$   & $0.35211(11)$       & $0.5566(4)$                 & $0.6968(4)$                 & $260$     & $50$            \\
				$A15$         & $9.00$  & $-0.50$   & $-0.50$   & $0.35259(8)$        & $0.5382(7)$                 & $0.6781(4)$                 & $609$     & $50$            \\
				$A16$         & $9.00$  & $-0.55$   & $-0.55$   & $0.35364(7)$        & $0.4944(6)$                 & $0.6338(4)$                 & $584$     & $50$            \\             
				$A17$         & $7.30$  & $-0.1977$ & $-0.2490$ & $0.26021(7)$        & $0.9160(1)$                 & $0.9984(1)$                 & $416$     & $20$            \\
				$A18$         & $7.33$  & $-0.1948$ & $-0.2490$ & $0.26163(8)$        & $0.9172(1)$                 & $0.99725(8)$                & $416$     & $20$            \\
				$A19$         & $7.50$  & $-0.1538$ & $-0.2264$ & $0.26912(8)$        & $0.9461(2)$                 & $1.0109(1)$                 & $416$     & $20$            \\
				$A20$         & $7.75$  & $-0.1240$ & $-0.1977$ & $0.28051(8)$        & $0.9607(2)$                 & $1.0267(2)$                 & $416$     & $20$            \\     
				$A21$         & $7.10$  & $-0.2043$ & $-0.2750$ & $0.25123(7)$        & $0.9193(2)$                 & $0.98270(9) $               & $416$     & $30$            \\
				$A22$         & $7.50$  & $-0.2043$ & $-0.2750$ & $0.26914(7)$        & $0.8992(1)$                 & $0.96701(9)$                & $416$     & $30$            \\
				$A23$         & $7.20$  & $-0.2100$ & $-0.2800$ & $0.25585(6)$        & $0.9098(2)$                 & $0.9743(1)$                 & $416$     & $30$            \\
				$A24$         & $7.75$  & $-0.2043$ & $-0.2750$ & $0.28099(8)$        & $0.88752(6)$                & $0.95639(6)$                & $416$     & $30$            \\
				$A25$         & $7.30$  & $-0.2043$ & $-0.2750$ & $0.26028(6)$        & $0.9102(2)$                 & $0.9749(1)$                 & $416$     & $30$            \\
				$A26$         & $7.33$  & $-0.2043$ & $-0.2750$ & $0.26163(8)$        & $0.9085(2)$                 & $0.9736(1)$                 & $416$     & $30$            \\ 
				$A30$         & $10.00$ & $-0.45$   & $-0.45$   & $0.5379(4)$         & $0.1968(12)$                & $0.3507(10)$                & $352$     & $50$            \\ 
				$A32$         & $9.50$  & $-0.40$   & $-0.40$   & $0.3898(2)$         & $0.5684(3)$                 & $0.7132(4)$                 & $260$     & $50$            \\ 
				$A33$         & $9.50$  & $-0.50$   & $-0.50$   & $0.3934(3)$         & $0.4772(6)$                 & $0.6201(7)$                 & $236$     & $50$            \\ 
				$A34$         & $9.50$  & $-0.55$   & $-0.55$   & $0.3954(2)$         & $0.4317(3)$                 & $0.5738(3)$                 & $221$     & $50$            \\ 
				$A35$         & $9.70$  & $-0.45$   & $-0.45$   & $0.3954(2)$         & $0.4822(6)$                 & $0.6277(7)$                 & $237$     & $50$            \\ 
				$A36$         & $9.70$  & $-0.55$   & $-0.55$   & $0.4215(7)$         & $0.3846(16)$                & $0.5272(15)$                & $167$     & $50$            \\
				\bottomrule
			\end{tabular}
		\end{center}
		\caption{Table run, volume $(8^3\times16)a^4$, plaquette gauge action and fermionic Wilson-clover $N_f=2+2$ action. Runs $A17-A26$ use the same bare parameters as in ref.~\cite{DeGrand:2015lna}, however a direct comparison cannot be done, since in this work we use a different gauge action with respect to ref.~\cite{DeGrand:2015lna}. Nevertheless, the tension between our results and the ones in ref.~\cite{DeGrand:2015lna} seems to indicate a surprisingly relevant shift of the line of constant physics due to the smearing procedure.}
		\label{tab:runtable1}
	\end{scriptsize}
\end{table}

\clearpage

\begin{table}[htbp]
	\begin{scriptsize}
		\centering
		\begin{center}
			\begin{tabular}{cllllllcc}
				\toprule
				\rm{Ensemble} & $\beta$ & $am_{4}$ & $am_{6}$ & $\langle P \rangle$ & $|\lambda^{(4)}_{\rm min}|$ & $|\lambda^{(6)}_{\rm min}|$ & \rm{cnfg} & $\Delta\rm{MD}$ \\
				\midrule
				$B1$          & $11.00$ & $-0.45$  & $-0.45$  & $0.60909(5)$        & $0.03763(56)$               & $0.17799(21)$               & $210$     & $80$            \\
				$C1$          & $10.70$ & $-0.45$  & $-0.45$  & $0.59263(4)$        & $0.06646(12)$               & $0.21411(13)$               & $265$     & $50$            \\
				$D1$          & $10.50$ & $-0.50$  & $-0.50$  & $0.58341(5)$        & $0.04051(23)$               & $0.19050(18)$               & $124$     & $80$            \\
				$E1$          & $10.30$ & $-0.50$  & $-0.50$  & $0.56996(5)$        & $0.06896(38)$               & $0.22217(30)$               & $172$     & $80$            \\
				$W1$          & $10.20$ & $-0.52$  & $-0.52$  & $0.56406(4)$        & $0.06453(23)$               & $0.21773(22)$               & $229$     & $80$            \\
				$X1$          & $10.10$ & $-0.55$  & $-0.55$  & $0.55868(5)$        & $0.04876(57)$               & $0.20371(33)$               & $153$     & $80$            \\
				$F1$          & $10.00$ & $-0.50$  & $-0.50$  & $0.54392(7)$        & $0.12849(37)$               & $0.28444(25)$               & $270$     & $50$            \\
				$F2$          & $10.00$ & $-0.55$  & $-0.55$  & $0.54977(9)$        & $0.06990(36)$               & $0.22517(53)$               & $220$     & $50$            \\
				$F3$          & $10.00$ & $-0.58$  & $-0.58$  & $0.55226(99)$       & $0.0443(61)$                & $0.1928(27)$                & $256$     & $80$            \\
				$G1$          & $9.70$  & $-0.55$  & $-0.55$  & $0.42166(7)$        & $0.37414(30)$               & $0.51579(42)$               & $350$     & $50$            \\
				$G2$          & $9.70$  & $-0.58$  & $-0.58$  & $0.4355(86)$        & $0.299(27)$                 & $0.441(27)$                 & $473$     & $80$            \\
				$G3$          & $9.70$  & $-0.60$  & $-0.60$  & $0.461(14)$         & $0.229(36)$                 & $0.372(35)$                 & $281$     & $80$            \\
				$G4$          & $9.70$  & $-0.62$  & $-0.62$  & $0.459(13)$         & $0.176(36)$                 & $0.319(35)$                 & $289$     & $80$            \\
				$G5$          & $9.70$  & $-0.66$  & $-0.66$  & $0.40744(11)$       & $0.2501(15)$                & $0.3869(14)$                & $252$     & $100$           \\
				$H1$          & $9.50$  & $-0.50$  & $-0.50$  & $0.39315(5)$        & $0.46836(21)$               & $0.60984(18)$               & $369$     & $50$            \\
				$H2$          & $9.50$  & $-0.55$  & $-0.55$  & $0.39518(5)$        & $0.42320(30)$               & $0.56550(34)$               & $362$     & $50$            \\
				$H3$          & $9.50$  & $-0.58$  & $-0.58$  & $0.4006(42)$        & $0.3871(92)$                & $0.5287(93)$                & $239$     & $40$            \\
				$H4$          & $9.50$  & $-0.60$  & $-0.60$  & $0.4048(72)$        & $0.358(21)$                 & $0.499(20)$                 & $245$     & $40$            \\
				$H5$          & $9.50$  & $-0.62$  & $-0.62$  & $0.4080(66)$        & $0.325(19)$                 & $0.464(20)$                 & $206$     & $80$            \\
				$H6$          & $9.50$  & $-0.66$  & $-0.66$  & $0.40145(7)$        & $0.32337(80)$               & $0.46156(57)$               & $84$      & $80$            \\
				$H7$          & $9.50$  & $-0.70$  & $-0.70$  & $0.39712(10)$       & $0.28820(66)$               & $0.42525(82)$               & $81$      & $80$            \\ 
				$I1$          & $9.00$  & $-0.50$  & $-0.50$  & $0.35265(4)$        & $0.53103(34)$               & $0.67267(24)$               & $253$     & $50$            \\
				$I2$          & $9.00$  & $-0.55$  & $-0.55$  & $0.35364(5)$        & $0.48762(26)$               & $0.62735(26)$               & $375$     & $50$            \\ 
				$I3$          & $9.00$  & $-0.58$  & $-0.58$  & $0.35430(10)$       & $0.46203(31)$               & $0.60068(39)$               & $365$     & $50$            \\
				$I4$          & $9.00$  & $-0.60$  & $-0.60$  & $0.29260(3)$        & $0.54014(19)$               & $0.68787(18)$               & $379$     & $50$            \\
				$I5$          & $9.00$  & $-0.62$  & $-0.62$  & $0.35534(3)$        & $0.42734(25)$               & $0.56563(30)$               & $398$     & $40$            \\
				$I6$          & $9.00$  & $-0.66$  & $-0.66$  & $0.34476(4)$        & $0.39496(47)$               & $0.53155(40)$               & $392$     & $40$            \\
				$I7$          & $9.00$  & $-0.70$  & $-0.70$  & $0.35529(5)$        & $0.36097(31)$               & $0.49591(40)$               & $86$      & $80$            \\
				$J1$          & $8.70$  & $-0.50$  & $-0.50$  & $0.33356(5)$        & $0.55701(25)$               & $0.69583(20)$               & $277$     & $40$            \\
				$J2$          & $8.70$  & $-0.55$  & $-0.55$  & $0.33427(5)$        & $0.51298(41)$               & $0.65164(28)$               & $270$     & $40$            \\
				$J3$          & $8.70$  & $-0.58$  & $-0.58$  & $0.33483(3)$        & $0.48786(25)$               & $0.62515(21)$               & $402$     & $40$            \\
				$J4$          & $8.70$  & $-0.60$  & $-0.60$  & $0.33526(4)$        & $0.47040(51)$               & $0.60809(23)$               & $399$     & $40$            \\
				$J5$          & $8.70$  & $-0.62$  & $-0.62$  & $0.33559(7)$        & $0.45422(30)$               & $0.59004(38)$               & $268$     & $40$            \\
				$J6$          & $8.70$  & $-0.66$  & $-0.66$  & $0.32763(3)$        & $0.42178(22)$               & $0.55645(29)$               & $398$     & $40$            \\
				$J7$          & $8.70$  & $-0.70$  & $-0.70$  & $0.33593(4)$        & $0.38681(33)$               & $0.52120(33)$               & $88$      & $80$            \\
				$J8$          & $8.70$  & $-0.75$  & $-0.75$  & $0.33319(3)$        & $0.34661(32)$               & $0.47763(40)$               & $75$      & $80$            \\
				$K1$          & $8.50$  & $-0.50$  & $-0.50$  & $0.32189(2)$        & $0.57159(25)$               & $0.70954(21)$               & $273$     & $40$            \\
				$K2$          & $8.50$  & $-0.55$  & $-0.55$  & $0.32258(3)$        & $0.52880(33)$               & $0.66528(23)$               & $275$     & $40$            \\
				$K3$          & $8.50$  & $-0.58$  & $-0.58$  & $0.32302(3)$        & $0.50294(29)$               & $0.63816(27)$               & $271$     & $40$            \\
				$K4$          & $8.50$  & $-0.60$  & $-0.60$  & $0.32327(4)$        & $0.48584(30)$               & $0.62112(22)$               & $271$     & $40$            \\
				$K5$          & $8.50$  & $-0.62$  & $-0.62$  & $0.32368(3)$        & $0.46872(28)$               & $0.60359(20)$               & $406$     & $40$            \\
				$K6$          & $8.50$  & $-0.66$  & $-0.66$  & $0.31696(2)$        & $0.43595(46)$               & $0.57044(20)$               & $267$     & $40$            \\
				$K7$          & $8.50$  & $-0.70$  & $-0.70$  & $0.29659(3)$        & $0.40651(37)$               & $0.53884(31)$               & $89$      & $80$            \\
				$K8$          & $8.50$  & $-0.75$  & $-0.75$  & $0.32632(5)$        & $0.36125(32)$               & $0.49169(46)$               & $79$      & $80$            \\
				\bottomrule
			\end{tabular}
		\end{center}
		\caption{Table run, volume $(16^3\times32)a^4$, plaquette gauge action and fermionic Wilson-clover $N_f=2+2$ action.}
		 \label{tab:runtable2}
	\end{scriptsize}
\end{table}

\begin{table}[htbp]
	\begin{scriptsize}
		\centering
		\begin{center}
			\begin{tabular}{llll|llll}
				\toprule
				\multirow{3}{*}{\rm{Ensemble}} & \multirow{3}{*}{$aM^{(4)}_{P}$} & $aM^{(4)}_{V_1}$ & \multirow{3}{*}{$aM^{(4)}_{\rm PCAC}$} & \multirow{3}{*}{\rm{Ensemble}} & \multirow{3}{*}{$aM^{(4)}_{P}$} & $aM^{(4)}_{V_1}$ & \multirow{3}{*}{$aM^{(4)}_{\rm PCAC}$} \\                                 
				                               &                                   & $aM^{(4)}_{V_2}$ &                                        &                                &                                   & $aM^{(4)}_{V_2}$ &                                        \\
				                               &                                   & $aM^{(4)}_{V_3}$ &                                        &                                &                                   & $aM^{(4)}_{V_3}$ &                                        \\
				\midrule
				\multirow{3}{*}{$A1$}          & \multirow{3}{*}{$1.9528(32)$}     & $2.0212(32)$        & \multirow{3}{*}{$1.1057(35)$}          & \multirow{3}{*}{$A16$}         & \multirow{3}{*}{$1.8835(8)$}      & $1.9731(16)$        & \multirow{3}{*}{$0.9661(12)$}          \\       
				                               &                                   & $2.0222(35)$        &                                        &                                &                                   & $1.9718(19)$        &                                        \\                     
				                               &                                   & $2.0219(35)$        &                                        &                                &                                   & $1.9747(8)$         &                                        \\                            
				\midrule  
				\multirow{3}{*}{$A2$}          & \multirow{3}{*}{$1.1826(24)$}     & $1.2659(27)$        & \multirow{3}{*}{$0.3668(12)$}          & \multirow{3}{*}{$A17$}         & \multirow{3}{*}{$2.3258(10)$}     & $2.3712(10)$        & \multirow{3}{*}{$1.8175(21)$}          \\   
				                               &                                   & $1.2668(27)$        &                                        &                                &                                   & $2.3706(10)$        &                                        \\                     
				                               &                                   & $1.2679(27)$        &                                        &                                &                                   & $2.3696(12)$        &                                        \\                            
				\midrule  
				\multirow{3}{*}{$A3$}          & \multirow{3}{*}{$0.8971(52)$}     & $0.9625(77)$        & \multirow{3}{*}{$0.2043(52)$}          & \multirow{3}{*}{$A18$}         & \multirow{3}{*}{$2.3230(11)$}     & $2.3653(14)$        & \multirow{3}{*}{$1.8129(21)$}          \\
				                               &                                   & $0.9867(61)$        &                                        &                                &                                   & $2.3655(14)$        &                                        \\                     
				                               &                                   & $0.9788(71)$        &                                        &                                &                                   & $2.3659(14)$        &                                        \\                            
				\midrule  
				\multirow{3}{*}{$A4$}          & \multirow{3}{*}{$0.9264(37)$}     & $0.9859(48)$        & \multirow{3}{*}{$0.2447(9)$}           & \multirow{3}{*}{$A19$}         & \multirow{3}{*}{$2.3506(6)$}      & $2.3929(8)$         & \multirow{3}{*}{$1.8809(19)$}          \\ 
				                               &                                   & $0.9753(47)$        &                                        &                                &                                   & $2.3923(7)$         &                                        \\                     
				                               &                                   & $0.9838(46)$        &                                        &                                &                                   & $2.3928(8)$         &                                        \\                            
				\midrule  
				\multirow{3}{*}{$A5$}          & \multirow{3}{*}{$0.8756(90)$}     & $0.9564(102)$       & \multirow{3}{*}{$0.1975(10)$}          & \multirow{3}{*}{$A20$}         & \multirow{3}{*}{$2.3608(9)$}      & $2.4012(9)$         & \multirow{3}{*}{$1.9042(17)$}          \\
				                               &                                   & $0.9465(116)$       &                                        &                                &                                   & $2.4007(9)$         &                                        \\                     
				                               &                                   & $0.9427(127)$       &                                        &                                &                                   & $2.4015(8)$         &                                        \\                            
				\midrule  
				\multirow{3}{*}{$A6$}          & \multirow{3}{*}{$0.8342(78)$}     & $0.9161(86)$        & \multirow{3}{*}{$0.1719(9)$}           & \multirow{3}{*}{$A21$}         & \multirow{3}{*}{$2.3289(7)$}      & $2.3724(8)$         & \multirow{3}{*}{$1.8263(17)$}          \\
				                               &                                   & $0.9153(123)$       &                                        &                                &                                   & $2.3730(9)$         &                                        \\                     
				                               &                                   & $0.9184(111)$       &                                        &                                &                                   & $2.3732(9)$         &                                        \\                            
				\midrule  
				\multirow{3}{*}{$A7$}          & \multirow{3}{*}{$0.8366(87)$}     & $0.9344(114)$       & \multirow{3}{*}{$0.1713(16)$}          & \multirow{3}{*}{$A22$}         & \multirow{3}{*}{$2.3093(8)$}      & $2.3531(10)$        & \multirow{3}{*}{$1.7788(15)$}          \\
				                               &                                   & $0.9140(87)$        &                                        &                                &                                   & $2.3538(9)$         &                                        \\                     
				                               &                                   & $0.9267(90)$        &                                        &                                &                                   & $2.3539(11)$        &                                        \\                            
				\midrule  
				\multirow{3}{*}{$A8$}          & \multirow{3}{*}{$0.8129(119)$}    & $0.9395(161)$       & \multirow{3}{*}{$0.1274(11)$}          & \multirow{3}{*}{$A23$}         & \multirow{3}{*}{$2.3181(7)$}      & $2.3617(9)$         & \multirow{3}{*}{$1.8019(16)$}          \\
				                               &                                   & $0.8928(153)$       &                                        &                                &                                   & $2.3613(8)$         &                                        \\                     
				                               &                                   & $0.9066(126)$       &                                        &                                &                                   & $2.3632(9)$         &                                        \\                            
				\midrule  
				\multirow{3}{*}{$A9$}          & \multirow{3}{*}{$0.7335(67)$}     & $0.8050(91)$        & \multirow{3}{*}{$0.0959(6)$}           & \multirow{3}{*}{$A24$}         & \multirow{3}{*}{$2.3026(8)$}      & $2.3489(10)$        & \multirow{3}{*}{$1.7594(18)$}          \\
				                               &                                   & $0.8231(58)$        &                                        &                                &                                   & $2.3484(11)$        &                                        \\                     
				                               &                                   & $0.9222(112)$       &                                        &                                &                                   & $2.3487(12)$        &                                        \\                            
				\midrule
				\multirow{3}{*}{$A10$}         & \multirow{3}{*}{$0.6927(120)$}    & $0.8200(123)$       & \multirow{3}{*}{$0.0891(12)$}          & \multirow{3}{*}{$A25$}         & \multirow{3}{*}{$2.3207(10)$}     & $2.3657(10)$        & \multirow{3}{*}{$1.8051(16)$}          \\
				                               &                                   & $0.7892(150)$       &                                        &                                &                                   & $2.3646(10)$        &                                        \\                     
				                               &                                   & $0.7869(163)$       &                                        &                                &                                   & $2.3657(10)$        &                                        \\                            
				\midrule  
				\multirow{3}{*}{$A12$}         & \multirow{3}{*}{$2.0405(13)$}     & $2.1094(15)$        & \multirow{3}{*}{$1.2193(16)$}          & \multirow{3}{*}{$A26$}         & \multirow{3}{*}{$2.3197(7)$}      & $2.3631(10)$        & \multirow{3}{*}{$1.8033(15)$}          \\
				                               &                                   & $2.1096(17)$        &                                        &                                &                                   & $2.3630(10)$        &                                        \\                     
				                               &                                   & $2.1104(17)$        &                                        &                                &                                   & $2.3637(11)$        &                                        \\                            
				\midrule  
				\multirow{3}{*}{$A13$}         & \multirow{3}{*}{$1.9906(4)$}      & $2.0647(8)$         & \multirow{3}{*}{$1.1353(16)$}          & \multirow{3}{*}{$A30$}         & \multirow{3}{*}{$0.9737(38)$}     & $1.0575(51)$        & \multirow{3}{*}{$0.2586(10)$}          \\
				                               &                                   & $2.0645(15)$        &                                        &                                &                                   & $1.0536(51)$        &                                        \\                     
				                               &                                   & $2.0636(6)$         &                                        &                                &                                   & $1.0518(51)$        &                                        \\                            
				\midrule  
				\multirow{3}{*}{$A14$}         & \multirow{3}{*}{$1.9607(10)$}     & $2.0398(12)$        & \multirow{3}{*}{$1.0847(26)$}          & \multirow{3}{*}{$A32$}         & \multirow{3}{*}{$1.9644(12)$}     & $2.0401(14)$        & \multirow{3}{*}{$1.0944(14)$}          \\
				                               &                                   & $2.0392(17)$        &                                        &                                &                                   & $2.0416(14)$        &                                        \\                     
				                               &                                   & $2.0399(15)$        &                                        &                                &                                   & $2.0420(15)$        &                                        \\                            
				\midrule
				\multirow{3}{*}{$A15$}         & \multirow{3}{*}{$1.9412(14)$}     & $2.0231(18)$        & \multirow{3}{*}{$1.0524(20)$}          & \multirow{3}{*}{$A33$}         & \multirow{3}{*}{$1.8410(19)$}     & $1.9298(24)$        & \multirow{3}{*}{$0.9100(23)$}          \\
				                               &                                   & $2.0232(16)$        &                                        &                                &                                   & $1.9297(23)$        &                                        \\                     
				                               &                                   & $2.0217(16)$        &                                        &                                &                                   & $1.9294(21)$        &                                        \\                            
				\bottomrule
			\end{tabular}
		\end{center}
		\caption{This table reports the value of the masses for the pseudoscalar ($M_{P}$) and the vector ($M_{V}$) states, together with the PCAC fermion mass. We note, as a consistency check, that the vector particle masses evaluated from correlators of their three different components are compatible with each other.}
		\label{tab:runtable_meas1}
	\end{scriptsize}
\end{table}

\begin{table}[htbp]
	\begin{scriptsize}
		\centering
		\begin{center}
			\begin{tabular}{llll}
				\toprule
				\multirow{3}{*}{\rm{Ensemble}} & \multirow{3}{*}{$aM^{(4)}_{P}$} & $aM^{(4)}_{V_1}$ & \multirow{3}{*}{$aM^{(4)}_{\rm PCAC}$} \\                 
				                               &                                   & $aM^{(4)}_{V_2}$ &                                        \\
				                               &                                   & $aM^{(4)}_{V_3}$ &                                        \\
				\midrule  
				\multirow{3}{*}{$A34$}         & \multirow{3}{*}{$1.7797(14)$}     & $1.8782(20)$        & \multirow{3}{*}{$0.8254(13)$}          \\
				                               &                                   & $1.8776(19)$        &                                        \\                     
				                               &                                   & $1.8789(20)$        &                                        \\                            
				\midrule  
				\multirow{3}{*}{$A35$}         & \multirow{3}{*}{$1.8388(19)$}     & $1.9284(29)$        & \multirow{3}{*}{$0.9097(21)$}          \\
				                               &                                   & $1.9273(22)$        &                                        \\                     
				                               &                                   & $1.9267(23)$        &                                        \\                            
				\midrule  
				\multirow{3}{*}{$A36$}         & \multirow{3}{*}{$1.6974(21)$}     & $1.8045(17)$        & \multirow{3}{*}{$0.7285(23)$}          \\
				                               &                                   & $1.8058(26)$        &                                        \\                     
				                               &                                   & $1.8042(22)$        &                                        \\                            
																								
				\bottomrule
			\end{tabular}
		\end{center}
		\caption{Table~\ref{tab:runtable_meas1}, continued.}
		\label{tab:runtable_meas1b}
	\end{scriptsize}
\end{table}

\begin{table}[htbp]
	\begin{scriptsize}
		\centering
		\begin{center}
			\begin{tabular}{llllllll}
				\toprule
				\multirow{3}{*}{\rm{Ensemble}} & \multirow{3}{*}{$aM^{(4)}_{P}$} & $aM^{(4)}_{V_1}$ & \multirow{3}{*}{$aM^{(4)}_{\rm PCAC}$} & \multirow{3}{*}{$aM^{(6)}_{P}$} & $aM^{(6)}_{V_1}$ & \multirow{3}{*}{$aM^{(6)}_{\rm PCAC}$} & \multirow{3}{*}{$t_0/a^2$}  \\                 
				                               &                                   & $aM^{(4)}_{V_2}$ &                                        &                                   & $aM^{(6)}_{V_2}$ &                                        &                             \\
				                               &                                   & $aM^{(4)}_{V_3}$ &                                        &                                   & $aM^{(6)}_{V_3}$ &                                        &                             \\
				\midrule     
				\multirow{3}{*}{$B1$}          & \multirow{3}{*}{$0.3412(35)$}     & $0.4411(47)$        & \multirow{3}{*}{$0.0448(3)$}           & \multirow{3}{*}{$0.7726(24)$}     & $0.7946(25)$        & \multirow{3}{*}{$0.2258(2)$}           & \multirow{3}{*}{$11.48(3)$} \\
				                               &                                   & $0.4553(46)$        &                                        &                                   & $0.7909(29)$        &                                        &                             \\ 
				                               &                                   & $0.4279(33)$        &                                        &                                   & $0.7957(37)$        &                                        &                             \\         \midrule
				\multirow{3}{*}{$C1$}          & \multirow{3}{*}{$0.4062(23)$}     & $0.4702(39)$        & \multirow{3}{*}{$0.0810(2)$}           & \multirow{3}{*}{$0.9115(14)$}     & $0.9376(19)$        & \multirow{3}{*}{$0.2799(3)$}           & \multirow{3}{*}{$8.74(5)$}  \\
				                               &                                   & $0.4626(31)$        &                                        &                                   & $0.9425(16)$        &                                        &                             \\ 
				                               &                                   & $0.4801(41)$        &                                        &                                   & $0.9405(16)$        &                                        &                             \\          \midrule  
				\multirow{3}{*}{$D1$}          & \multirow{3}{*}{$0.3144(31)$}     & $0.4201(36)$        & \multirow{3}{*}{$0.0517(2)$}           & \multirow{3}{*}{$0.8709(12)$}     & $0.9091(16)$        & \multirow{3}{*}{$0.2530(2)$}           & \multirow{3}{*}{$5.96(9)$}  \\
				                               &                                   & $0.4205(52)$        &                                        &                                   & $0.9087(18)$        &                                        &                             \\ 
				                               &                                   & $0.4172(51)$        &                                        &                                   & $0.9073(14)$        &                                        &                             \\           \midrule  
				\multirow{3}{*}{$E1$}          & \multirow{3}{*}{$0.4463(31)$}     & $0.5136(36)$        & \multirow{3}{*}{$0.0846(3)$}           & \multirow{3}{*}{$0.9805(23)$}     & $1.0175(27)$        & \multirow{3}{*}{$0.3014(5)$}           & \multirow{3}{*}{$6.58(12)$} \\
				                               &                                   & $0.5041(49)$        &                                        &                                   & $1.0147(24)$        &                                        &                             \\ 
				                               &                                   & $0.5051(43)$        &                                        &                                   & $1.0162(27)$        &                                        &                             \\           \midrule  
				\multirow{3}{*}{$W1$}          & \multirow{3}{*}{$0.4401(42)$}     & $0.5165(53)$        & \multirow{3}{*}{$0.0794(3)$}           & \multirow{3}{*}{$0.9938(22)$}     & $1.0359(32)$        & \multirow{3}{*}{$0.3010(6)$}           & \multirow{3}{*}{$7.00(17)$} \\
				                               &                                   & $0.5231(48)$        &                                        &                                   & $1.0374(45)$        &                                        &                             \\ 
				                               &                                   & $0.5270(50)$        &                                        &                                   & $1.0367(40)$        &                                        &                             \\           \midrule  
				\multirow{3}{*}{$X1$}          & \multirow{3}{*}{$0.3951(29)$}     & $0.4822(83)$        & \multirow{3}{*}{$0.0629(4)$}           & \multirow{3}{*}{$0.9775(13)$}     & $1.0241(24)$        & \multirow{3}{*}{$0.2861(5)$}           & \multirow{3}{*}{$5.36(4)$}  \\
				                               &                                   & $0.4959(54)$        &                                        &                                   & $1.0261(23)$        &                                        &                             \\ 
				                               &                                   & $0.4756(106)$       &                                        &                                   & $1.0255(30)$        &                                        &                             \\           \midrule  
				\multirow{3}{*}{$F1$}          & \multirow{3}{*}{$0.7228(20)$}     & $0.7908(28)$        & \multirow{3}{*}{$0.1672(5)$}           & \multirow{3}{*}{$1.2333(21)$}     & $1.2862(27)$        & \multirow{3}{*}{$0.4217(8)$}           & \multirow{3}{*}{$2.83(3)$}  \\
				                               &                                   & $0.7871(28)$        &                                        &                                   & $1.2845(29)$        &                                        &                             \\ 
				                               &                                   & $0.7918(21)$        &                                        &                                   & $1.2846(27)$        &                                        &                             \\          \midrule
				\multirow{3}{*}{$F2$}          & \multirow{3}{*}{$0.4867(30)$}     & $0.5665(49)$        & \multirow{3}{*}{$0.0895(3)$}           & \multirow{3}{*}{$1.0544(17)$}     & $1.1067(26)$        & \multirow{3}{*}{$0.3228(5)$}           & \multirow{3}{*}{$4.61(8)$}  \\
				                               &                                   & $0.5715(56)$        &                                        &                                   & $1.1065(29)$        &                                        &                             \\ 
				                               &                                   & $0.5751(54)$        &                                        &                                   & $1.1056(28)$        &                                        &                             \\           \midrule  
				\multirow{3}{*}{$F3$}          & \multirow{3}{*}{$0.3541(91)$}     & $0.4686(122)$       & \multirow{3}{*}{$0.0466(7)$}           & \multirow{3}{*}{$0.9417(18)$}     & $0.9896(32)$        & \multirow{3}{*}{$0.2689(10)$}          & \multirow{3}{*}{$5.92(8)$}  \\
				                               &                                   & $0.4737(105)$       &                                        &                                   & $0.9908(31)$        &                                        &                             \\ 
				                               &                                   & $0.4753(180)$       &                                        &                                   & $0.9893(31)$        &                                        &                             \\                  
				\bottomrule
			\end{tabular}
		\end{center}
		\caption{Same as in table~\ref{tab:runtable_meas1}, but including also meson-like states constructed from fermions in the sextet representation. We observe that at fix bare fermion mass these states are 
		heavier than the ones built from fundamental fermions. This observation is consistently supported by the value of the pseudoscalar-state masses, PCAC fermion masses, as well as the average smallest eigenvalue of the Dirac-Wilson operator. In the last column, we also report the value of the scale-setting parameter $t_0/a^2$.} 
		\label{tab:runtable_meas2}
	\end{scriptsize}
\end{table}

\bibliography{paper}

\providecommand{\href}[2]{#2}\begingroup\begin{thebibliography}{10}

\bibitem{Aad:2012tfa}
{\bf ATLAS} Collaboration, G.~Aad {\em et.~al.}, {\it {Observation of a new
  particle in the search for the Standard Model Higgs boson with the ATLAS
  detector at the LHC}},  {\em Phys. Lett.} {\bf B716} (2012) 1--29,
  [\href{https://arxiv.org/abs/1207.7214}{{\ttfamily 1207.7214}}].

\bibitem{Chatrchyan:2012xdj}
{\bf CMS} Collaboration, S.~Chatrchyan {\em et.~al.}, {\it {Observation of a
  new boson at a mass of 125 GeV with the CMS experiment at the LHC}},  {\em
  Phys. Lett.} {\bf B716} (2012) 30--61,
  [\href{https://arxiv.org/abs/1207.7235}{{\ttfamily 1207.7235}}].

\bibitem{Fukuda:1998mi}
{\bf Super-Kamiokande} Collaboration, Y.~Fukuda {\em et.~al.}, {\it {Evidence
  for oscillation of atmospheric neutrinos}},  {\em Phys. Rev. Lett.} {\bf 81}
  (1998) 1562--1567, [\href{https://arxiv.org/abs/hep-ex/9807003}{{\ttfamily
  hep-ex/9807003}}].

\bibitem{Ahmad:2002jz}
{\bf SNO} Collaboration, Q.~R. Ahmad {\em et.~al.}, {\it {Direct evidence for
  neutrino flavor transformation from neutral current interactions in the
  Sudbury Neutrino Observatory}},  {\em Phys. Rev. Lett.} {\bf 89} (2002)
  011301, [\href{https://arxiv.org/abs/nucl-ex/0204008}{{\ttfamily
  nucl-ex/0204008}}].

\bibitem{Ade:2015xua}
{\bf Planck} Collaboration, P.~A.~R. Ade {\em et.~al.}, {\it {Planck 2015
  results. XIII. Cosmological parameters}},  {\em Astron. Astrophys.} {\bf 594}
  (2016) A13, [\href{https://arxiv.org/abs/1502.01589}{{\ttfamily
  1502.01589}}].

\bibitem{Dawson:2018dcd}
S.~Dawson, C.~Englert, and T.~Plehn, {\it {Higgs Physics: It ain't over till
  it's over}},  \href{https://arxiv.org/abs/1808.01324}{{\ttfamily
  1808.01324}}.

\bibitem{Wess:1974tw}
J.~Wess and B.~Zumino, {\it {Supergauge Transformations in Four-Dimensions}},
  {\em Nucl. Phys.} {\bf B70} (1974) 39--50. [, 24 (1974)].

\bibitem{Coleman:1967ad}
S.~R. Coleman and J.~Mandula, {\it {All Possible Symmetries of the S Matrix}},
  {\em Phys. Rev.} {\bf 159} (1967) 1251--1256.

\bibitem{Kaplan:1983fs}
D.~B. Kaplan and H.~Georgi, {\it {SU(2) x U(1) Breaking by Vacuum
  Misalignment}},  {\em Phys. Lett.} {\bf 136B} (1984) 183--186.

\bibitem{Kaplan:1983sm}
D.~B. Kaplan, H.~Georgi, and S.~Dimopoulos, {\it {Composite Higgs Scalars}},
  {\em Phys. Lett.} {\bf 136B} (1984) 187--190.

\bibitem{Hill:2002ap}
C.~T. Hill and E.~H. Simmons, {\it {Strong dynamics and electroweak symmetry
  breaking}},  {\em Phys. Rept.} {\bf 381} (2003) 235--402,
  [\href{https://arxiv.org/abs/hep-ph/0203079}{{\ttfamily hep-ph/0203079}}].
  [Erratum: Phys. Rept. {\bf 390}, 553 (2004)].

\bibitem{Peskin:1990zt}
M.~E. Peskin and T.~Takeuchi, {\it {A New constraint on a strongly interacting
  Higgs sector}},  {\em Phys. Rev. Lett.} {\bf 65} (1990) 964--967.

\bibitem{Kaplan:1991dc}
D.~B. Kaplan, {\it {Flavor at SSC energies: A New mechanism for dynamically
  generated fermion masses}},  {\em Nucl. Phys.} {\bf B365} (1991) 259--278.

\bibitem{Contino:2010rs}
R.~Contino, {\it {The Higgs as a Composite Nambu-Goldstone Boson}},  in {\em
  {Physics of the large and the small, TASI 09, proceedings of the Theoretical
  Advanced Study Institute in Elementary Particle Physics, Boulder, Colorado,
  USA, 1-26 June 2009}}, pp.~235--306, 2011.
\newblock \href{https://arxiv.org/abs/1005.4269}{{\ttfamily 1005.4269}}.

\bibitem{Ferretti:2013kya}
G.~Ferretti and D.~Karateev, {\it {Fermionic UV completions of Composite Higgs
  models}},  {\em JHEP} {\bf 03} (2014) 077,
  [\href{https://arxiv.org/abs/1312.5330}{{\ttfamily 1312.5330}}].

\bibitem{Ferretti:2014qta}
G.~Ferretti, {\it {UV Completions of Partial Compositeness: The Case for a
  SU(4) Gauge Group}},  {\em JHEP} {\bf 06} (2014) 142,
  [\href{https://arxiv.org/abs/1404.7137}{{\ttfamily 1404.7137}}].

\bibitem{Dugan:1984hq}
M.~J. Dugan, H.~Georgi, and D.~B. Kaplan, {\it {Anatomy of a Composite Higgs
  Model}},  {\em Nucl. Phys.} {\bf B254} (1985) 299--326.

\bibitem{Ayyar:2017qdf}
V.~Ayyar, T.~DeGrand, M.~Golterman, D.~C. Hackett, W.~I. Jay, E.~T. Neil,
  Y.~Shamir, and B.~Svetitsky, {\it {Spectroscopy of SU(4) composite Higgs
  theory with two distinct fermion representations}},  {\em Phys. Rev.} {\bf
  D97} (2018), no.~7 074505,
  [\href{https://arxiv.org/abs/1710.00806}{{\ttfamily 1710.00806}}].

\bibitem{Ayyar:2018zuk}
V.~Ayyar, T.~Degrand, D.~C. Hackett, W.~I. Jay, E.~T. Neil, Y.~Shamir, and
  B.~Svetitsky, {\it {Baryon spectrum of SU(4) composite Higgs theory with two
  distinct fermion representations}},  {\em Phys. Rev.} {\bf D97} (2018),
  no.~11 114505, [\href{https://arxiv.org/abs/1801.05809}{{\ttfamily
  1801.05809}}].

\bibitem{Ayyar:2018ppa}
V.~Ayyar, T.~DeGrand, D.~C. Hackett, W.~I. Jay, E.~T. Neil, Y.~Shamir, and
  B.~Svetitsky, {\it {Finite-temperature phase structure of SU(4) gauge theory
  with multiple fermion representations}},  {\em Phys. Rev.} {\bf D97} (2018),
  no.~11 114502, [\href{https://arxiv.org/abs/1802.09644}{{\ttfamily
  1802.09644}}].

\bibitem{Ayyar:2019exp}
V.~Ayyar, M.~F.~L. Golterman, D.~C. Hackett, W.~Jay, E.~T. Neil, Y.~Shamir, and
  B.~Svetitsky, {\it {Radiative Contribution to the Composite-Higgs Potential
  in a Two-Representation Lattice Model}},  {\em Phys. Rev.} {\bf D99} (2019),
  no.~9 094504, [\href{https://arxiv.org/abs/1903.02535}{{\ttfamily
  1903.02535}}].

\bibitem{Mykkanen:2012ri}
A.~Mykk{\"a}nen, M.~Panero, and K.~Rummukainen, {\it {Casimir scaling and
  renormalization of Polyakov loops in large-N gauge theories}},  {\em JHEP}
  {\bf 1205} (2012) 069, [\href{https://arxiv.org/abs/1202.2762}{{\ttfamily
  1202.2762}}].

\bibitem{Golterman:2015zwa}
M.~Golterman and Y.~Shamir, {\it {Top quark induced effective potential in a
  composite Higgs model}},  {\em Phys. Rev.} {\bf D91} (2015), no.~9 094506,
  [\href{https://arxiv.org/abs/1502.00390}{{\ttfamily 1502.00390}}].

\bibitem{Verbaarschot:2000dy}
J.~J.~M. Verbaarschot and T.~Wettig, {\it {Random matrix theory and chiral
  symmetry in QCD}},  {\em Ann. Rev. Nucl. Part. Sci.} {\bf 50} (2000)
  343--410, [\href{https://arxiv.org/abs/hep-ph/0003017}{{\ttfamily
  hep-ph/0003017}}].

\bibitem{Verbaarschot:1994qf}
J.~J.~M. Verbaarschot, {\it {The Spectrum of the QCD Dirac operator and chiral
  random matrix theory: The Threefold way}},  {\em Phys. Rev. Lett.} {\bf 72}
  (1994) 2531--2533, [\href{https://arxiv.org/abs/hep-th/9401059}{{\ttfamily
  hep-th/9401059}}].

\bibitem{Bruckmann:2008xr}
F.~Bruckmann, S.~Keppeler, M.~Panero, and T.~Wettig, {\it {Polyakov loops and
  spectral properties of the staggered Dirac operator}},  {\em Phys. Rev.} {\bf
  D78} (2008) 034503, [\href{https://arxiv.org/abs/0804.3929}{{\ttfamily
  0804.3929}}].

\bibitem{Follana:2006zz}
{\bf UKQCD, HPQCD} Collaboration, E.~Follana, C.~T.~H. Davies, and A.~Hart,
  {\it {Improved staggered eigenvalues and epsilon regime universality in
  SU(2)}},  {\em PoS} {\bf LAT2006} (2006) 051.

\bibitem{Boyle:2016lbp}
P.~A. Boyle, G.~Cossu, A.~Yamaguchi, and A.~Portelli, {\it {Grid: A next
  generation data parallel C++ QCD library}},  {\em PoS} {\bf LATTICE2015}
  (2016) 023.

\bibitem{DelDebbio:2008zf}
L.~Del~Debbio, A.~Patella, and C.~Pica, {\it {Higher representations on the
  lattice: Numerical simulations. SU(2) with adjoint fermions}},  {\em Phys.
  Rev.} {\bf D81} (2010) 094503,
  [\href{https://arxiv.org/abs/0805.2058}{{\ttfamily 0805.2058}}].

\bibitem{Musberg:2013foa}
S.~Musberg, G.~M{\"u}nster, and S.~Piemonte, {\it {Perturbative calculation of
  the clover term for Wilson fermions in any representation of the gauge group
  SU(N)}},  {\em JHEP} {\bf 05} (2013) 143,
  [\href{https://arxiv.org/abs/1304.5741}{{\ttfamily 1304.5741}}].

\bibitem{Bali:2013kia}
G.~S. Bali, F.~Bursa, L.~Castagnini, S.~Collins, L.~Del~Debbio, B.~Lucini, and
  M.~Panero, {\it {Mesons in large-N QCD}},  {\em JHEP} {\bf 06} (2013) 071,
  [\href{https://arxiv.org/abs/1304.4437}{{\ttfamily 1304.4437}}].

\bibitem{Leinweber:2004it}
D.~B. Leinweber, W.~Melnitchouk, D.~G. Richards, A.~G. Williams, and J.~M.
  Zanotti, {\it {Baryon spectroscopy in lattice QCD}},  {\em Lect. Notes Phys.}
  {\bf 663} (2005) 71--112,
  [\href{https://arxiv.org/abs/nucl-th/0406032}{{\ttfamily nucl-th/0406032}}].
  [,71(2004)].

\bibitem{Luscher:2010iy}
M.~L{\"u}scher, {\it {Properties and uses of the Wilson flow in lattice QCD}},
  {\em JHEP} {\bf 1008} (2010) 071,
  [\href{https://arxiv.org/abs/1006.4518}{{\ttfamily 1006.4518}}].

\bibitem{Ce:2016awn}
M.~C{\`e}, M.~Garc{\'{\i}}a~Vera, L.~Giusti, and S.~Schaefer, {\it {The
  topological susceptibility in the large-{$N$} limit of SU({$N$}) Yang-Mills
  theory}},  {\em Phys. Lett.} {\bf B762} (2016) 232--236,
  [\href{https://arxiv.org/abs/1607.05939}{{\ttfamily 1607.05939}}].

\bibitem{DeGrand:2015lna}
T.~DeGrand, Y.~Liu, E.~T. Neil, Y.~Shamir, and B.~Svetitsky, {\it {Spectroscopy
  of SU(4) gauge theory with two flavors of sextet fermions}},  {\em Phys.
  Rev.} {\bf D91} (2015) 114502,
  [\href{https://arxiv.org/abs/1501.05665}{{\ttfamily 1501.05665}}].

\bibitem{Lucini:2012gg}
B.~Lucini and M.~Panero, {\it {SU(N) gauge theories at large N}},  {\em Phys.
  Rept.} {\bf 526} (2013) 93--163,
  [\href{https://arxiv.org/abs/1210.4997}{{\ttfamily 1210.4997}}].

\bibitem{Gertov:2019yqo}
H.~Gertov, A.~E. Nelson, A.~Perko, and D.~G.~E. Walker, {\it {Lattice-Friendly
  Gauge Completion of a Composite Higgs with Top Partners}},  {\em JHEP} {\bf
  02} (2019) 181, [\href{https://arxiv.org/abs/1901.10456}{{\ttfamily
  1901.10456}}].

\bibitem{Katz:2005au}
E.~Katz, A.~E. Nelson, and D.~G.~E. Walker, {\it {The Intermediate Higgs}},
  {\em JHEP} {\bf 08} (2005) 074,
  [\href{https://arxiv.org/abs/hep-ph/0504252}{{\ttfamily hep-ph/0504252}}].

\bibitem{Bennett:2017kga}
E.~Bennett, D.~K. Hong, J.-W. Lee, C.~J.~D. Lin, B.~Lucini, M.~Piai, and
  D.~Vadacchino, {\it {Sp(4) gauge theory on the lattice: towards SU(4)/Sp(4)
  composite Higgs (and beyond)}},  {\em JHEP} {\bf 03} (2018) 185,
  [\href{https://arxiv.org/abs/1712.04220}{{\ttfamily 1712.04220}}].

\bibitem{Ayyar:2018glg}
V.~Ayyar, T.~DeGrand, D.~C. Hackett, W.~I. Jay, E.~T. Neil, Y.~Shamir, and
  B.~Svetitsky, {\it {Partial compositeness and baryon matrix elements on the
  lattice}},  {\em Phys. Rev.} {\bf D99} (2019), no.~9 094502,
  [\href{https://arxiv.org/abs/1812.02727}{{\ttfamily 1812.02727}}].

\bibitem{Witzel:2019jbe}
O.~Witzel, {\it {Review on Composite Higgs Models}},  in {\em {36th
  International Symposium on Lattice Field Theory (Lattice 2018) East Lansing,
  MI, United States, July 22-28, 2018}}, 2019.
\newblock \href{https://arxiv.org/abs/1901.08216}{{\ttfamily 1901.08216}}.

\end{thebibliography}\endgroup

\end{document}